\documentclass[aps,prb,preprint,superscriptaddress,nofootinbib]{revtex4-2}

\usepackage{amsmath,amssymb}
\usepackage{graphicx}
\usepackage{bm}
\usepackage[colorlinks=true,linkcolor=blue,citecolor=blue,urlcolor=blue]{hyperref}
\usepackage{orcidlink}
\usepackage{braket}
\usepackage{xcolor}
\newcommand{\bk}{\mathbf{k}}   \newcommand{\br}{\mathbf{r}}
\newcommand{\kk}{\mathbf{k}}
\newcommand{\qq}{\mathbf{q}}
\newcommand{\rr}{\mathbf{r}}
\newcommand{\kh}{\hat{\kk}}
\newcommand{\bs}{\bm{\sigma}}
\newcommand{\Tc}{T_{c}}
\newcommand{\avg}[1]{\left\langle #1 \right\rangle}
\newcommand{\half}{\tfrac12}
\newcommand{\Oc}{O}
\newcommand{\dd}{\mathbf{d}}

\begin{document}

\title{Disorder-tuned crossing of monopole and conventional pairing
instabilities in multi-Weyl semimetals}

\author{Enrique Mu\~noz~\orcidlink{0000-0003-4457-0817} }
\email{ejmunozt@uc.cl}
\affiliation{Facultad de F\'isica, Pontificia Universidad Cat\'olica de Chile, Vicu\~{n}a Mackenna 4860, Santiago, Chile}
\affiliation{Center for Nanotechnology and Advanced Materials CIEN-UC, Avenida Vicuña Mackenna 4860, Santiago, Chile}

\author{Rodrigo Soto-Garrido~\orcidlink{0000-0002-6593-5443} }
\email{rodsoto@uc.cl}
\affiliation{Facultad de F\'isica, Pontificia Universidad Cat\'olica de Chile, Vicu\~{n}a Mackenna 4860, Santiago, Chile}

\date{\today}

\begin{abstract}
{We study how non-magnetic impurity scattering affects the emergence and possible
coexistence of pairing instabilities in a two-node multi-Weyl semimetal.}
Within an explicitly specified projected impurity kernel---{valley diagonal and
momentum independent across each Fermi pocket, the leading behaviour of non-magnetic disorder
in the small-pocket window $q^{\max}_{\rm intra}\xi_{\rm dis}\ll1\ll|2\mathbf Q|\xi_{\rm dis}$,}
treated at leading order in Born and Abrikosov--Gor'kov theory---quenched disorder tunes the
leading pairing instability from a topologically nontrivial monopole channel to a
conventional ($s$-wave) one, {extending our earlier clean-system analysis into the
disordered regime}.
In an orbital-pseudospin model the inter-node monopole gap
$f_m=\cos\psi\,e^{iJ\phi}\propto(k_x+ik_y)^J$ arises from a charge-parity-dependent pairing
matrix, tied to the anti-unitarity $\Theta_J^2=-1$ (odd $J$) / $+1$ (even $J$). A Born
self-energy calculation in the chiral band basis then gives:
(i) {a band-isotropic conventional channel that is Anderson protected against
intra-node scalar disorder ($\eta_s=1$), introduced phenomenologically at the projected-band
level}; solving the competition for
arbitrary $\eta_s$ yields the crossing criterion
$(1-\eta_s)/(1-\eta_m)<T_{c0}^{(s)}/T_{c0}^{(m)}$, so the mechanism tolerates substantial
loss of conventional-channel protection; and
(ii) a rank-one monopole sector fixing $f_m$ as the exact eigenfunction with
{$\eta_m(J)=1/(J+2)$, exact given that kernel}. The crossing location in
$\Gamma_N/T_{c0}^{(m)}$ is set by $\eta_m(J)$, $\eta_s$, and $r=T_{c0}^{(s)}/T_{c0}^{(m)}$.
From the clean projected BdG Hamiltonian the pure monopole nodes carry Berry charge $\pm J$,
distinct from the gapped conventional solution; the clean nodal thermodynamics is
charge-dependent, $N_{\rm SC}(E)\propto E^{2/J}$ and $C\propto T^{1+2/J}$, and the residual
density of states shows a threshold only for $J=1$. The crossing lies in the moderately
metallic regime, {$\mu/\Gamma_N\simeq11$--$14$ for the illustrative
$T_{c0}^{(m)}/\mu=0.133$ used in the figures ($\simeq75$--$90$ at weak coupling
$T_{c0}^{(m)}/\mu=0.02$)}.
\end{abstract}

\maketitle


\section{Introduction}
Weyl and multi-Weyl semimetals are three-dimensional gapless topological phases
whose band-touching points act as monopoles of Berry curvature, carrying an
integer topological (monopole) charge $J$; the linear ($J=1$) Weyl node generalizes
to anisotropic double- and triple-Weyl nodes with $J=2,3$ at which the dispersion
is linear along one axis and grows as $k_\perp^{J}$ in the
plane, {thus violating the full Lorentz symmetry}~\cite{ArmitageMeleVishwanath2018}.
 
{These phases are neither hypothetical nor exotic. The Weyl semimetal was first
identified theoretically in the pyrochlore iridates~\cite{WanTurnerVishwanathSavrasov2011}
and in magnetically doped topological-insulator
multilayers~\cite{BurkovBalents2011}, and was then observed directly by
angle-resolved photoemission in the noncentrosymmetric monopnictide
TaAs~\cite{XuBelopolskiTaAs2015,LvTaAs2015}, following the material-specific
predictions of Refs.~\cite{WengFangFangBernevigDai2015,HuangTaAsClass2015}. The defining
consequences of the Berry monopole---the surface Fermi arcs, and the chiral anomaly
inherited from the Adler--Bell--Jackiw physics of continuum Weyl
fermions~\cite{NielsenNinomiya1983,SonSpivak2013}---are by now standard diagnostics of the
phase. Crucially for the present work, crystalline point-group symmetry permits monopole
charges larger than one: $C_{4,6}$ symmetry protects double-Weyl nodes with quadratic
in-plane dispersion and $C_6$ protects triple-Weyl nodes with cubic
dispersion~\cite{FangGilbertDaiBernevig2012}, with the ferromagnet
HgCr$_2$Se$_4$~\cite{XuWengWangDaiFang2011} and SrSi$_2$~\cite{HuangSrSi2_2016} identified
as candidate double-Weyl materials. Higher monopole charge is thus a genuine material
degree of freedom, and more broadly the classification of band degeneracies in crystals
extends well beyond the Dirac and Weyl cases~\cite{BradlynCano2016}. Throughout this work
$J$ enters quantitatively rather than as a label, through the anisotropy of the Fermi
surface it induces.}
 
{Superconductivity in Weyl and Dirac systems has correspondingly become a subject in its
own right. Proposals range from engineered
superconductor/topological-insulator multilayers realizing Bogoliubov--Weyl
nodes~\cite{MengBalents2012} and ferromagnetic superconductors with Majorana surface
arcs~\cite{SauTewari2012}, to intrinsic three-dimensional Dirac and Weyl
superconductors~\cite{YangPanZhang2014}, phonon-mediated pairing in doped Weyl
metals~\cite{BednikZyuzinBurkov2015}, time-reversal-invariant topological
pairing~\cite{HosurDaiFangQi2014}, and odd-parity
channels~\cite{WeiChaoAji2014}. What these treatments share is that the pairing must be
assembled from Bloch states carrying Berry curvature, so the admissible gap functions are
constrained by topology and not by point-group symmetry alone.}
 
Doping such a semimetal and, {at the same time,} adding an
attractive channel, produces a distinctive superconducting problem: because the
normal-state Bloch {eigenstates} inherit the Berry monopole, the natural Cooper pair
amplitudes (here the zero-momentum inter-node channel) are not ordinary spherical
harmonics but \emph{monopole harmonics}, carrying their own topological charge and an unavoidable nodal
structure~\cite{ChoBardarsonLuMoore2012,LiHaldane2018}. A microscopic theory of
the competition between such a topologically nontrivial ``monopole'' channel and
a conventional, fully gapped $s$-wave channel has been {previously developed by us} in the
\emph{clean} limit~\cite{MunozSotoGarridoJuricic2020,MunozEsparza2024,TapiaMunoz2026},
yielding explicit phase diagrams, critical temperatures, a ``topological
repulsion'' between the two order parameters, and specific-heat signatures that
can serve as a thermodynamic probe of the chirality index.
 
Real materials are disordered, and disorder {affects} the two channels very
differently. Anderson's theorem guarantees that the conventional {BCS}~\cite{Bardeen1957} $s$-wave state
is insensitive to weak non-magnetic disorder, while nodal and sign-changing order
parameters are generically fragile, their transition temperature obeying an
Abrikosov--Gor'kov pair-breaking
law~\cite{Anderson1959,AbrikosovGorkov1960,SigristUeda1991}. {This contrast is what makes disorder a potential \emph{tuning parameter} rather than
merely a nuisance. Non-magnetic defects that would be almost invisible in a conventional
$s$-wave material suppress $T_c$ dramatically in an unconventional one, a sensitivity that
has become one of the sharpest experimental discriminators of pairing
symmetry~\cite{Balatsky2006,AlloulBobroffGabayHirschfeld2009}, and the corresponding
pair-breaking theory is well developed for $d$-wave~\cite{Hirschfeld1993,Kogan2009},
strongly anisotropic~\cite{GolubovMazin1997}, and multiband $s_\pm$
gaps~\cite{Efremov2011}. The last of these is the closest precedent for what follows: there,
non-magnetic scattering alone can drive a transition between distinct pairing states, in a
setting with no topology involved.}
 
{The dichotomy is nevertheless} not
the whole story: when the pairing carries internal (orbital, sublattice, or
chiral) structure, a generalized Anderson theorem---formulated through the notion
of superconducting fitness---can shield even a nodal gap from scattering far in
excess of the gap scale, as established for topological-insulator-derived and
doped-semimetal superconductors~\cite{AndersenRamiresAndo2020,TimmonsPdTe2_2020}.
Whether the monopole channel of a multi-Weyl superconductor is protected,
pair-broken, or something in between is therefore a genuine question that must be
settled by the chiral structure of its Bloch states.
 
A second, independent disorder effect complicates the picture. The multi-Weyl
\emph{normal} state is itself disorder-sensitive: beyond a critical strength it
undergoes a non-Anderson, disorder-driven transition to a diffusive metal in
which the zero-energy density of states turns on as an order
parameter~\cite{BeraSauRoy2016,SyzranovRadzihovsky2018,KlierGornyiMirlin2019}, {and where
disorder and interactions are strongly intertwined~\cite{Serbyn2013}}, a transition recently
observed by ARPES through the disappearance of the topological Fermi arcs in
NdAlSi~\cite{LiNdAlSi2025}. Higher monopole charge makes this normal-state
criticality \emph{more} easily reached, for instance, a double-Weyl node already yields a
finite residual density of states at weak
disorder~\cite{BeraSauRoy2016}. Any claim of a disorder-driven \emph{pairing}
transition must therefore demonstrate that it occurs while the host is still a
good metal, well below the scale at which the underlying Weyl
band-structure itself dissolves.
 
In this work, {we pose the following} question:
\emph{does scalar disorder select pairing topology} in a multi-Weyl
superconductor, tuning a \emph{crossing} of the monopole and conventional
linearized pairing instabilities?
{Our answer turns out to be affirmative, as exposed by the arguments presented in this article.} {Here is a summary of the main results}: (i) From a single Born self-energy calculation in the chiral band
basis, we obtain the pair-breaking asymmetry: \emph{conditional on the single-band
anomalous-kernel assumption} (Sec.~\ref{sec:model}), the conventional channel is
Anderson protected because the Bloch Berry phase that dresses the normal
scattering rate cancels against the anomalous self-energy, while the monopole
channel is pair-broken at a rate softened by the chiral suppression of
backscattering, with a protection factor $\eta_m(J)=1/(J+2)$ on the
Fermi surface, fixed \emph{exactly} by
the rank-one $m=\pm J$ sector of the chiral coherence kernel (within the projected
intra-node scalar model). (ii) A
pair-breaking Abrikosov--Gor'kov (AG) solution, with a separately evaluated {self-consistent Lorentzian spectral broadening of the} density of states, locates the instability crossing, which moves to a smaller pair-breaking rate $\Gamma_N^\ast/T_{c0}^{(m)}$ with increasing monopole charge $J$, and supplies
its thermodynamic fingerprints (specific-heat, residual density of states).
(iii) The two competing { pairings possess} distinct nodal topology---the
monopole solution carries BdG nodal charges $\pm J$, {whereas}
the conventional solution is fully gapped. (iv) A
disorder-symmetry classification shows that the selection is specific to
intra-node potential disorder within the projected kernel; the operative requirement is differential pair-breaking (Eq.~\eqref{eq:crosscrit}) rather than the disorder label, and magnetic or chirality-mixing disorder requires a separate vertex calculation (Sec.~\ref{sec:dissym}). {(v) A continuum Ginzburg--Landau estimate disfavors simple homogeneous coexistence,
but does not fix the order of the ordered-state transition, which we leave open
(App.~\ref{app:gl}).} (vi) Throughout, the transition
sits in the moderately metallic regime, {$\mu/\Gamma_N\simeq11$--$14$ for the enhanced
illustrative parameter set used in the figures and appreciably larger for weaker coupling
(Sec.~\ref{sec:scba})}, separated from the Weyl
disorder-criticality of the normal state. We deliberately
quote \emph{no} residual-resistivity or impurity-concentration values
(Sec.~\ref{sec:anchor}); the only material-level target given is the conditional
quantum lifetime $\tau_q=\hbar/(2\Gamma_N^\ast)$. The growing
list of Weyl superconductors with resolved topological surface and pairing
structure---most recently the superconducting Fermi arcs of
PtBi$_2$~\cite{KuibarovBorisenko2024}---makes this an {experimentally testable
proof-of-principle}
prediction.
\section{Model}\label{sec:model}
We consider a pair of multi-Weyl nodes of opposite chirality at $\pm\mathbf{Q}$
with monopole charge $J\in\{1,2,3\}$. Measuring momenta from each node,
\begin{equation}
h_\pm(\qq) = \pm v_z q_z\,\sigma_z
            + \alpha\big(q_-^{\,J}\sigma_+ + q_+^{\,J}\sigma_-\big),
\qquad q_\pm = q_x \pm i q_y ,
\label{eq:hpm}
\end{equation}
with $E_\pm(\qq)=\pm\sqrt{v_z^2 q_z^2 + \alpha^2 q_\perp^{2J}}$, $\sigma_\pm=(\sigma_x\pm i\sigma_y)/2$ (where $\sigma_j$, with $j=x,y,z$, are the Pauli matrices) and low-energy
density of states (DOS) \emph{per node and per pseudospin band}
\begin{equation}
N(\varepsilon) \propto |\varepsilon|^{2/J},
\qquad
N_0 \equiv N(\mu) = N_J\,|\mu|^{2/J},
\label{eq:dos}
\end{equation}
{where the coefficient $N_J = \frac{\alpha^{-2/J}}{J\, v_z (2\pi)^2}B\left(\frac{1}{J},\frac{1}{2}\right)$, for $B(x,y)$ the Beta function (see Appendix~\ref{app:dos} for details).}

{\paragraph*{Density-of-states multiplicities.} Because several distinct densities of
states could be meant here, we fix the convention once and use it throughout. $N(\varepsilon)$
in Eq.~\eqref{eq:dos} is the DOS \emph{of a single node, for the single conduction
pseudospin band}, and $N_0\equiv N(\mu)$ is its value at the Fermi level. This same $N_0$ is
the one that enters (i) the dimensionless couplings $\lambda_a=g_aN_0\avg{|f_a|^2}$ of
Eq.~\eqref{eq:tcclean} and (ii) the Born rates $\Gamma_b$ and $\Gamma_N$ of
Sec.~\ref{sec:coh}, since both the pairing and the intra-node impurity self-energy are
evaluated on one node's conduction Fermi surface. The \emph{total} normal-state DOS of the
two-node system is $N_{\rm tot}=2N_0$ (times an additional factor of two if physical spin is
retained as a degenerate spectator), and it is $N_{\rm tot}$---not $N_0$---that would enter a
measured Sommerfeld coefficient. We use $N_0$ everywhere below and flag explicitly the one
place, the normal-state specific heat $C_n$ of App.~\ref{app:cv}, where the multiplicity
cancels in the ratio $\Delta C/C_n$ and is therefore immaterial.}

\paragraph*{Impurity model: smooth, intra-node-dominated disorder.} {The disorder is a
Gaussian, zero-mean, nonmagnetic random potential specified by its correlator in momentum
space,}
\begin{equation}
{\overline{V(\rr)}=0,\qquad
\overline{V_{\qq}\,V_{\qq'}}=(2\pi)^3\,\delta(\qq+\qq')\,W(\qq),}
\label{eq:correlator}
\end{equation}
{where $W(\qq)=n_{\rm imp}|V(\qq)|^2\ge0$ is the disorder power spectrum of scatterers of
density $n_{\rm imp}$, treated in the Born limit. We take the potential to be \emph{smooth
on the scale of the internodal separation}, i.e.\ of finite range $\xi_{\rm dis}$ with
$\xi_{\rm dis}\,|2\mathbf Q|\gg1$, so that}
\begin{equation}
{\frac{W(2\mathbf Q)}{W(0)}\ll1 .}
\label{eq:smoothcond}
\end{equation}
{Equation~\eqref{eq:smoothcond} suppresses inter-node scattering, but by itself it is
\emph{not} enough to justify the projected kernel used below. A correlator smooth enough to
kill $2\mathbf Q$ processes is, generically, also forward-peaked \emph{within} a single
pocket, in which case the intra-node kernel would carry an extra factor
$W(\kk_F-\kk_F')/W(0)$, the normal eigenvalue would no longer be $\kappa_0=\tfrac12$, and
$f_m$ would not remain an exact eigenfunction. A second condition is therefore required:
the correlation length must be short compared with the \emph{intra-pocket} momentum
transfers, so that $W(\kk_F-\kk_F')\simeq W(0)$ across one Fermi surface. The two
requirements together define a \emph{two-scale window},}
\begin{equation}
{q^{\max}_{\rm intra}\,\xi_{\rm dis}\ll1\ll |2\mathbf Q|\,\xi_{\rm dis},}
\label{eq:twoscale}
\end{equation}
{where $q^{\max}_{\rm intra}$ is the largest momentum transfer between two states on the
\emph{same} pocket and $2\mathbf Q$ is the actual (vector) node separation. We take the
nodes at $\pm Q\hat{\mathbf z}$, i.e.\ separated along the rotation axis, as in the
standard multi-Weyl realizations; the condition is then purely geometric,
$q^{\max}_{\rm intra}\ll2Q$. Because the pocket is anisotropic there is no single $k_F$,
and it is bounded by the two Fermi momenta}
\begin{equation}
{k_{F,z}=\frac{\mu}{v_z},\qquad
k_{F,\perp}=\Big(\frac{\mu}{\alpha}\Big)^{1/J},\qquad
q^{\max}_{\rm intra}\le2\sqrt{k_{F,z}^2+k_{F,\perp}^2}\;,}
\label{eq:qFS}
\end{equation}
{which we use only as an order-of-magnitude scale, not as a sharp bound. The window
Eq.~\eqref{eq:twoscale} therefore requires simply}
\begin{equation}
{k_{F,z}\ll Q\qquad\text{and}\qquad k_{F,\perp}\ll Q ,}
\label{eq:smallpocket}
\end{equation}
{i.e.\ Fermi pockets small compared with the node separation---a low-doping,
small-pocket regime. We emphasize that Eq.~\eqref{eq:smallpocket} is a statement about
\emph{momenta}, not about energies: with $\mathbf Q\parallel\hat{\mathbf z}$ the combination
$\alpha Q^{J}$ is not an inter-valley energy of the band structure. Which of the two conditions in Eq.~\eqref{eq:smallpocket} binds more tightly depends on
the material parameters $v_z$, $\alpha$ and $Q$ and on the doping. If instead $\mathbf Q$ points in a generic direction, the inequalities
should be imposed component-wise on $Q_z$ and $Q_\perp$, or directly as
$q^{\max}_{\rm intra}\ll|2\mathbf Q|$.}

{Within this window the disorder acts, to leading order, as a \emph{valley-diagonal,
intra-pocket momentum-independent} random potential, and it is precisely this projected
ensemble---not generic smooth disorder---for which the kernel Eq.~\eqref{eq:kernel} and the
eigenvalue $\eta_m=1/(J+2)$ are exact. Two departures are left uncomputed and are not
claimed: generic short-range disorder ($W(2\mathbf Q)\sim W(0)$), which requires the full
valley-space anomalous ladder (Sec.~\ref{sec:dissym}); and generic
forward-peaked disorder ($q^{\max}_{\rm intra}\xi_{\rm dis}\gtrsim1$), which requires diagonalizing
the finite-range kernel $W(\kk_F-\kk_F')\,|\Oc|^2/W(0)$ and would change $\kappa_0$,
$\eta_s$ and $\eta_m$ as functions of $\xi_{\rm dis}$. With Eq.~\eqref{eq:twoscale} the
intra-node Born rate is set by the zero-momentum weight, $\Gamma_b=\pi N_0W(0)$; we write
$W(0)\equiv n_{\rm imp}|V|^2$ throughout, understanding $n_{\rm imp}|V|^2$ as the constant
intra-pocket value in this scale-separated limit, so that the rate conventions of
Eq.~\eqref{eq:ratelist} retain their familiar form.}

{A further distinction should be kept in mind, since the two are easily conflated: the
\emph{internal matrix structure} of the impurity potential (valley-even $\tau_0$,
valley-odd $\tau_z$, intervalley $\tau_{x,y}$) and the \emph{momentum transfer} it
supplies are independent properties. A valley-even scalar potential $V(\rr)\tau_0$ has
inter-node matrix elements whenever its Fourier support extends to $|\qq|\simeq2Q$; it is
Eq.~\eqref{eq:smoothcond}, restricting that support, and not the matrix structure alone,
that suppresses inter-node processes here.}

{Superconductivity is introduced in the four-component Nambu spinor built from the
\emph{orbital-pseudospin} states of the two nodes,}
$\Psi_\kk=(c_{+,\kk 1},c_{+,\kk 2},
c^\dagger_{-,-\kk 1},c^\dagger_{-,-\kk 2})^{\!\top}$, {where $1,2$ label the orbital
(sublattice) pseudospin components of Eq.~\eqref{eq:hpm} and \emph{not} physical spin,
{which is deliberately omitted for notational simplicity} throughout the main text; it is reintroduced explicitly, with matrices
$s_i$, only in App.~\ref{app:spin_singlet_realization}.} This gives the
BdG block
\begin{equation}
\mathcal{H}_{\rm BdG}(\kk)=
\begin{pmatrix}
h_+(\kk)-\mu & \hat{\Delta}_a(\kk)\\[4pt]
\hat{\Delta}_a^\dagger(\kk) & -\,h_-^\ast(-\kk)+\mu
\end{pmatrix},
\qquad
{\hat{\Delta}_m(\kk)=\Delta_m\,M_m^{(J)} .}
\label{eq:bdg}
\end{equation}
{The two pairing channels enter this block on a different footing, and we write them
differently. For the
\emph{monopole} channel the gap is the constant, charge-parity-dependent pseudospin matrix
$M_m^{(J)}$ of Eq.~\eqref{eq:pairmat}, whose conduction-band projection generates the form
factor $f_m$; this is a genuine matrix in the orbital space. For the \emph{band-isotropic
conventional} channel we do \emph{not} posit a constant matrix $M_s^{(J)}$: as discussed
below and in App.~\ref{app:orbital_nogo}, that channel is defined directly at the projected-band
level by its form factor $f_s(\kh)=1$, entering the linearized gap equation through
$\Delta_sf_s(\kh)$ rather than through Eq.~\eqref{eq:bdg}. Where a gap matrix is needed for
it, the momentum-dependent, gauge-covariant band-sewing construction
$M_s(\kk)=u_+(\kk)u_-^{T}(-\kk)$ of App.~\ref{app:band_sewing_gap} is used.}
The two-component structure of Eq.~\eqref{eq:hpm} is an orbital (sublattice)
pseudospin, not physical spin. Because the Cooper pair is \emph{inter-node}
(node ``$+$" at $\kk$, with node ``$-$" at $-\kk$), the pairing lives off-diagonal in
node space; in the full node$\,\otimes~$pseudospin space the gap is
$\hat\Delta_m(\kk)=\tau_+\!\otimes\!\delta_m(\kk)+\tau_-\!\otimes\!\delta'_m(\kk)$ with
$\delta_m(\kk)=\Delta_m\,g(\kk)\,M_m^{(J)}$, and fermionic antisymmetry fixes
$\delta'(\kk)=-\delta^{\rm T}(-\kk)$, i.e.\ $\hat\Delta_m(\kk)=-\hat\Delta_m^{\rm T}(-\kk)$
{(see App.~\ref{app:conventional_sewing})}. The pseudospin pairing matrix that yields
each conduction channel is \emph{charge-parity dependent}: projecting onto the conduction bands,
$f_a(\kh)=u_+^{\dagger}(\kk)\,M_a^{(J)}\,u_-^{*}(-\kk)$, one finds the monopole gap
$f_m=\cos\psi\,e^{iJ\phi}\propto(k_x+ik_y)^J$ from
\begin{equation}
M_m^{(J)}=\begin{cases}{i\sigma_y},& J\ \text{odd},\\[2pt] {\sigma_x},& J\ \text{even},\end{cases}
\label{eq:pairmat}
\end{equation}
{whereas the complementary matrix projects to \emph{zero}: $\sigma_x$ gives zero for odd
$J$, and $i\sigma_y$ gives zero for even $J$ (App.~\ref{app:spinors}). This assignment is
the one required by the antiunitary bookkeeping of Eq.~\eqref{eq:antiunit}: the
antisymmetric $i\sigma_y$ pairs $\Theta_J$-partners precisely when $\Theta_J^2=-1$, i.e.\
for odd $J$.} For the conventional channel, we take the isotropic nodeless conduction-band gap
$f_s(\kh)=1$ as an \emph{effective-model input}, inserted at the projected level
(in contrast to the monopole gap, which we derived microscopically above from
$M_m^{(J)}$). We are explicit that this is an asymmetry of the construction: we do
not claim the conventional channel arises from the same orbital-pseudospin
projection. The single-band Abrikosov--Gor'kov treatment used throughout dresses
both the normal and the anomalous self-energy with one coherence factor
$|\Oc|^2=\tfrac12(1+\dd\!\cdot\!\dd')$. Within it, the results are exact: a
\emph{constant} $f_s=1$ is the top eigenfunction with $\eta_s=1$ identically (for
$f_s=1$ the normal- and anomalous-line averages coincide term by term), and the
monopole $f_m$ occupies the rank-one $m=\pm J$ sector with $\eta_m=1/(J+2)$.

Through out this work we assume that the conventional
projected channel is Anderson protected, $\eta_s=1$, i.e.\ that its anomalous
impurity vertex equals the normal one (see App.~\ref{app:conventional_sewing} for details). 
Our
assumption is precisely that the analogous node-space contraction reduces to
$|\Oc|^2$. 
{We state once, here, the logical standing of
each ingredient; 
\emph{(i) Assumption-free.} Results that follow from the clean projected Hamiltonian
alone: the angular moments, the nodal laws $N_{\rm SC}\propto E^{2/J}$ and
$C\propto T^{1+2/J}$, the clean specific-heat jumps, and the pure-state BdG nodal charges
of magnitude $J$.
\emph{(ii) Exact given the projected kernel.} $\eta_m(J)=1/(J+2)$ is the exact rank-one
eigenvalue of the valley-diagonal, intra-pocket momentum-independent kernel
Eq.~\eqref{eq:kernel} in the sector $m=\pm J$.
\emph{(iii) Assumed.} The equality of the normal and anomalous coherence kernels, hence
$\eta_s=1$. App.~\ref{app:conventional_sewing} constructs a band-sewing gap matrix
$M_s(\bk)=u_+(\bk)u_-^{T}(-\bk)$ with $f_s\equiv1$ and exact fermionic antisymmetry, and
derives a \emph{sufficient} condition---that the impurity potential preserves the anti-unitary
sewing relation, $\mathcal A\widehat V_{\bk\bk'}\mathcal A^{-1}=\widehat V_{-\bk,-\bk'}$,
which holds for valley-even scalar disorder and fails for valley-odd ($\tau_z$) and
intervalley ($\tau_{x,y}$) components. We do \emph{not} claim a local orbital realization:
no momentum-independent orbital sewing matrix exists for Eq.~\eqref{eq:hpm}
(App.~\ref{app:orbital_nogo}), and a spectator spin singlet does not supply one
(App.~\ref{app:spin_singlet_realization}). The band-isotropic channel is therefore a
projected-band input, $\eta_s$ is treated as a parameter, and the consequences of
$\eta_s<1$ are quantified by Eq.~\eqref{eq:crosscrit}.}

{Let us now analyze the anti-unitarity}. Writing the transverse terms as
$\alpha\,\mathrm{Re}(k_x{+}ik_y)^J\sigma_x+\alpha\,\mathrm{Im}(k_x{+}ik_y)^J\sigma_y$
and using $\mathcal{H}_0(\kk)=\mathrm{diag}\big(h_+(\kk),h_-(\kk)\big)$, the
operator {$\Theta_J=\sigma_{a(J)}K$, for
\begin{equation}
a(J)=\begin{cases}y,& J\ \text{odd}\\ x,& J\ \text{even}\end{cases}
\end{equation}
satisfies the properties
\begin{equation}
\Theta_J\,h_\chi(\kk)\,\Theta_J^{-1}=h_\chi(-\kk),\quad \Theta_J^2=\begin{cases}-1,& J\ \text{odd}\\ +1,& J\ \text{even}\end{cases}
\label{eq:antiunit}
\end{equation}
and hence is an exact anti-unitary} symmetry of \emph{each} node ($\chi=\pm$) for
$J=1,2,3$. {(For an anti-unitary $\Theta=U\mathcal K$, $\Theta^2=UU^{*}$ is
phase-convention independent, and $\sigma_y\sigma_y^{*}=-1$ while
$\sigma_x\sigma_x^{*}=+1$.)} The Kramers-like square for odd $J$
is the origin of the parity dependence of $M_m^{(J)}$, since
the antisymmetric $i\sigma_y$ pairs $\Theta$-partners only when $\Theta^2=-1$. We use $\Theta_J$ only to organize the
disorder classes: intra-node scalar disorder $\tau_{0,z}\otimes\sigma_0$ preserves
$\Theta_J$, while chiral $\sigma_z$ disorder does not.
{An important distinction must be drawn here, since $\Theta_J$ is a
\emph{within-node} pseudospin anti-unitary, whereas the operator that controls the
conventional-channel protection is the \emph{pairing} antiunitary $\mathcal A$ that
exchanges the two paired nodes (App.~\ref{app:conventional_sewing}). These are not
the same operator, and they classify disorder differently: the valley-even scalar
potential $\tau_0\otimes\sigma_0$ is invariant under both, but the valley-odd
$\tau_z\otimes\sigma_0$, while it preserves $\Theta_J$, obeys
$\mathcal A\,\widehat V_z\,\mathcal A^{-1}=-\widehat V_z$ and therefore does
\emph{not} protect the conventional channel. Because it is $\mathcal A$ and not
$\Theta_J$ that enters the anomalous vertex, we accordingly restrict the exact
protection claim to \emph{valley-even} intra-node scalar disorder
$\tau_0\otimes\sigma_0$ (Sec.~\ref{sec:coh} and
App.~\ref{app:conventional_sewing}); for $\tau_z$ disorder one expects
$\eta_s<1$ and $\Gamma^{\rm pb}_s>0$.}
Two pairing channels, distinguished by the form factor $f_a$:
a {\emph{band-isotropic conventional} channel} $f_s(\kh)=1$, $\avg{f_s}_{\rm FS}=1$, fully gapped;
and a topologically nontrivial \emph{monopole} inter-node channel with
$\avg{f_m}_{\rm FS}=0$ and point nodes. For a genuine multi-Weyl pair the nodes
carry Chern charges $C_\pm=\pm J$, so the zero-center-of-mass inter-node pairing
bundle has monopole charge $q=(C_+-C_-)/2=J$, {following the standard inter-node
monopole-pairing construction~\cite{LiHaldane2018,MunozSotoGarridoJuricic2020}; the
consequence is that the pair amplitude is a section of a charge-$J$ monopole bundle and is
therefore expanded in monopole harmonics rather than ordinary spherical harmonics, which is
why $f_m$ cannot be a nonvanishing constant. We do not rely on this bundle-level bookkeeping
anywhere below: the topological statement we actually use is the \emph{local} BdG nodal
charge, derived from the Bogoliubov Hamiltonian itself by a degree argument in
Sec.~\ref{sec:topo}, which is both rigorous and free of inter-node gauge conventions.} The corresponding lowest inter-node
monopole gap is
\begin{equation}
f_m(\kh)=\cos\psi\,e^{iJ\phi}
=\frac{\alpha\,(k_x+ik_y)^J}{\mu}\qquad
(\,\to\sin\theta\,e^{i\phi}\ \text{for the isotropic }J=1\text{ node}\,),
\label{eq:fm}
\end{equation}
obtained by projecting the constant inter-node pairing with the
\emph{charge-parity-dependent} matrix $M_m^{(J)}$ of Eq.~\eqref{eq:pairmat} onto the
conduction band,
$f_m(\kk)\propto u_+^{\dagger}(\kk)\,M_m^{(J)}\,u_-^{*}(-\kk)$. Explicitly (with
conduction spinors $u_\pm$ of the two nodes): {for odd $J$, $M_m^{(J)}=i\sigma_y$ gives
$|u_+^\dagger (i\sigma_y) u_-^*|=\cos\psi$ with azimuthal winding $J$ (while $\sigma_x$
projects to \emph{zero}); for even $J$, $M_m^{(J)}=\sigma_x$ gives the same
$\cos\psi$ and winding $J$ (and $i\sigma_y$ projects to zero). The explicit spinors and the
evaluation are given in App.~\ref{app:spinors}.} {Two facts about this gap, both established below, are central}.
(i) One must distinguish the \emph{pseudospin texture degree} from the
\emph{pairing line-bundle Chern number}. As a section of the Bloch bundle,
$f_m=\cos\psi\,e^{iJ\phi}$ resembles a degree-one texture in the abstract
$\hat{\mathbf d}$ frame, but the physical object is the pairing line bundle over
the momentum Fermi surface, whose obstruction is set by the $J$-fold winding
$f_m\propto(k_x+ik_y)^J$. {Correspondingly, the Berry charge on a small sphere
enclosing each BdG point node is $\pm J$: this \emph{is} the genuine charge-$J$ monopole
gap, with $\mathcal{C}_{\rm BdG}=J$ (Sec.~\ref{sec:topo}). We write ``correspondingly''
rather than ``equivalently'' deliberately. The two are distinct invariants on distinct base
spaces---the pairing-bundle Chern class is carried by the pair section over the Fermi
surface, whereas $\mathcal{C}_{\rm BdG}$ is a Chern number of Bogoliubov eigenstates over a
small momentum sphere around one node---and they are related as two manifestations of the
same $J$-fold winding obstruction rather than as the same object. Only the local BdG charge
is used below, and it is established independently by the degree argument of
Sec.~\ref{sec:topo}.} (ii) $f_m$ is not a variational trial state but the
\emph{exact} eigenfunction of the projected impurity ladder in the rank-one
azimuthal sector $m=\pm J$, with a closed-form pair-breaking eigenvalue
$\eta_m(J)=1/(J+2)$ (Sec.~\ref{sec:coh}). The remaining modeling input is the
pairing symmetry class itself, an inter-node channel with the parity-dependent
matrix $M_m^{(J)}$ and the antiunitary $\Theta_J$ of Eq.~\eqref{eq:antiunit},
classified in the full node$\,\otimes~$pseudospin space rather than by the
pseudospin matrix alone, not the choice of harmonic. All
Fermi-surface averages use the single measure
$\avg{\cdots}_{\rm FS}=\big[\!\int\! d\psi\,(\cos\psi)^{2/J-1}(\cdots)\big]/
\big[\!\int\! d\psi\,(\cos\psi)^{2/J-1}\big]$ (see Appendix~\ref{app:angavg}), with $\psi\in[-\tfrac\pi2,\tfrac\pi2]$. We use the \emph{unnormalized} convention
$f_s=1$, $f_m=\cos\psi\,e^{iJ\phi}$ throughout (so $\Delta_a$ is the gap amplitude,
the monopole gap magnitude being $\Delta_m\cos\psi$), with $\avg{|f_s|^2}=1$,
$\avg{|f_m|^2}=2/(J{+}2)$, and $\avg{|f_m|^4}=(2/J)(2/J{+}2)/[(2/J{+}1)(2/J{+}3)]$ {(see Appendix~\ref{app:angavg} for details)}; effective couplings $\lambda_a\equiv g_a N_0\avg{|f_a|^2}$
absorb this measure. This single convention is used in the gap, AG, heat-capacity,
and residual-DOS equations and in all plots. In the clean weak-coupling limit each
channel obeys
\begin{equation}
T_{c0}^{(a)}=\displaystyle\frac{2e^{\gamma_E}}{\pi}\,\Lambda\,e^{-1/\lambda_a},
\qquad \lambda_a=g_a N_0\avg{|f_a|^2}_{\rm FS},
\label{eq:tcclean}
\end{equation}
so that the regime of interest,
\begin{equation}
T_{c0}^{(m)}>T_{c0}^{(s)}\ \Longleftrightarrow\ \lambda_m>\lambda_s
\ \Longleftrightarrow\ g_m\avg{|f_m|^2}>g_s\avg{|f_s|^2},
\label{eq:premise}
\end{equation}
is a condition on the effective channel couplings $\lambda_a$, and the effective couplings $\lambda_a=g_aN_0\avg{|f_a|^2}$ differ from the bare attractions through the second moment $\avg{|f_a|^2}$ (which is $1$ for $s$--wave but $2/(J{+}2)$ for the monopole), so a given difference in $\lambda$ is not the same difference in microscopic pairing strength. {The asymmetry needed is modest. For the couplings
\emph{actually used} in the numerics: at the
gap-equation cutoff $\Lambda=10\,\mu$, inverting Eq.~\eqref{eq:tcclean} for the clean
scales $T_{c0}^{(m)}=0.133\,\mu$ and $T_{c0}^{(s)}=0.083\,\mu$ gives}
\begin{equation}
{\lambda_m\simeq0.225,\qquad \lambda_s\simeq0.203,}
\label{eq:lambdavals}
\end{equation}
{i.e.\ a $\simeq11\%$ stronger monopole coupling, which yields
$T_{c0}^{(m)}/T_{c0}^{(s)}=1.60$ and hence $r=0.625$. These are the values used in
all figures and tables below.} This is a \emph{chosen,
plausible} regime, whenever pairing is dominated by inter-node, large-momentum
($\sim\!2\mathbf{Q}$) processes: the chiral Bloch structure projects such
processes onto the charge-$J$ monopole harmonic, while the
conventional $s$-wave channel is additionally weakened by the Coulomb
pseudo-potential $\mu^\ast$, which is expected to suppress $\lambda_s$ but not the
higher-$\ell$ monopole channel, as in unconventional superconductors. 
This is the clean-limit regime established for monopole pairing in multi-Weyl
systems by
Refs.~\cite{MunozSotoGarridoJuricic2020,MunozEsparza2024,TapiaMunoz2026}; we take
it as the starting point and ask how disorder resolves the competition.

\section{Disorder self-energy in the chiral basis: Anderson protection
and monopole pair-breaking}\label{sec:coh}
We treat both protection and pair-breaking on the same footing by computing the
Born self-energy in the chiral (conduction-band) basis. For $\mu>0$ each node
contributes a conduction Fermi surface with Bloch spinor $|u(\kh)\rangle$ (with pseudospin texture $\dd(\kh)$). Scalar disorder $V(\rr)\sigma_0$ scatters band states with the
chiral coherence factor
\begin{equation}
\Oc(\kh,\kh') = \langle u(\kh)|u(\kh')\rangle,
\qquad
|\Oc(\kh,\kh')|^2 = \tfrac12\big(1+\dd\!\cdot\!\dd'\big),
\label{eq:coh}
\end{equation}
where $\dd(\kh)$ is the pseudospin (Anderson) unit vector, \emph{not} the Euclidean
momentum direction; $\Oc\to0$ at $\dd'=-\dd$ (a \emph{pseudospin} reversal, which
for the multi-Weyl texture need not coincide with physical momentum
backscattering). {Within this projected single-band kernel the eigenvalues below are exact and
parity-independent. Both $\eta_s=1$ and $\eta_m=1/(J+2)$ are properties of the
\emph{same} kernel Eq.~\eqref{eq:kernel}: the first is an input (Sec.~\ref{sec:model};
a sufficient condition for it is derived in App.~\ref{app:conventional_sewing}), the
second the exact rank-one eigenvalue of that input (for the valley-diagonal, intra-pocket momentum-independent kernel defined by the
two-scale window Eq.~\eqref{eq:twoscale}).}

\paragraph*{Self-consistent Born equations.}
{We use the \emph{normalized} quasiclassical propagator, $\check g^2=\check 1$, so that
the density of states appears once and only once in the self-energy:}
\begin{equation}
{\check g(\kh',i\omega_n)=\frac{i\tilde\omega_n\tau_0
+\tilde\Delta(\kh')\tau_1}{\sqrt{\tilde\omega_n^2+\tilde\Delta(\kh')^2}}\,,}
\label{eq:gnorm}
\end{equation}
(written in a \emph{local gauge patch}; globally the complex gap occupies $\tau_1,\tau_2$, with a momentum-dependent connection, which does not affect the angle-averaged rates, and the $\tau_3$ term integrates to zero). {With this normalization the disorder self-energy is}
\begin{equation}
\check\Sigma(\kh,i\omega_n)={\pi N_0\,n_{\rm imp}|V|^2}\!\int\! d\mu_{\rm FS}(\kh')\,
\mathcal{C}(\kh,\kh')\,\check g(\kh',i\omega_n){{}=\Gamma_b\!\int\! d\mu_{\rm FS}(\kh')\,
\mathcal{C}(\kh,\kh')\,\check g(\kh',i\omega_n)},
\label{eq:Sigma}
\end{equation}
{with $\Gamma_b=\pi N_0W(0)$ the bare rate of Eq.~\eqref{eq:ratelist}, {and $d\mu_{\rm FS}(\kh')$ is the anisotropic density-of-states measure
$\propto(\cos\psi')^{2/J-1}d\psi'\,d\phi'/2\pi$ (see Appendix~\ref{app:angavg}); it reduces
to $d\kh'/4\pi$ only for the isotropic $J=1$ node.} We fix the
normalization convention once here: $\check g$ is dimensionless and obeys
$\check g^2=\check 1$, the density of states appears solely through the explicit prefactor
$\pi N_0$, and in the normal state $\check g\to i\,{\rm sgn}(\omega_n)\tau_0$. Other
common conventions absorb a factor $-i\pi N_0$ into $\check g$; the rates below are
unchanged provided the factor is counted once.}

The crucial point is the coherence structure $\mathcal{C}$. The normal (particle)
line carries $\Oc(\kh,\kh')$ and the anomalous (Cooper) line pairs the
$\Theta$-conjugate partner, which carries $\Oc^\ast(\kh,\kh')$. \emph{Under the
single-band anomalous-kernel assumption} (stated in Sec.~\ref{sec:model}: the
conventional channel's anomalous vertex equals its normal vertex), both the
$\tau_0$ and $\tau_1$ channels are weighted by the same real kernel
\begin{equation}
\mathcal{C}(\kh,\kh')=|\Oc(\kh,\kh')|^2=\tfrac12\big(1+\dd(\kh)\!\cdot\!\dd(\kh')\big),
\label{eq:kernel}
\end{equation}
\paragraph*{Spectral decomposition.}
{Let us first consider the \emph{isotropic $J=1$} node} (the anisotropic $J>1$ kernel is
treated below), expanding the kernel in Legendre polynomials,
$\mathcal{C}(\kh\!\cdot\!\kh')=\tfrac12 P_0+\tfrac12 P_1$. Its eigenvalues $\kappa_\ell$ (kernel eigenvalues, distinct from the pairing couplings $\lambda_a$) on angular-momentum-$\ell$ form factors are $\kappa_\ell=c_\ell/(2\ell+1)$, i.e.
\begin{equation}
\kappa_0=\tfrac12,\qquad \kappa_1=\tfrac16,\qquad \kappa_{\ell\ge2}=0 ,
\label{eq:lambdas}
\end{equation}
so that $\int(d\kh'/4\pi)\,\mathcal{C}(\kh,\kh')f_\ell(\kh')=\kappa_\ell
f_\ell(\kh)$. The normal scattering rate is set by $\kappa_0$, {as follows}
\begin{equation}
\frac{\hbar}{2\tau}=\pi N_0\,n_{\rm imp}|V|^2\,\kappa_0
=\tfrac12\,\pi N_0\,n_{\rm imp}|V|^2 ,
\label{eq:rate}
\end{equation}
{with} the factor $\kappa_0=\tfrac12$ being the familiar chiral halving of the quantum
lifetime. The pair-breaking rate for a gap of angular momentum $\ell_a$ is the
\emph{mismatch} between the normal ($\kappa_0$) and anomalous ($\kappa_{\ell_a}$)
renormalizations,
\begin{equation}
\boxed{\;\Gamma^{\rm pb}_a=\pi N_0\,n_{\rm imp}|V|^2\,(\kappa_0-\kappa_{\ell_a})
=\frac{\hbar}{2\tau}\,(1-\eta_a),\qquad
\eta_a=\frac{\kappa_{\ell_a}}{\kappa_0}.\;}
\label{eq:alpha}
\end{equation}

\paragraph*{Consequences.}
For the \emph{conventional} channel ($\ell_s=0$) the gap is the leading
eigenfunction, $\kappa_{\ell_s}=\kappa_0$, hence $\eta_s=1$ and
$\Gamma^{\rm pb}_s=0$: scalar disorder does not pair-break the conventional channel even
in the Weyl semimetal. This is Anderson's theorem \emph{as realized within the assumed kernel} (Sec.~\ref{sec:model}), not a theorem derived for the full node-space vertex: the Berry phase that
halves $1/\tau$ [Eq.~\eqref{eq:rate}] cancels identically between numerator and
denominator of the pairing renormalization and drops out. For the
\emph{monopole} channel ($\ell_m=1$),
\begin{equation}
\eta_m=\frac{\kappa_1}{\kappa_0}=\frac{1/6}{1/2}=\frac13,
\qquad
\Gamma^{\rm pb}_m=\frac23\,\frac{\hbar}{2\tau},
\label{eq:etam}
\end{equation}
i.e.\ the chiral kernel reduces the pair-breaking rate to \emph{two thirds} of the
unprotected value $\Gamma^{\rm pb}=\Gamma_N$ (the rate for a channel with vanishing
anomalous eigenvalue, $\eta=0$) at the same normal scattering rate---the chiral
suppression of pseudospin reversal partially shields even the topological channel,
but cannot protect it fully. (Eq.~\eqref{eq:etam} is the isotropic value for the
spherical $J=1$ node; the Fermi-surface anisotropy of the $J=2,3$ nodes is
treated below and gives $\eta_m(J)=1/(J+2)$.)

{\paragraph*{Monopole-harmonic check.}
Equation~\eqref{eq:etam} uses the ordinary-harmonic addition theorem. One might
worry that the genuine monopole-harmonic ($q=1$) addition theorem, which governs the
projection of $Y_{1,1,m}$ onto the $q=0$ kernel Eq.~\eqref{eq:kernel}, shifts the
result. It does not: in this projected kernel the Berry-connection phase of
$\Oc(\kh,\kh')$ cancels against that of its partner, so the monopole harmonic
reproduces the value already obtained in closed form above. The cancellation is
specific to the assumed $\Theta$-compatible kernel; for a $\Theta$-broken
(triplet/chiral) vertex the Berry phase survives and even the isotropic component
is incompletely protected. The remaining shift to $\eta_s$ comes from inter-node
scattering (momentum transfer $2\mathbf Q$, with weight
{$W(2\mathbf Q)/W(0)$, small by assumption in the smooth-disorder regime of
Eq.~\eqref{eq:smoothcond} but not in the generic short-range case {(Sec.~\ref{sec:dissym}))}}. The consequences
of any resulting $\eta_s<1$ are quantified by Eq.~\eqref{eq:crosscrit}.}

We therefore diagonalize the projected impurity ladder on the \emph{actual}
anisotropic Fermi surface. Parametrizing it by
$v_zq_z=\mu\sin\psi$, $\alpha q_\perp^J=\mu\cos\psi$, the pseudospin is
$\dd=(\cos\psi\cos J\phi,\cos\psi\sin J\phi,\sin\psi)$, the chiral kernel is
$|\Oc|^2=\tfrac12(1+\dd\!\cdot\!\dd')$, and the density-of-states measure is
$w(\psi)\propto(\cos\psi)^{2/J-1}$ (Appendix~\ref{app:dos}). The pair-breaking eigenproblem is
$\langle|\Oc(\kh,\kh')|^2 f(\kh')\rangle_{\kh'}=\lambda_f\,f(\kh)$, with
$\eta_f=\lambda_f/\langle|\Oc|^2\rangle_{\rm FS}=2\lambda_f$ and pair-breaking rate
$\Gamma^{\rm pb}_f=\Gamma_N(1-\eta_f)$. Crucially, because
$\dd\!\cdot\!\dd'=\sin\psi\sin\psi'+\cos\psi\cos\psi'\cos J(\phi-\phi')$ depends on
azimuth only through $\cos J(\phi-\phi')$, the kernel is \emph{block-diagonal in
the azimuthal quantum number} $m$ (eigenfunctions $\propto e^{im\phi}g(\psi)$),
and only three sectors are populated:
\begin{itemize}\itemsep2pt
\item $m=0$, $f=\mathrm{const}$: $\kappa=\tfrac12$, so $\eta_s=1$ \emph{exactly
within the projected anomalous kernel of Eq.~\eqref{eq:kernel}} and, given that
kernel, independently of Fermi-surface shape (Anderson protection of the
conventional channel; the microscopic equality of normal and anomalous vertices
remains the assumption of Sec.~\ref{sec:model});
\item $m=0$, $f=\sin\psi$: a \emph{line}-nodal member (the gap vanishes on the
$q_z=0$ equator), with {$\eta=\langle\sin^2\psi\rangle_{\rm FS}=J/(J+2)$}
({Appendix~\ref{app:angavg}});
\item $m=\pm J$, $f=\cos\psi\,e^{\pm iJ\phi}$: the \emph{point}-nodal monopole gap.
This sector is \emph{rank one}
[$K\propto\cos\psi\cos\psi'$], so $f_m$ is the exact eigenfunction, with kernel eigenvalue {$\kappa_m=\tfrac14\langle\cos^2\psi\rangle_{\rm FS} = 1/(2(J+2))$ (Appendix~\ref{app:angavg})}.
\end{itemize}
{The two $m=0$ entries and the $m=\pm J$ entry carry different numerical weights
for a simple reason:
All $\eta$'s are normalized to the constant mode, $\eta=\kappa/\kappa_0$ with
$\kappa_0=\tfrac12$. In the $m=0$ sector the azimuthal average of
$\dd\!\cdot\!\dd'$ retains $\sin\psi\sin\psi'$ with weight $\tfrac12$, giving
$\kappa=\tfrac12\langle\sin^2\psi\rangle$ and hence
$\eta=\langle\sin^2\psi\rangle$. In the $m=\pm J$ sector the transverse term
$\cos\psi\cos\psi'\cos J(\phi-\phi')$ contributes only half of its weight upon
projection onto $e^{\pm iJ\phi}$, giving the extra factor $\tfrac12$:
$\kappa_m=\tfrac14\langle\cos^2\psi\rangle$ and $\eta_m=\tfrac12\langle\cos^2\psi\rangle$.
With $\langle\cos^2\psi\rangle=2/(J{+}2)$ and $\langle\sin^2\psi\rangle=J/(J{+}2)$
this yields $\eta_m=1/(J{+}2)$ and $\eta_{\rm line}=J/(J{+}2)$.}
For an isotropic ($J=1$) node these three states are the degenerate $\ell=1$
monopole triplet ($\eta=\tfrac13$); the multi-Weyl anisotropy \emph{splits} the
triplet, pushing the topological $m=\pm J$ member down to ({see Appendix~\ref{app:angavg}})
{
\begin{eqnarray}
\eta_m(J)&=&\tfrac12\,\langle\cos^2\psi\rangle_{\rm FS}
=\frac{1}{J + 2}.
\label{eq:etaJ}
\end{eqnarray}
Therefore} \
$\eta_m=\tfrac13,\tfrac14,\tfrac15$ for $J=1,2,3$. This is an \emph{exact} eigenvalue of the projected ladder for the
physical point-nodal monopole gap $f_m=\cos\psi\,e^{iJ\phi}$
(Eq.~\eqref{eq:fm}), not a trial-state expectation value: the rank-one structure
of the $m=\pm J$ sector forces $f_m$ to be the eigenfunction. Two consequences
follow. First, $\eta_s=1$ is exact within this projected scalar kernel. Second,
the anisotropy \emph{reduces} $\eta_m$ with increasing $J$, i.e.\ strengthens the
monopole pair-breaking $\Gamma^{\rm pb}_m=(1-\eta_m)\Gamma_N$; this sharpens, and does not weaken, the central trend that the crossing rate $\Gamma_N^\ast/T_{c0}^{(m)}$ decreases with $J$
(Fig.~\ref{fig:univpd}). 

\paragraph*{Two-rate structure.}
A useful way to read Eqs.~\eqref{eq:rate}--\eqref{eq:etam} is that scalar
disorder enters the monopole channel through \emph{two distinct rates}: the
normal (frequency) renormalization is set by $\kappa_0=\tfrac12$ while the
anomalous (gap) renormalization is set by $\kappa_1=\tfrac16$. The dimensionless
disorder strength is $\gamma\equiv\tfrac12 n_{\rm imp}|V|^2$, so that the bare rate
is $\Gamma_b\equiv\pi N_0 n_{\rm imp}|V|^2=2\pi\gamma N_0$, the normal rate is
$\Gamma_N=\Gamma_b/2=\pi\gamma N_0=\hbar/2\tau$ (fixed self-consistently below by
$N_0\to N(\mu;\Gamma_N)$), the anomalous rate is $\Gamma_b\kappa_m=\Gamma_b/[2(J{+}2)]$ ($\Gamma_b/6$ at $J{=}1$), and their
difference---the channel-$m$ pair-breaking rate
$\Gamma^{\rm pb}_m$ of Eq.~\eqref{eq:etam}, here abbreviated
$\Gamma_{\rm pb}$---is $\Gamma_{\rm pb}=(1-\eta_m)\Gamma_N=\Gamma_b\,(J{+}1)/[2(J{+}2)]$
($\Gamma_b/3$ at $J{=}1$). {Since several rates appear in this work, we define the
complete set here, in the order in which they are constructed:}

\begin{align}
\Gamma_b&=\pi n_{\rm imp}|V|^2N_0=2\pi\gamma N_0
&&\text{(bare Born rate)},\nonumber\\
\Gamma_N&=\kappa_0\,\Gamma_b=\tfrac12\Gamma_b=\pi\gamma N_0
&&\text{(normal/frequency rate)},\nonumber\\
\Gamma_{{\rm an},a}&=\kappa_a\,\Gamma_b
&&\text{(anomalous/gap rate)},\nonumber\\
\Gamma^{\rm pb}_a&=\Gamma_N-\Gamma_{{\rm an},a}=(1-\eta_a)\,\Gamma_N
&&\text{(pair breaking)},\nonumber\\
\eta_a&=\kappa_a/\kappa_0
&&\text{(protection factor)} .
\label{eq:ratelist}
\end{align}
{Throughout, $n_{\rm imp}|V|^2\equiv W(0)$ denotes the constant intra-pocket value of
the disorder power spectrum in the scale-separated limit Eq.~\eqref{eq:twoscale}; no
separate pointlike-impurity assumption is made or needed. Here $N_0$ is the per-node,
per-band DOS fixed in Sec.~\ref{sec:model},
$\kappa_0=\tfrac12$ always, $\kappa_m=1/[2(J{+}2)]$ for the monopole channel, and
$\kappa_s=\kappa_0$ in the Anderson-protected limit $\eta_s=1$. Two further rates appear
later and are \emph{not} members of this family: $\Gamma_0$, the residual zero-energy
broadening in the ordered state (App.~\ref{app:resid}), and $\tau_q=\hbar/(2\Gamma_N)$,
the quantum lifetime used for material conversions (Sec.~\ref{sec:anchor}).}
This is the qualitative distinction from isotropic magnetic
Abrikosov--Gor'kov, where the frequency and pair-breaking rates coincide. The
self-consistent Born equations then carry an angle-independent
$\tilde\omega_n=\omega_n+\Gamma_N\,\omega_n^{-1}$-type renormalization while the
gap retains its $\cos\psi\,e^{iJ\phi}$ winding under the anomalous rate $\Gamma_b/[2(J{+}2)]$, a
separation we use in the thermodynamics below.

\section{Generalized Abrikosov--Gor'kov gap equation}
With $\Gamma^{\rm pb}_a$ from Eq.~\eqref{eq:alpha}, each channel obeys (derived, with the rate conventions, in Appendix~\ref{app:ag}, {and discussed in Appendix~\ref{app_criterion}})
\begin{equation}
\ln\!\frac{T_{c0}^{(a)}}{\Tc^{(a)}}
=\psi\!\Big(\half+\frac{\Gamma^{\rm pb}_a}{2\pi k_B \Tc^{(a)}}\Big)-\psi\!\big(\half\big),
\label{eq:ag}
\end{equation}
{with $\psi(z)$ the Digamma function, and hence it is} destroyed at $\Gamma^{\rm pb}_a=\Gamma_c\simeq0.882\,k_B T_{c0}^{(a)}$. With $\Gamma^{\rm pb}_s=0$
the conventional channel is unaffected by pair-breaking \emph{in the $\eta_s=1$ projected-kernel limit} (for general $\eta_s$ see Eq.~\eqref{eq:crosscrit}) and changes only through
the disorder-dressed DOS $N(\gamma)$ entering $T_{c0}^{(s)}$; the monopole
channel is suppressed per Eq.~\eqref{eq:ag} with
$\Gamma^{\rm pb}_m=(1-\eta_m)\Gamma_N=\tfrac{J+1}{J+2}\,\Gamma_N$.

\paragraph*{Dimensionless form of the crossing.} The AG crossing is most
naturally stated in dimensionless variables. Since
$\Gamma^{\rm pb}_m/(2\pi k_B T_c^{(m)})$ depends on disorder only through
$\Gamma_N/k_B T_{c0}^{(m)}$ and the charge only through $\eta_m(J)$, the crossing
condition $T_c^{(m)}(\gamma)=T_c^{(s)}(\gamma)$ is governed by the two universal
inputs $\{\Gamma_N/k_B T_{c0}^{(m)},\,\eta_m(J)\}$ together with the clean ratio
$T_{c0}^{(s)}/T_{c0}^{(m)}$. This is the dimensionless projected-model content (at fixed $r$ and $\eta_s$) of the
charge trend: at fixed clean ratio, larger $J$ (smaller $\eta_m$) reaches the
crossing at smaller $\Gamma_N/k_B T_{c0}^{(m)}$. The conversion to a bare disorder
coordinate $\gamma$ (and hence any bare-coordinate values) additionally
carries the DOS normalization $N_J$, the cutoff, $\mu$, and the fixed-$\mu$ vs.\
fixed-density choice, and is therefore \emph{not} automatically comparable across
$J$. {All disorder strengths are therefore quoted as $\Gamma_N/k_BT_{c0}^{(m)}$ or
$\Gamma_N/\mu$.}

\paragraph*{AG crossing boundary for general $\eta_s$.} Because {we started with} $\eta_s=1$ {as a working assumption}
(Sec.~\ref{sec:model}), we now formulate the competition for
\emph{arbitrary} $\eta_s$ and identify $\eta_s=1$ as the Anderson-protected limit.
Both channels obey the same digamma law with their own rates
\begin{equation}
\Gamma^{\rm pb}_s=(1-\eta_s)\Gamma_N,\qquad
\Gamma^{\rm pb}_m=(1-\eta_m)\Gamma_N ,
\label{eq:tworate}
\end{equation}
and each is destroyed when its own $\Gamma^{\rm pb}$ reaches~\cite{AbrikosovGorkov1960,SkalskiBetbederWeiss1964} {the critical limit}
$\Gamma_c=0.882\,k_BT_{c0}$. Hence, the monopole channel dies at
$\Gamma_N^{(m0)}=0.882\,k_BT_{c0}^{(m)}/(1-\eta_m)$ and the conventional one at
$\Gamma_N^{(s0)}=0.882\,k_BT_{c0}^{(s)}/(1-\eta_s)$, {respectively}. A crossing at \emph{finite} $T_c$ exists---necessary and sufficient given the verified monotonicity of the AG scaling function (see below)---precisely when the conventional channel outlives the monopole channel,
$\Gamma_N^{(s0)}>\Gamma_N^{(m0)}$, i.e.
\begin{equation}
\boxed{\;\frac{1-\eta_s}{1-\eta_m}\;<\;r\equiv\frac{T_{c0}^{(s)}}{T_{c0}^{(m)}}\;}
\qquad\Longleftrightarrow\qquad \eta_s>\eta_s^c\equiv1-r\,(1-\eta_m).
\label{eq:crosscrit}
\end{equation}

\begin{figure}[t]
\centering
\includegraphics[width=0.99\columnwidth]{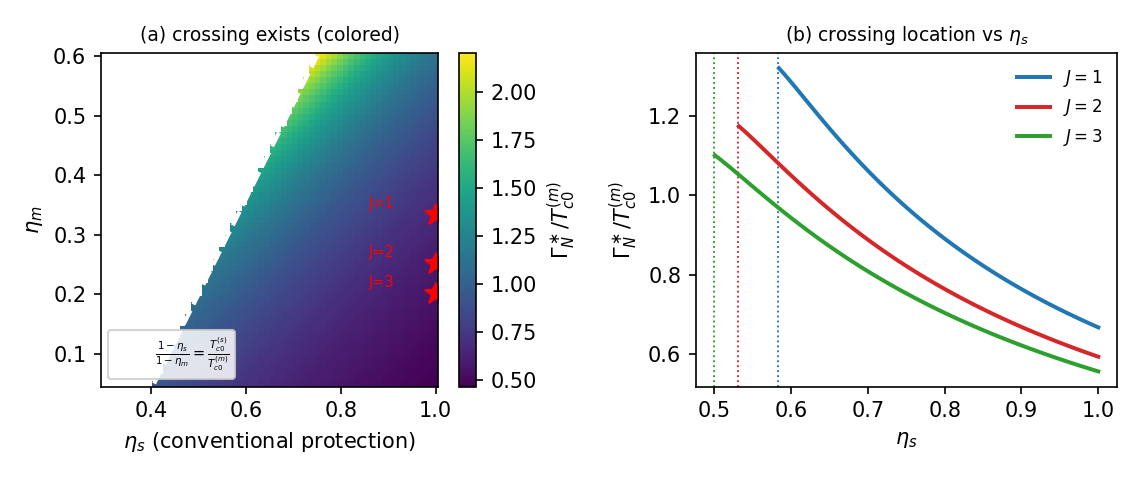}
\caption{Crossing criterion for general channel protection, clean ratio
$r=T_{c0}^{(s)}/T_{c0}^{(m)}=0.625$. (a) Crossing location
$\Gamma_N^\ast/T_{c0}^{(m)}$ over the $(\eta_s,\eta_m)$ plane; white region: no crossing at finite $T_c$, i.e.\ the strict inequality in Eq.~\eqref{eq:crosscrit} fails (the equality line itself belongs to the no-crossing side, where the two curves meet only at $T_c=0$). Dashed line: the AG boundary, Eq.~\eqref{eq:crosscrit}. Stars mark the assumed $\eta_s=1$ with the derived
$\eta_m=1/(J+2)$. {Since both protection factors are microscopically unresolved, the
stars should be read as projected-kernel benchmark points, not as predictions for generic
scalar disorder; the plane itself is the substantive content.} (b) Crossing location versus $\eta_s$ at fixed $\eta_m$;
dotted lines mark $\eta_s^c$. The crossing persists down to $\eta_s^c$ and disappears below it. {Grid: $60\times55$ in $(\eta_s,\eta_m)$, each point solving the two digamma
equations by bisection to a precision $10^{-10}$; the dashed boundary is
Eq.~\eqref{eq:crosscrit}.}}
\label{fig:etamap}
\end{figure}

Two caveats on scope. Eq.~\eqref{eq:crosscrit} is derived at \emph{fixed} clean
ratio $r$; if disorder-induced DOS feedback is retained, $r\to r(\Gamma_N)$ and the
criterion becomes an implicit local condition [cf.\ Eq.~\eqref{eq:lnr}]. {Moreover, at} $\eta_s=1$ the conventional curve is flat, so the crossing temperature is simply
$T^\ast=T_{c0}^{(s)}$, i.e.\ $T^\ast/T_{c0}^{(m)}=r=0.625$ for every $J$; the
charge dependence resides entirely in $\Gamma_N^\ast$. Three consequences
matter for the scope of this paper. (i) The criterion is a statement about the
\emph{ratio} of pair-breaking rates, not about exact protection: the mechanism
requires only that the conventional channel be pair-broken \emph{sufficiently more
slowly} than the monopole channel. (ii) With the clean ratio $r=0.625$ used below
and the derived $\eta_m=1/(J+2)$, the crossing survives for all
$\eta_s>\eta_s^c=0.583,\,0.531,\,0.500$ ($J=1,2,3$), with $\Gamma_N^\ast$ shifting
upward continuously as $\eta_s$ decreases toward $\eta_s^c$
[Fig.~\ref{fig:etamap}(b)]. We quote these critical values directly rather than
describing them as a fraction of ``protection,'' since $\eta_s$ is a kernel
eigenvalue ratio and not a linear measure of protection. (iii) Conversely, if the
unresolved node-space vertex were to drive $\eta_s$ below $\eta_s^c$, the crossing
would disappear, and {in this case Eq.~\eqref{eq:crosscrit} becomes a criterion to quantify} how much protection is needed (as the AG boundary, conditional on the
verified monotonicity and fixed $r$, {see Appendix~\ref{app_criterion}}). All crossing values quoted later are
conditional in this precise sense.

{The scattering rate is itself fixed self-consistently within a phenomenological
Lorentzian spectral-broadening scheme, $\Gamma_N=\pi\gamma N(\mu;\Gamma_N)$. Because that
convolution retains only the imaginary self-energy, is ultraviolet sensitive for $J=1,2$,
and is \emph{not} propagated into the phase diagrams presented here, we develop it
separately in App.~\ref{app:dosfeedback}, where we also verify that the instability
crossing survives when the feedback is coupled to the linearized $T_c$ equations. All
results in the main text are at fixed $\mu$ with the feedback disabled.}

\section{Thermodynamics}\label{sec:thermo}
\paragraph*{Clean specific-heat jump.}
The gap magnitude follows from the gap equation
{
\begin{eqnarray}
1=g_a N_0\int_0^{\Lambda}d\xi\,\langle |f_a|^2\tanh(E/2T)/E\rangle_{\kh},
\label{eq:BCSGAP}
\end{eqnarray}}
with
$E=\sqrt{\xi^2+\Delta_a^2|f_a(\kh)|^2}$. The jump at $\Tc$ is derived in
Appendix~\ref{app:cv}, {and is given by the explicit expression
\begin{equation}
\Delta C/C_n=\tfrac{12}{7\zeta(3)}\avg{|f_a|^2}^2/\avg{|f_a|^4},
\label{eq:DeltaC}
\end{equation}
with $C_n$ the specific heat of the normal state at the critical temperature.}
The gapped conventional channel gives the BCS value {$\Delta C/C_n = 1.43$}, while {for the nodal
monopole channel we find $1.19,\,0.95,\,0.79$ for $J=1,2,3$, respectively}. The more anisotropic gap at higher charge giving a smaller jump.

\begin{figure}[t]
\centering
\includegraphics[width=0.99\columnwidth]{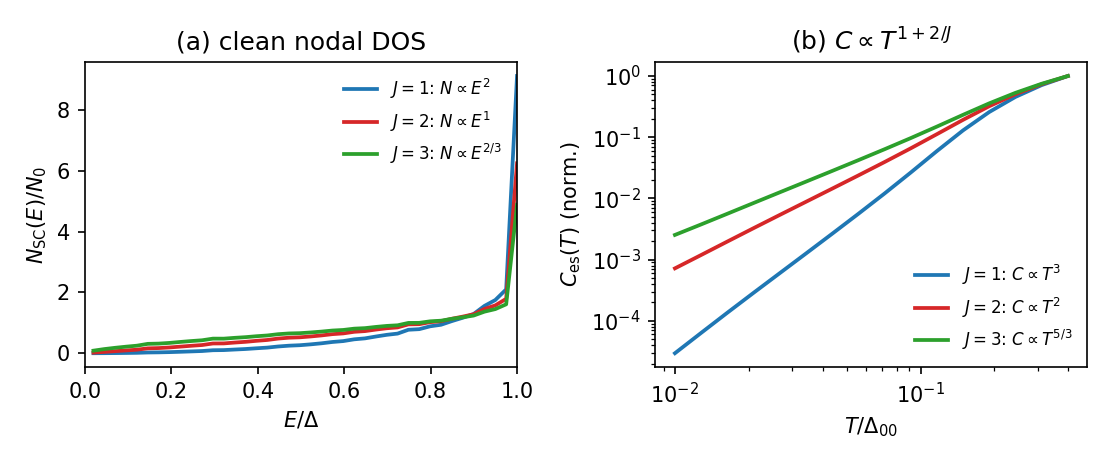}
\caption{Clean nodal thermodynamics of the monopole channel.
(a) Quasiparticle density of states $N_{\rm SC}(E)/N_0$; the point node gives
$N_{\rm SC}\propto E^{2/J}$ ($E^2,E^1,E^{2/3}$ for $J=1,2,3$)---not a hard gap.
(b) Electronic specific heat $C_{\rm es}\propto T^{1+2/J}$ ($T^3,T^2,T^{5/3}$),
a charge-sensitive low-temperature fingerprint. Curves computed from the
anisotropic gap $|f_m|=\cos\psi$ with the multi-Weyl measure.}
\label{fig:ag1}
\end{figure}

\paragraph*{Low-energy spectrum: a charge-dependent power law.}
The monopole gap $|f_m|=\cos\psi$ has point nodes at the two Fermi-surface poles
($\cos\psi=0$). There $|f_m|$ vanishes linearly in the polar angle $\vartheta\equiv
\tfrac\pi2-|\psi|$, while the density-of-states measure scales as
$(\cos\psi)^{2/J-1}\sim\vartheta^{2/J-1}$. The fraction of the Fermi surface with
gap below $E$ therefore scales as $\int_0^{E/\Delta}\vartheta^{2/J-1}d\vartheta
\propto E^{2/J}$, so the quasiparticle density of states and the electronic
specific heat obey
\begin{equation}
N_{\rm SC}(E)\propto E^{2/J},\qquad C_{\rm es}(T)\propto T^{\,1+2/J},
\label{eq:nodalpower}
\end{equation}
i.e.\ $C_{\rm es}\propto T^3,\,T^2,\,T^{5/3}$ for $J=1,2,3$ [Fig.~\ref{fig:ag1}]. Only
the isotropic ($J=1$) node gives the familiar point-node law $C\sim T^3$; the
higher-charge nodes are ``softer,'' and the charge-dependent exponent $1+2/J$ is
itself a low-temperature fingerprint of the monopole charge. 

\begin{figure}[t]
\centering
\includegraphics[width=0.7\columnwidth]{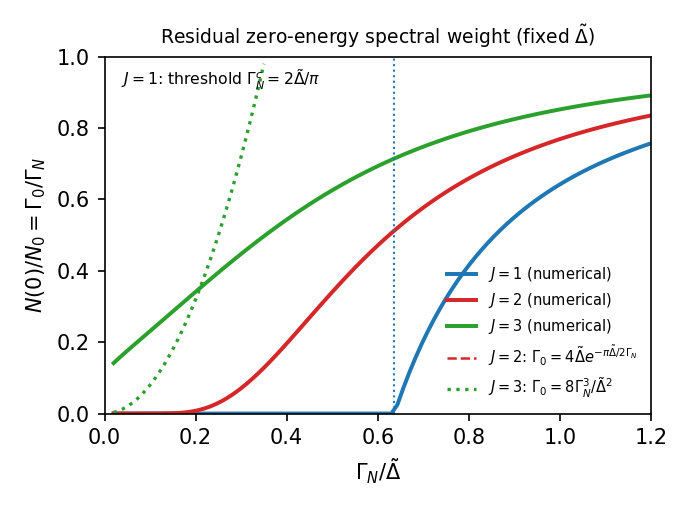}
\caption{Residual zero-energy DOS $N(0)/N_0=\Gamma_0/\Gamma_N$ from the
angle-resolved Born self-consistency Eq.~\eqref{eq:resid}, versus normal
scattering rate; $\tilde\Delta$ is a \emph{fixed effective spectral amplitude} (no nonlinear gap equation is solved in Appendix~\ref{app:resid}). The point-node threshold
$\Gamma_N^c=\frac{2\tilde\Delta}{\pi}$ exists \emph{only} for $J=1$ (dotted line;
$N(0)=0$ strictly below). For $J=2$ the onset is exponential,
$\Gamma_0=4\tilde\Delta e^{-\pi\tilde\Delta/(2\Gamma_N)}$, and for $J=3$ algebraic,
$\Gamma_0=8\Gamma_N^3/\tilde\Delta^2$ (dashed and dotted overlays; exact prefactors
derived in Appendix~\ref{app:resid}); both have \emph{no} threshold. A residual
Sommerfeld term therefore accompanies any disorder for $J\ge2$ \emph{within this
fixed-$\tilde\Delta$ Born theory} (see main text for the scope of that qualifier). Curves are the full numerical solutions of
Eq.~\eqref{eq:resid}; the analytic forms are shown only as asymptotic checks.}
\label{fig:ag2}
\end{figure}

\paragraph*{Residual density of states under disorder: a charge-dependent threshold.}
The low-energy behavior is governed by the point node and is more subtle than a
naive node-filling estimate. Scalar Born disorder generates a zero-energy
scattering rate $\Gamma_0=\mathrm{Im}\,\tilde\omega^R(0)$. Retaining \emph{both}
the normal and anomalous self-energies in the real-frequency continuation of the
angle-resolved equations of Appendix~\ref{app:ag}, and using that the rank-one
$m=\pm J$ sector preserves the $\cos\psi$ angular structure of the gap {at a
fixed effective spectral amplitude $\tilde\Delta$ (App.~\ref{app:resid})}, the
$\omega\to0$ limit reduces to
\begin{equation}
1=\Gamma_N\big\langle(\tilde\Delta^2\cos^2\psi+\Gamma_0^2)^{-1/2}\big\rangle_{\rm FS},
\qquad \frac{N(0)}{N_0}=\frac{\Gamma_0}{\Gamma_N},
\label{eq:resid}
\end{equation}
whose $\Gamma_0\!\to\!0$ limit is controlled by $\langle1/|f_m|\rangle=
\langle1/\cos\psi\rangle_{\rm FS}$, i.e.\ by $\int_0 d\vartheta\,\vartheta^{2/J-2}$
near a pole. \emph{The convergence of this integral, and hence the existence of a
threshold, depends on the charge}:
\begin{itemize}\itemsep2pt
\item $J=1$ ($2/J-2=0$): $\langle1/\cos\psi\rangle=\pi/2$ is finite, so $N(0)=0$
\emph{exactly} below a threshold $\Gamma_N^c=\frac{2\tilde\Delta}{\pi}$
(equivalently $\Gamma^{\rm pb}_c=(1-\eta_1)\frac{2\tilde\Delta}{\pi}=\frac{4\tilde\Delta}{3\pi}\simeq0.42\,\tilde\Delta$). The clean point-node thermodynamics survives up to this rate.
\item $J=2$ ($2/J-2=-1$): $\langle1/\cos\psi\rangle$ is \emph{logarithmically
divergent}, so there is \emph{no} threshold; arbitrarily weak disorder generates
an exponentially small residual scale, with the \emph{complete} asymptotic $\Gamma_0=4\tilde\Delta\,e^{-\pi\tilde\Delta/(2\Gamma_N)}$ (prefactor and exponent both analytic; see Appendix~\ref{app:resid}).
\item $J=3$ ($2/J-2=-4/3$): $\langle1/\cos\psi\rangle$ is \emph{power divergent},
so again no threshold, with the algebraic residual $\Gamma_0=8\,\Gamma_N^{3}/\tilde\Delta^{2}$ (coefficient exact; App.~\ref{app:resid}).
\end{itemize}
{All three cases are solved from Eq.~\eqref{eq:resid} (see Fig.~\ref{fig:ag2})}. Thus a strictly vanishing residual DOS over a finite disorder window is
special to $J=1$; for $J\ge2$ a residual Sommerfeld term $C_{\rm es}/T\to\text{const}$ is present at any nonzero disorder \emph{within the fixed-$\tilde\Delta$ Born theory used here}, though it is
exponentially (\,$J=2$\,) or algebraically (\,$J=3$\,) small at weak disorder. {That
qualifier is essential rather than formal: at the exponentially (or algebraically) small
scales involved, rare-event and localization physics, finite-size effects, and the full
self-consistent solution of the gap equation may all modify the experimentally accessible
asymptotics, and none of these is contained in Eq.~\eqref{eq:resid}.} {We stress that the threshold is \emph{not} universal across $J$: using the spherical
$J=1$ average $\langle1/\sin\theta\rangle=\pi/2$ for all $J$ would incorrectly predict a
finite threshold in every case, whereas the correct multi-Weyl measure makes
$\avg{1/|f_m|}$ divergent for $J\ge2$.} The threshold/exponential/algebraic \emph{classification} is fixed by
the convergence of $\int_0 d\vartheta\,\vartheta^{2/J-2}$ near the node and is
therefore robust {within the momentum-independent projected Born kernel used here---not
for arbitrary finite-range Born kernels, for which the angular average
$\avg{1/|f_m|}$ is reweighted}; only the numerical prefactors (threshold coefficient,
exponential constant $c$, algebraic prefactor) inherit the disorder-renormalized
amplitude $\tilde\Delta$ through Eq.~\eqref{eq:resid}. {Table~\ref{tab:thermo} collects
the charge-dependent thermodynamic signatures together with their differing logical
status.}

\paragraph*{Disorder dependence of the gap and jump.}
Away from $T_c$ the anisotropic nodal state is \emph{not} described by the
isotropic Abrikosov--Gor'kov spectral formulas (there is no hard BCS gap edge to
be driven downward); the full $\Delta(T;\gamma)$ would require the angle-resolved nonlinear Green's function, which we do \emph{not} solve; the residual-DOS results below are obtained at a \emph{fixed} renormalized amplitude $\tilde\Delta$ and presented as functions of $\Gamma_N/\tilde\Delta$.
The rigorous, disorder-independent anchors are the clean jumps
($0.95$ for the $J=2$ monopole, {and $1.43$ for the conventional channel, see
Appendix~\ref{app:cv})} and the linearized digamma law $\Tc(\Gamma^{\rm pb})$ of
Eq.~\eqref{eq:ag}. Under regular thermodynamic behavior the \emph{absolute}
heat-capacity discontinuity vanishes as $T_c^{(m)}\to0$; we assert no monotonic
trend for the \emph{normalized} jump $\Delta C/C_n$, which would require the
disorder-dressed quartic coefficient {as defined in Eq.~\eqref{eq:DeltaC}}. Anderson protection removes \emph{pair-breaking} from the
conventional channel ($\Gamma^{\rm pb}_s=0$), so its clean jump structure is
disorder-independent; its $T_c$, however, still shifts through the normal-state
DOS $N(\mu;\Gamma_N)$ entering $T_{c0}^{(s)}$ (Sec.~\ref{sec:scba}), and is not
``pinned'' in that sense.

\begin{table}[t]
\caption{Charge-dependent summary. The columns have \emph{different} logical status.
$\Delta C/C_n$, $N_{\rm SC}(E)$ and $C_{\rm es}(T)$ are clean projected-Hamiltonian
results (assumption-free). $\eta_m$ is the exact eigenvalue of the \emph{assumed}
{valley-diagonal, intra-pocket momentum-independent} projected impurity kernel Eq.~\eqref{eq:kernel}. The residual-DOS onset is a
zero-energy Born asymptotic at fixed $\tilde\Delta$ on the reflection-symmetric
branch (Appendix~\ref{app:resid}). Order of limits: for $J\ge2$ the residual DOS is
nonzero at any disorder {within the fixed-$\tilde\Delta$ Born theory}, so at fixed $\Gamma_N\ne0$ the Born-theory $T\to0$ law is
$C/T\to$ const; the clean power law $C\propto T^{1+2/J}$ is observable only in the
intermediate window above the (exponentially or algebraically small) residual
scale.}
\label{tab:thermo}
\begin{ruledtabular}
\begin{tabular}{cccccc}
& projected kernel & \multicolumn{3}{c}{clean projected Hamiltonian} & Born, fixed $\tilde\Delta$ \\
$J$ & $\eta_m=\tfrac{1}{J+2}$ & $\Delta C/C_n$ & $N_{\rm SC}(E)$ & $C_{\rm es}(T)$ & residual-DOS onset \\
\hline
1 & $1/3$ & $1.19$ & $E^{2}$   & $T^{3}$    & threshold $\Gamma_N^c=\frac{2\tilde\Delta}{\pi}$ \\
2 & $1/4$ & $0.95$ & $E^{1}$   & $T^{2}$    & none; $\Gamma_0=4\tilde\Delta e^{-\pi\tilde\Delta/(2\Gamma_N)}$ \\
3 & $1/5$ & $0.79$ & $E^{2/3}$ & $T^{5/3}$  & none; $\Gamma_0=8\Gamma_N^{3}/\tilde\Delta^{2}$ \\
\end{tabular}
\end{ruledtabular}
\end{table}

\section{Nodal charge of the pure-state solutions}\label{sec:topo}
The strongest topological statement is a property of the two
\emph{clean} pure-state solutions, and we present it as such. The monopole state
is \emph{nodal} (point nodes on the Fermi surface), so it has no global
three-dimensional gapped BdG invariant; what is well defined is the Berry (Chern)
charge on a small sphere enclosing each BdG point node, computed directly from the
Hermitian clean projected BdG Hamiltonian $h(\kk)=\dd(\kk)\!\cdot\!\bs$. For the
disorder statement, we add the observation that, {within this broadening scheme}, a scalar
identity self-energy $\mathrm{Re}\,\Sigma(0)\propto\sigma_0$ shifts $\mu$ and the
gap magnitude, but does not rotate the eigenvectors carrying the Berry charge, so
the pure-state nodal charges are unchanged by such disorder so long as the nodes remain isolated (the shift does not merge or gap them) and the projected two-band description remains valid, i.e.\ $\Gamma_N\ll\mu$. 

Projected onto the chiral Fermi surface, $h=\mathbf{n}(\kk)\!\cdot\!\bm{\tau}$ with
$\mathbf{n}=(\mathrm{Re}\,\Delta,-\mathrm{Im}\,\Delta,\xi)$ and $\xi=v(|\kk|-k_F)$ (we write $\mathbf{n}$ for the \emph{BdG} vector to distinguish it from the normal-state pseudospin texture $\dd$ of Sec.~\ref{sec:coh}). A BdG
point node occurs where $\xi=0$ and $\Delta(\kh)=0$. The conventional gap
$\Delta=\Delta_s$ never vanishes on the Fermi surface, so the spectrum is fully
gapped and there are no BdG nodes (trivial nodal content, which we label
$\mathcal{C}_{\rm BdG}=0$). The monopole gap $f_m=\cos\psi\,e^{iJ\phi}\propto
(k_x+ik_y)^J$ [Eq.~\eqref{eq:fm}] vanishes at the two Fermi-surface poles, where
the BdG spectrum hosts a pair of point nodes. The Berry charge on a small sphere
enclosing each node is $\pm J$, which we now establish
{by a local-degree argument}. 

\begin{figure}[t]
\centering
\includegraphics[width=0.86\columnwidth]{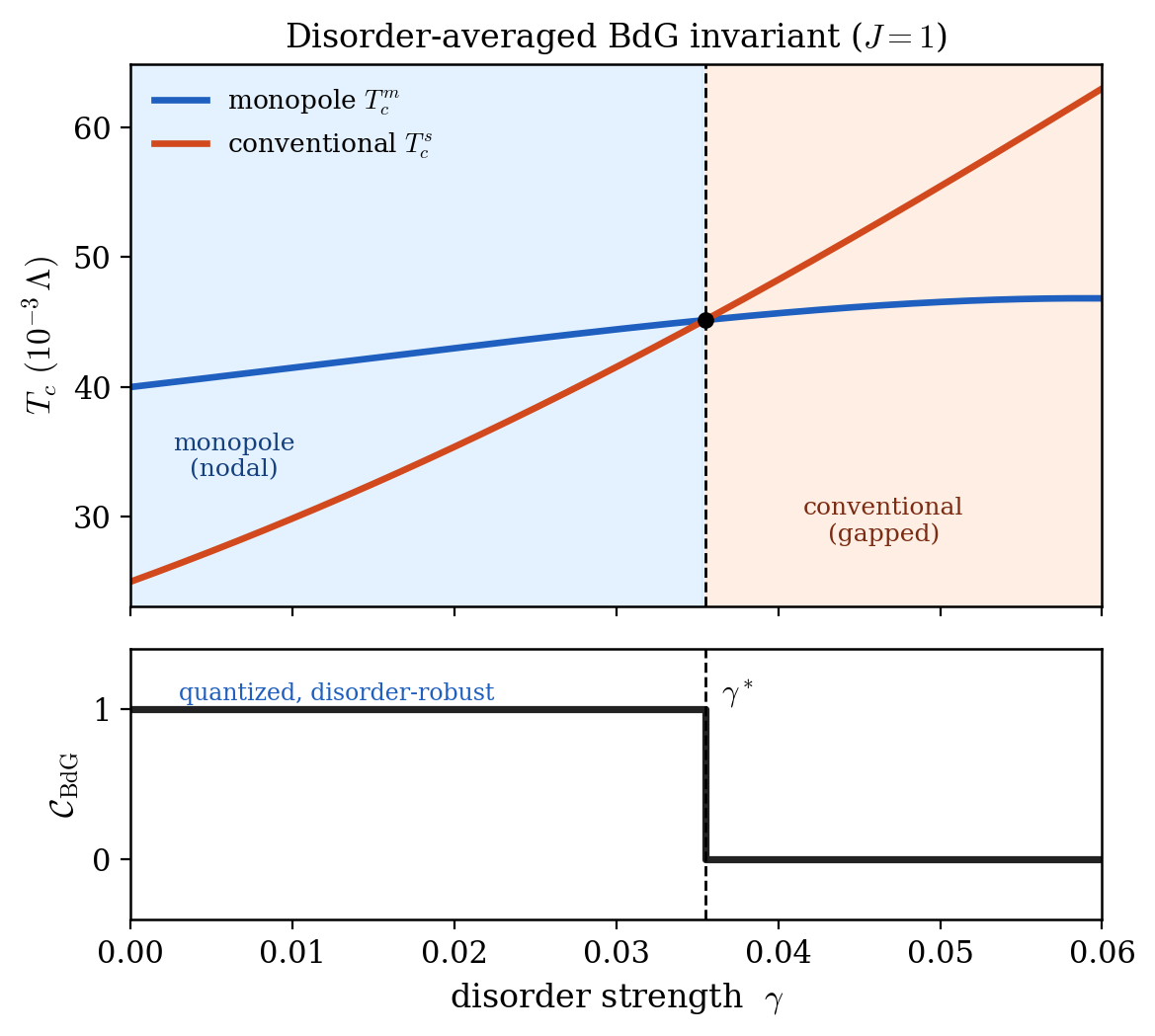}
\caption{The two \emph{pure-state} gap structures, shown side by side with
\emph{no disorder axis} so that no jump is implied. (a) The monopole solution has
point nodes on the Fermi surface, each carrying BdG Berry charge
$\mathcal{C}_{\rm BdG}=\pm J$ (Sec.~\ref{sec:topo}). (b) The conventional solution
is fully gapped, $\mathcal{C}_{\rm BdG}=0$. This is a comparison of two clean
solutions; the topology along any physical disorder path depends on the unresolved
nonlinear phase diagram (Sec.~\ref{sec:order}).}
\label{fig:topo}
\end{figure}

{Consider the north node at $\kk_0=(0,0,k_F)$ and expand in the local displacement
$\delta\kk=\kk-\kk_0$. Writing the clean projected BdG Hamiltonian as
$h=\hat{\mathbf n}\cdot\bm\tau$ with
$\mathbf n=(\mathrm{Re}\,\Delta,-\mathrm{Im}\,\Delta,\xi)$, the two transverse components
inherit the gap winding while the longitudinal component is the normal dispersion,}
\begin{equation}
{n_x+i\,n_y \;\propto\; (\delta k_x+i\,\delta k_y)^{J},\qquad
n_z=\xi\simeq v_\parallel\,\delta k_\parallel ,}
\label{eq:localdeg}
\end{equation}
{so that the node is a multi-Weyl point of azimuthal degree $J$. Parameterize a small
enclosing sphere of radius $R$ by $\delta\kk=R(\sin\theta\cos\phi,\sin\theta\sin\phi,
\cos\theta)$. Along the equator the transverse part gives
$\arg(n_x+in_y)=J\phi$, i.e.\ the image winds $J$ times as $\phi$ runs once around;
meanwhile $n_z\propto\cos\theta$ changes sign exactly once between the north
($\theta=0$) and south ($\theta=\pi$) poles of the enclosing sphere, so the image of
$\hat{\mathbf n}$ sweeps from one pole of the target sphere to the other. The map
$\hat{\mathbf n}:S^2\to S^2$ therefore has degree $J$ (up to orientation) and the Chern
number is $\mathcal C=+J$ at the north node, with the
orientation of the enclosing sphere fixed by the outward normal in $\delta\kk$.

{We fix the convention by working in the \emph{same} global Cartesian frame at both
poles, so that the transverse coordinates---and hence the winding
$\arg(n_x+in_y)=+J\phi$---are unchanged. The only difference is then the sign of the radial
dispersion, $\xi\simeq\pm v_\parallel\delta k_z$ at the north/south pole, since the outward
Fermi-surface normal is $+\hat z$ at one and $-\hat z$ at the other. Reversing $n_z$ at
fixed transverse winding composes $\hat{\mathbf n}$ with a reflection, so the degree changes
sign and $\mathcal C_{\rm north}=-\mathcal C_{\rm south}=J$, consistent with vanishing total
Berry flux over the closed Fermi surface.} The $J$-fold covering thus
resides in the \emph{image} of the map, not in the enclosing surface, which is an ordinary
sphere in physical momentum traversed once; no rescaled coordinates and no branch or
covering ambiguity are involved.} The two competing \emph{pure-state} solutions therefore
have distinct nodal topology: the monopole solution carries BdG nodal charges
$\pm J$, while the conventional solution is fully gapped {(Fig.~\ref{fig:topo})},
\begin{equation}
{|\mathcal{C}_{\rm BdG}|=
\begin{cases}
\;J\ \text{per node}, & \text{pure monopole solution (point-nodal)},\\[2pt]
\;0, & \text{pure conventional solution (gapped)},
\end{cases}}
\label{eq:cbdg}
\end{equation}
{where $\mathcal{C}_{\rm BdG}$ is the Berry charge of \emph{one} node of the clean
projected $2\times2$ nodal Hamiltonian.} We
stress the scope. First, this is a property of the two pure states, computed from
the clean projected nodal Hamiltonian; the topological-Hamiltonian construction
above shows only that a static scalar $\mathrm{Re}\,\Sigma(0)\propto\sigma_0$ does not change the local eigenvector map while the enclosing surface remains gapped. Second, and importantly, we do \emph{not} claim a
sharp ``jump'' of a bulk invariant along the physical disorder path: at a mere
crossing of linearized instabilities the ordered-state gap need not change
discontinuously, and whether the nodes reconstruct, split, or gap---and in what
order relative to the linearized crossing---is part of the unresolved nonlinear
phase diagram (Sec.~\ref{sec:order}), in particular whether an $s+m$ coexistence
window intervenes. What is rigorous is the contrast between the two pure-state
nodal structures ($\pm J$ vs.\ gapped); the topology along the actual ordered-state
boundary depends on that nonlinear analysis. This is the disorder-driven analog of
the clean topological-to-trivial pairing
changes of our previous works Refs.~\cite{MunozSotoGarridoJuricic2020,TapiaMunoz2026}.

\section{Beyond intra-node scalar disorder: scope}\label{sec:dissym}
The disorder calculation treated in this paper concerns \emph{intra-node scalar} disorder within the projected kernel of Eq.~\eqref{eq:kernel}
(Table~\ref{tab:dissym}), for which the projected kernel gives $\eta_m=1/(J+2)$
and, under the assumption of Sec.~\ref{sec:model}, $\eta_s=1$. Two other classes
matter physically and we delimit them rather than model them.

\begin{table}[t]
\caption{Pair-breaking rates for \emph{intra-node scalar} disorder---the disorder class analyzed in the projected kernel Eq.~\eqref{eq:kernel}. Rates are quoted separately rather than as a ratio (the
conventional rate vanishes under the assumption). $\eta_s=1$ is the assumption of
Sec.~\ref{sec:model}; $\eta_m=1/(J+2)$ is exact within the projected kernel. For
general $\eta_s$ see Eq.~\eqref{eq:crosscrit}.}
\label{tab:dissym}
\begin{ruledtabular}
\begin{tabular}{lcccc}
disorder & $\Theta_J$ preserved? & $\eta_s$ & $\Gamma^{\rm pb}_s/\Gamma_N$ & $\Gamma^{\rm pb}_m/\Gamma_N$ \\
\hline
scalar intra-node & preserved & $1$ (assumed) & $0$ & $\dfrac{J+1}{J+2}$ \\
\end{tabular}
\end{ruledtabular}
\end{table}

\emph{Inter-node ($2\mathbf{Q}$) scattering.} Its strength relative to intra-node
scattering is set parametrically by the Fourier content of the impurity potential,
\begin{equation}
\frac{\Gamma_{2\mathbf Q}}{\Gamma_{0}}\sim\frac{W(2\mathbf Q)}{W(0)},
\label{eq:2Q}
\end{equation}
{which is the same ratio that defines the smooth-disorder condition
Eq.~\eqref{eq:smoothcond}: the regime studied in this paper is
$W(2\mathbf Q)/W(0)\ll1$, whereas short-ranged (in the limit, pointlike) disorder has
$W(2\mathbf Q)\simeq W(0)$ and mixes the valleys at leading order.} We do \emph{not} compute the inter-node channel
eigenvalues: that requires diagonalizing the full valley-space impurity ladder with
the anisotropic measure, including the possibility that inter-node scattering
\emph{also} pair-breaks the conventional channel ($\eta_s<1$). {The reason is structural:
for zero-center-of-mass inter-node pairing the anomalous ladder contains one impurity matrix
element from \emph{each} paired node,
$K^{\rm an}(\bk,\bk')\sim\overline{U_{+}(\bk,\bk')U_{-}^{*}(-\bk,-\bk')}$, which is not
generically equal to $\overline{|U_{+}(\bk,\bk')|^2}$. Replacing it by the normal kernel
$|\Oc|^2$ requires the sewing identity of App.~\ref{app:band_sewing_gap} and is
\emph{not} a consequence of the disorder being scalar; a full node-resolved treatment could
therefore shift \emph{both} $\eta_s$ away from unity and $\eta_m$ away from $1/(J+2)$. This is the central technical problem left open here, and we leave it for future work.} Equation~\eqref{eq:crosscrit} \emph{parameterizes the consequences} of an effective
$\eta_s<1$; it does not replace computing the inter-node vertex. 

\emph{Magnetic / chirality-mixing disorder.} This breaks the antiunitary
$\Theta_J$ and is expected to pair-break both channels. The actual requirement is not strictly a nonmagnetic host but the inequality Eq.~\eqref{eq:crosscrit}: the conventional channel must be pair-broken sufficiently more slowly than the monopole channel. Magnetic disorder is disfavored because it is expected to break both comparably, but the criterion, not the disorder class, is the operative condition.

\section{Nature of the transition: crossing of pairing instabilities}\label{sec:order}
At the crossing rate $\Gamma_N^\ast$ the two \emph{linearized} instabilities cross and the leading
pairing symmetry of the normal state changes. {A crossing of two $T_c(\Gamma_N)$ curves
is well defined within the projected model, but it is \emph{not} the ordered-state phase
boundary, and we do not treat it as one.} Whether the physical boundary is
first order, or whether an $s+m$ coexistence or an $s+im$ time-reversal-breaking
window intervenes, requires the two-component Ginzburg--Landau functional with the
disorder-dressed quartic vertices \emph{and the actual crystalline point group},
which we do not compute. {A continuum ($C_\infty$) estimate, given in
App.~\ref{app:gl}, yields a biquadratic ratio
$R_{\rm clean}=2\avg{|f_m|^2}/\!\sqrt{\avg{|f_m|^4}}=1.83,\,1.63,\,1.48$ for $J=1,2,3$, all
exceeding unity, so that simple homogeneous coexistence is disfavored within that estimate.
It does not determine the transition order: the continuum expansion drops the phase-locking
term $\propto e^{2iJ\phi}$ restored by a discrete $C_n$ whenever $n\mid2J$, and disorder
renormalizes all quartic coefficients. We therefore leave the order undetermined and make
no claims of hysteresis, latent heat, or rare-region rounding.}

\section{Validity of the projected Born treatment near Weyl disorder-criticality}\label{sec:scba}
A doped Weyl semimetal undergoes a disorder-driven semimetal-to-diffusive-metal
transition where the Born expansion becomes uncontrolled. It is therefore
essential that the superconducting transition occurs at weaker disorder. The Born
parameter is $\Gamma_N/\mu=\hbar/(2\tau\mu)=\pi\gamma N(\mu)/\mu$ (for the isotropic $J=1$ node this equals $1/(k_F\ell)$). The
monopole channel is \emph{destroyed} when $\Gamma^{\rm pb}_m=(1-\eta_m)\Gamma_N$ reaches $\Gamma_c=0.882\,k_BT_{c0}^{(m)}$, i.e.\ at $\Gamma_N^{(m0)}=\Gamma_c/(1-\eta_m)$. This is an \emph{upper} scale, distinct from the earlier instability crossing
$\Gamma_N^\ast<\Gamma_N^{(m0)}$. We do not use a rough estimate here: both scales are
taken directly from the AG solution and tabulated below, giving
{$\Gamma_N^\ast/\mu=0.089,\,0.079,\,0.074$ for $J=1,2,3$}. The entire phase diagram
therefore lives in a moderately metallic regime,
separated from the disorder-criticality (which requires $\Gamma_N\sim\mu$, i.e.\
$\mu/\Gamma_N\sim1$, or the node physics at $\mu=0$).

\paragraph*{Separation of scales for $J=1,2,3$.}
We make the separation quantitative by comparing the crossing rate
$\Gamma_N^\ast$ (of order $T_{c0}^{(m)}$) with the rate at which the broadening
reaches the Fermi energy, $\Gamma_N=\mu$, beyond which the Fermi surface dissolves
and the chiral-basis {Born treatment} loses control. {We do not quote a corresponding statement in the bare disorder coordinate
$\gamma$: as discussed in Sec.~\ref{sec:poc}, $\gamma$ carries the density-of-states
normalization $N_J$, the cutoff and the ensemble choice, and is not itself observable.}
\begin{center}
\begin{tabular}{ccccc}
\hline\hline
$J$ & $\eta_m$ & $\Gamma_N^\ast/T_{c0}^{(m)}$ & $\Gamma_N^{(m0)}/T_{c0}^{(m)}$ &
$\Gamma_N^\ast/\mu$\\
\hline
1 & $1/3$ & 0.667 & 1.32 & {0.089}\\
2 & $1/4$ & 0.593 & 1.18 & {0.079}\\
3 & $1/5$ & 0.556 & 1.10 & {0.074}\\
\hline\hline
\end{tabular}
\end{center}
{(the crossing temperature is $T^\ast/T_{c0}^{(m)}=r$ for every $J$ in the
$\eta_s=1$ limit).}
\paragraph*{Numerical parameters and cutoff.} All curves use fixed $\mu$ and the
same effective-model cutoff; the crossing is set by the
\emph{pair-breaking} asymmetry ($\eta_s=1$ vs.\ $\eta_m=1/(J{+}2)$), which is
cutoff-independent \emph{because the cutoff-sensitive DOS feedback is omitted}
(a property of the decomposition, not of the full model); so the tabulated
$\Gamma_N^\ast/T_{c0}^{(m)}$ and $\Gamma_N^\ast/\mu$ do not depend on the cutoff
ratio. The
cutoff enters only the separately reported DOS feedback $N(\mu;\Gamma_N)$, which we now treat
as UV-sensitive for $J=1$ (Sec.~\ref{sec:scba}); {the disorder-feedback data are computed at $\Lambda/\mu=10$, and their $J=1$ values carry the cutoff caveat}.
Fixing the carrier density rather than $\mu$ shifts $\mu(\gamma)$ and would modify
the $J=1$ numbers most; we quote the fixed-$\mu$ ensemble throughout.
The instability crossing sits at $\Gamma_N^\ast/\mu\simeq0.07$--$0.13$ for all $J$,
and the margin \emph{widens} with $J$: although the broadened DOS
generates a finite residual weight at the node already at weak disorder for
$J=2,3$ (the Bera--Sau--Roy non-Anderson rounding at $\varepsilon=0$), this is
parametrically suppressed for the pairing, which lives on the Fermi surface at $\varepsilon=\mu$, by powers of $\Gamma_N/\mu\ll1$ (rather than being strictly immaterial).
As long as $\Gamma_N\ll\mu$ the Fermi surface remains a sharp metallic sheet with
well-defined chiral Bloch states, so the node rounding (a separate phenomenon at
$\varepsilon=0$) has an influence on the finite-$\mu$ Fermi-surface pairing calculation that is parametrically suppressed by $\Gamma_N/\mu$, and correspondingly on the Abrikosov--Gor'kov
suppression; the relevant control parameter is $\Gamma_N^\ast/\mu\simeq0.07$--$0.13$,
not the node-DOS. The separation is supported by the weak-coupling hierarchy
{$T_{c0}^{(m)}/\mu\ll1$. We acknowledge how strong that hierarchy actually is: the numerical value $T_{c0}^{(m)}=0.133\,\mu$ corresponds in BCS weak coupling to
$\Delta_{00}/\mu=1.764\,T_{c0}^{(m)}/\mu\simeq0.24$, i.e.\ a gap of order a quarter of the
Fermi energy. This is \emph{not} parametrically small, but the value
is chosen so that the crossing and the destruction rate are graphically resolvable on the
same axes; it should be read as an illustrative, deliberately enhanced pairing scale, not
as a quantitatively weak-coupling parameter set. At $\Delta_{00}/\mu\sim0.24$ one expects
appreciable corrections from the energy dependence of the density of states, valence-band
and interband pairing components, and particle--hole asymmetry, none of which we retain.
The dimensionless results that do not reference $\mu$---the kernel eigenvalues, the
crossing criterion Eq.~\eqref{eq:crosscrit}, and the rates in units of
$T_{c0}^{(m)}$---are unaffected by this choice; only the conversions to $\mu$
(Sec.~\ref{sec:anchor}) inherit it, and they would improve for a genuinely weak-coupling
$T_{c0}/\mu\lesssim0.02$.}

Consequently the {projected Born treatment} is
a leading-order treatment in a moderately metallic regime, but only marginally so: the relative
nominal expansion parameter $\Gamma_N/\mu$ is {$\simeq7$--$9\%$} across $J=1,2,3$ (power counting, not an uncertainty bound)
{(largest, $\simeq9\%$, for $J=1$)---comparable to the $\simeq11\%$ clean coupling
asymmetry}, so the quantitative crossing values should be read with a commensurate uncertainty. We also caution that $\mu/\Gamma_N$ conflates the
quantum and transport lifetimes and, for $J>1$, an anisotropic velocity tensor
with no single scalar $k_F$; these enter the prefactors of any material estimate
(Sec.~\ref{sec:anchor}). Beyond-Born effects, including rare-region states that round the nominal semimetal-to-diffusive-metal transition into an avoided quantum-critical crossover \cite{PixleyHuseDasSarma2016}, and quantum-interference and interaction corrections \cite{LuShen2015}, are expected to be subleading at finite $\mu$, being suppressed by the finite Fermi-surface density of states or by $\Gamma_N/\mu$. We have not calculated these effects at finite doping, however, so this is a power-counting expectation rather than a demonstrated bound; a full accounting of vertex and crossed diagrams, rare regions, and fixed-density $\mu(\gamma)$ shifts is left for future work. The
one quantity that probes stronger disorder is the DOS \emph{enhancement}
$N(\gamma)/N_0$ relevant to the $J=1$ channel, which reaches $\mu/\Gamma_N\sim3$--$4$
at the largest disorder shown; its magnitude is therefore {broadening-scheme-level}
(qualitative), whereas the Abrikosov--Gor'kov suppression that drives the
transition is the leading pair-breaking result.

\section{Proof-of-concept results}\label{sec:poc}
We solve the Abrikosov--Gor'kov gap equation (with the {DOS-feedback broadening} treated as a \emph{separate}, decoupled correction---the pair-breaking phase diagram and the DOS-feedback panel are not one joint self-consistent solution) using the
\emph{derived} FS-anisotropy values $\eta_m(J)=1/(J+2)$ ($\tfrac13,\tfrac14,\tfrac15$
for $J=1,2,3$). We quote energies in units of $\mu$, with cutoff $\Lambda=10\,\mu$
(consistent with the disorder-feedback data {of Sec.~\ref{sec:scba}}) and clean
$T_{c0}^{(m)}=0.133\,\mu>T_{c0}^{(s)}=0.083\,\mu$. \emph{The phase-diagram
crossing is computed with the DOS feedback switched off} (pair-breaking only), so
$\Gamma_N^\ast/T_{c0}^{(m)}$ and $\Gamma_N^\ast/\mu$ are independent of the cutoff;
the cutoff enters only the
separately reported $N(\mu;\Gamma_N)$ feedback, which is not propagated into $T_c$. {The complete parameter set used in every figure is: fixed $\mu$, gap-equation
cutoff $\Lambda=10\,\mu$, couplings $\lambda_m=0.225$ and $\lambda_s=0.203$
[Eq.~\eqref{eq:lambdavals}], and DOS feedback disabled except where explicitly stated.}

\paragraph*{Charge dependence in the universal rate.}
In the universal variable the crossings sit at
$\Gamma_N^\ast/T_{c0}^{(m)}=0.667,\,0.593,\,0.556$ for $J=1,2,3$, with the
monopole channel destroyed at
$\Gamma_N^{(m0)}/T_{c0}^{(m)}=0.882/(1-\eta_m)=1.32,\,1.18,\,1.10$. Smaller $\eta_m$
(larger $J$) reaches the crossing at a smaller rate, $\eta_m=1/(J+2)$ being the exact eigenvalue of {the momentum-independent projected kernel} Eq.~\eqref{eq:kernel}.

\begin{figure}[t]
\centering
\includegraphics[width=0.72\columnwidth]{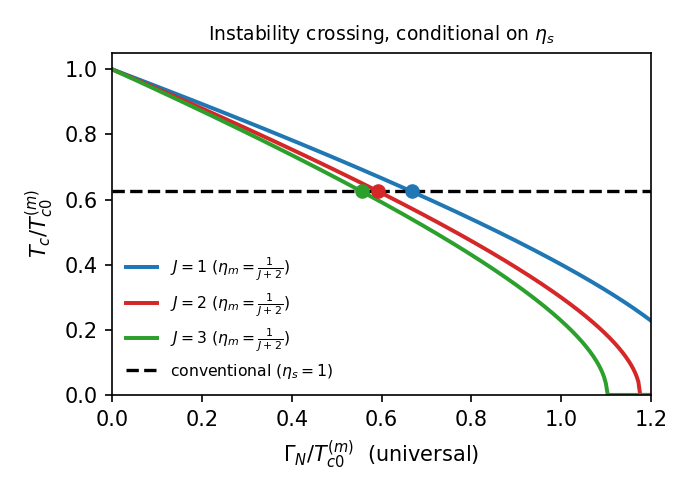}
\caption{Instability crossing in the \emph{universal} variable
$\Gamma_N/T_{c0}^{(m)}$ (primary presentation). Solid: monopole $T_c$ with the derived
$\eta_m=1/(J+2)$; dashed: conventional channel at the assumed $\eta_s=1$
(clean ratio $r=0.625$). Markers: crossings at
$\Gamma_N^\ast/T_{c0}^{(m)}=0.667,\,0.593,\,0.556$ for $J=1,2,3$. In this limit the
conventional curve is flat, so the crossing temperature is
$T^\ast/T_{c0}^{(m)}=r=0.625$ for every $J$---the charge dependence lies entirely in
$\Gamma_N^\ast$. {Pair-breaking only; DOS feedback
disabled.}}
\label{fig:univpd}
\end{figure}

We regard Fig.~\ref{fig:univpd} as the primary presentation of the crossing: it
uses the universal rate variable, is comparable across $J$, and involves no
DOS-feedback or cutoff input. {We do not present phase diagrams in the bare
coordinate $\gamma$, which is not directly
observable. All disorder strengths are therefore quoted as the
normal-state rate $\Gamma_N$, in units of either $T_{c0}^{(m)}$ or $\mu$.} Alongside $\Gamma_N^\ast$ we quote the
crossing temperature: in the $\eta_s=1$ limit $T^\ast/T_{c0}^{(m)}=r=0.625$
independently of $J$, so an experiment must resolve the change of pairing symmetry
at a temperature $\sim\!0.6\,T_{c0}^{(m)}$, not merely a $T_c$ suppression.

\paragraph*{Specific heat.}
The thermodynamic signatures were given quantitatively in
Figs.~\ref{fig:ag1}--\ref{fig:ag2}: the clean jumps are $1.19,0.95,0.79$
(monopole) and $1.43$ (conventional), and the low-energy laws are
$N_{\rm SC}\propto E^{2/J}$, $C\propto T^{1+2/J}$. A residual $C/T$ distinguishes
the nodal (monopole) from the gapped (conventional) state; for $J\ge2$ it is nonzero at any disorder \emph{within the fixed-$\tilde\Delta$
Born theory} [Eq.~\eqref{eq:resid}], but the $J=2$ scale is exponentially small,
$\Gamma_0=4\tilde\Delta e^{-\pi\tilde\Delta/2\Gamma_N}$. Concretely, at
$\Gamma_N/\tilde\Delta=0.3$ one finds $\Gamma_0\simeq2\times10^{-2}\tilde\Delta$,
already marginal, while at $\Gamma_N/\tilde\Delta=0.2$ it is
$\sim2\times10^{-3}\tilde\Delta$, {i.e.\ $\Gamma_0/k_B\simeq20~$mK for a spectral gap
$\tilde\Delta/k_B=10~$K (we convert the \emph{gap} amplitude, not $T_c$)---far below any
realistic base temperature.} The $J=2$ residual Sommerfeld term is therefore
\emph{masked by finite-temperature effects} in any practical experiment except very
close to $\Gamma_N^{(m0)}$; the residual term is realistically resolvable only near the monopole-destruction rate $\Gamma_N^{(m0)}$---the same class of observable proposed as a probe of topological
quantum criticality in the clean
theory~\cite{MunozSotoGarridoJuricic2020,MunozEsparza2024}.

\paragraph*{Robustness in \emph{both} channel parameters.}
The relevant robustness question is not how the crossing rate moves when $\eta_m$ alone
is varied, but whether the crossing survives when the assumed conventional-channel
protection is relaxed. That is answered within two-channel AG theory at fixed $r$, given the verified monotonicity, by Eq.~\eqref{eq:crosscrit} and Fig.~\ref{fig:etamap}: the crossing exists iff
$(1-\eta_s)/(1-\eta_m)<r$, i.e.\ for $\eta_s>1-r(1-\eta_m)$, which with $r=0.625$
and $\eta_m=1/(J+2)$ means $\eta_s>0.583,\,0.531,\,0.500$ for $J=1,2,3$. Varying
$\eta_m$ at fixed $\eta_s=1$ (the former one-parameter scan) is a strict subset of
this two-parameter statement and is not shown separately.

\section{{Model-level scattering-rate targets}}\label{sec:anchor}
The projected model defines an \emph{illustrative quantum-rate target}. The bare
$\gamma^\ast$ is not itself a laboratory observable, and the rate below depends on
$r$, $\eta_s$, the assumed kernel, fixed $\mu$, and disabled DOS feedback. With
those caveats, the {broadening scheme} fixes the dimensionless rate at the
\emph{instability crossing} (where $T_c^{(m)}=T_c^{(s)}$, not at the later
monopole-destruction rate $\Gamma_N^{(m0)}$),
\begin{equation}
\frac{\Gamma_N^\ast}{\mu}\simeq 0.089,\,0.079,\,0.074\quad(J=1,2,3),
\label{eq:kflstar}
\end{equation}
so {$\mu/\Gamma_N^\ast\simeq 11.3,\,12.7,\,13.5$ \emph{at the enhanced illustrative
value $T_{c0}^{(m)}/\mu=0.133$}: the crossing lies in a moderately
metallic regime for every charge, though $\mu/\Gamma_N^\ast\sim10$ is moderate rather than
parametrically large. Since $\Gamma_N^\ast/\mu$ is linear in $T_{c0}^{(m)}/\mu$, a
weak-coupling choice improves this substantially, to $\mu/\Gamma_N^\ast\simeq75$--$90$ at
$T_{c0}^{(m)}/\mu=0.02$. These figures refer to the decoupled pair-breaking calculation;
the flagged $J=2$, $\Lambda/\mu=20$ entry of {Table~\ref{tab:feedback} in
App.~\ref{app:dosfeedback}}, with
$\mu/\Gamma_N^\ast\simeq5.9$, is excluded from such statements}. (We write $\mu/\Gamma_N$ rather than $k_F\ell$, which is not a scalar for $J>1$ because the velocity tensor is anisotropic; for the isotropic $J=1$ node the identification $\mu/\Gamma_N=k_F\ell$ may be made parenthetically.) The one \emph{conditional model-level} laboratory target (conditional on $\eta_s$, the clean ratio $r$, the AG form, the projected kernel, fixed $\mu$, and disabled DOS feedback)
is the \emph{quantum scattering rate} at the crossing, $\Gamma_N^\ast/\mu$, or
equivalently the quantum lifetime $\tau_q=\hbar/(2\Gamma_N^\ast)$. For an
illustrative doping range $\mu\simeq20$--$50~$meV one has
$\Gamma_N^\ast/\mu\simeq0.089,\,0.079,\,0.074$ ($J=1,2,3$), i.e. {$\tau_q=\hbar/(2\Gamma_N^\ast)\simeq0.074$--$0.185~$ps
($J=1$), $0.083$--$0.209~$ps ($J=2$), and $0.089$--$0.223~$ps ($J=3$)}
[Table~\ref{tab:anchor}], the smaller $\eta_m$ at larger $J$ reaching the crossing
at a smaller rate and hence a longer lifetime. {These lifetimes inherit the deliberately
enhanced pairing scale $T_{c0}^{(m)}/\mu=0.133$ discussed in Sec.~\ref{sec:scba} and are
therefore not a robust material target: because $\Gamma_N^\ast/\mu$ is linear in
$T_{c0}^{(m)}/\mu$, a genuinely weak-coupling choice $T_{c0}^{(m)}/\mu=0.02$ gives
$\Gamma_N^\ast/\mu\simeq0.012$ for $J=2$ rather than $0.079$, and required lifetimes about
$6.6$ times longer (lower block of Table~\ref{tab:anchor}). What is
comparatively robust is the \emph{ordering} $\tau_q(J{=}1)<\tau_q(J{=}2)<\tau_q(J{=}3)$
and the ratios between charges, rather than the absolute picoseconds---and even that
ordering holds only at fixed $r$, fixed $T_{c0}^{(m)}/\mu$, and the benchmark kernel
$\eta_m=1/(J+2)$.} We deliberately stop here: converting
$\tau_q$ to a residual resistivity or impurity concentration would require a
consistent single-regime $T$-matrix and an anisotropic Boltzmann-vertex
calculation for the $J$-fold node (there is no single scalar $k_F$ or transport
time for $J>1$), which we do not carry out. The operative requirement is Eq.~\eqref{eq:crosscrit} rather than the disorder label---%
magnetic or chirality-mixing disorder requires a separate vertex calculation and
is expected on symmetry grounds to reduce the conventional-channel protection---so a candidate should have disorder for which the conventional channel is pair-broken sufficiently more slowly than the monopole channel [Eq.~\eqref{eq:crosscrit}]; we make no compound-specific claim, as identifying one
would need independent evidence of the node multiplicity, chemical potential,
nonmagnetic disorder, and pairing channels.

\begin{table}[t]
\caption{Conditional model-level targets at the crossing: the normal-state
(quantum) scattering rate and the corresponding quantum lifetime
$\tau_q=\hbar/(2\Gamma_N^\ast)$ for an illustrative doping range
$\mu=20$--$50~$meV. We deliberately quote no residual resistivity or impurity
concentration (Sec.~\ref{sec:anchor}). Values are conditional on $\eta_s=1$
[Eq.~\eqref{eq:crosscrit}] and computed with the DOS feedback disabled.
{These numbers are \emph{not} universal outputs of the crossing mechanism: since
$\Gamma_N^\ast/\mu=(\Gamma_N^\ast/T_{c0}^{(m)})\times(T_{c0}^{(m)}/\mu)$, they scale
linearly with the assumed $T_{c0}^{(m)}/\mu$, which is stated explicitly in the first
column. The upper block uses the deliberately enhanced, graphically convenient value
$T_{c0}^{(m)}/\mu=0.133$ used throughout the figures; the lower block gives a genuinely
weak-coupling benchmark, $T_{c0}^{(m)}/\mu=0.02$, for which the required lifetimes are
$6.6$ times longer.}}
\label{tab:anchor}
\begin{ruledtabular}
\begin{tabular}{cccc}
{$T_{c0}^{(m)}/\mu$} & $J$ & $\Gamma_N^\ast/\mu$ & $\tau_q$ (ps), $\mu=20$--$50~$meV \\
\hline
{0.133 (enhanced)} & 1 & {0.089} & {0.074--0.185} \\
{0.133 (enhanced)} & 2 & {0.079} & {0.083--0.209} \\
{0.133 (enhanced)} & 3 & {0.074} & {0.089--0.223} \\
\hline
{0.02 (weak coupling)} & {1} & {0.0133} & {0.49--1.23} \\
{0.02 (weak coupling)} & {2} & {0.0119} & {0.55--1.39} \\
{0.02 (weak coupling)} & {3} & {0.0111} & {0.59--1.48} \\
\end{tabular}
\end{ruledtabular}
\end{table}

{\section{Discussion and Conclusions
}}\label{sec:status}
{As stated in the Introduction, this work extends the clean monopole-pairing analysis of
Refs.~\cite{MunozSotoGarridoJuricic2020,MunozEsparza2024,TapiaMunoz2026} into the
disordered regime within a {Matsubara/Born} and generalized Abrikosov--Gor'kov
framework. A family of clean projected-Hamiltonian results---the angular moments,
the specific-heat jumps, the nodal laws $N_{\rm SC}\propto E^{2/J}$ and
$C\propto T^{1+2/J}$, and the pure-state BdG nodal charges $\pm J$---follows without
any disorder input, while $\eta_m(J)=1/(J+2)$ is the exact eigenvalue of the
{valley-diagonal, intra-pocket momentum-independent} projected kernel Eq.~\eqref{eq:kernel}. From these we derived the crossing of
linearized instabilities: its existence, its location $\Gamma_N^\ast/T_{c0}^{(m)}$,
and the charge trend (Sec.~\ref{sec:poc}), with the arbitrary-$\eta_s$ boundary
Eq.~\eqref{eq:crosscrit} requiring $\eta_s>1-r(1-\eta_m)$, i.e.\
$\eta_s\gtrsim0.5$--$0.58$ here. We verified the survival of the crossing under
{the coupled Lorentzian-DOS-plus-AG scheme} for $J=2,3$ {(App.~\ref{app:dosfeedback},
Table~\ref{tab:feedback})}, and obtained the
charge-dependent residual-DOS classification (threshold for $J=1$, exponential for
$J=2$, algebraic for $J=3$; App.~\ref{app:resid}). The precise logical status of
each of these results is set out below.

Our treatment is leading order in a \emph{moderately} metallic regime,
{$\mu/\Gamma_N^\ast\simeq11$--$14$ at the enhanced illustrative $T_{c0}^{(m)}/\mu=0.133$
(larger, and hence better controlled, for weaker coupling; the flagged $J=2$,
$\Lambda/\mu=20$ coupled entry of {Table~\ref{tab:feedback} (App.~\ref{app:dosfeedback})} is excluded, having
$\mu/\Gamma_N^\ast\simeq5.9$).} We do not claim a controlled error bar:
{$\Gamma_N/\mu\simeq7$--$9\%$} is power counting, and the clean coupling hierarchy
that generates the competition is itself {an $\simeq11\%$} effect, so vertex, crossed-diagram
and fixed-density corrections of comparable size could shift the numerical crossing rate appreciably. The \emph{existence} of the crossing is governed by
Eq.~\eqref{eq:crosscrit} rather than by these specific numbers.

Sharpening a materials estimate for a named compound, via \emph{ab initio}
band structures that fix $\mu,v_z,\alpha$ and the node multiplicity, is the natural
next step toward direct experimental contact.

{\paragraph*{Status of the results.} The logical standing of each ingredient was set out
in Sec.~\ref{sec:model} and is not repeated here. We recall only the two boundaries that
matter for reading the numbers above. First, every disorder result is exact \emph{given}
the valley-diagonal, intra-pocket momentum-independent kernel of the small-pocket window
Eq.~\eqref{eq:twoscale}, and is not established for generic forward-peaked or intervalley
disorder. Second, the residual-DOS \emph{classification} is robust within that projected
Born kernel, whereas its explicit coefficients are conditional on fixed $\tilde\Delta$ and
the reflection-symmetric branch. The principal open problem remains the full node-resolved
anomalous impurity vertex.}

Beyond the bulk thermodynamics of
this work, the disorder robustness of the Majorana/Fermi-arc surface states---which
the present disorder-averaged bulk self-energy does not address---and the
non-equilibrium (Keldysh) transport signatures are natural follow-up studies.}

\section{Acknowledgments 
}\label{sec:acknowledgments}

This work is supported by Fondecyt (Chile) Grants  No. 1230440 (E.M.) and 1241033 (R.S.-G.). 


{
\section{Declaration of generative AI and AI-assisted technologies in the manuscript preparation process}

During the preparation of this work, the authors used Anthropic’s Claude Opus to assist with coding, generation of plots and manuscript preparation. After using this tool/service, the authors reviewed and edited the content as needed and take full responsibility for the content of the published article.

}

\bibliography{refs}

@article{Bardeen1957,
  author  = {Bardeen, J. and Cooper, L. N. and Schrieffer, J. R.},
  title   = {Theory of Superconductivity},
  journal = {Phys. Rev.},
  volume  = {108},
  pages   = {1175},
  year    = {1957},
  doi     = {10.1103/PhysRev.108.1175}
}

@article{Anderson1959,
  author  = {Anderson, P. W.},
  title   = {Theory of dirty superconductors},
  journal = {J. Phys. Chem. Solids},
  volume  = {11},
  pages   = {26},
  year    = {1959},
  doi     = {10.1016/0022-3697(59)90036-8}
}

@article{AbrikosovGorkov1960,
  author  = {Abrikosov, A. A. and Gor'kov, L. P.},
  title   = {Contribution to the theory of superconducting alloys with paramagnetic impurities},
  journal = {Sov. Phys. JETP},
  volume  = {12},
  pages   = {1243},
  year    = {1961},
  note    = {[Zh. Eksp. Teor. Fiz. 39, 1781 (1960)]}
}

@article{SkalskiBetbederWeiss1964,
  author  = {Skalski, S. and Betbeder-Matibet, O. and Weiss, P. R.},
  title   = {Properties of superconducting alloys containing paramagnetic impurities},
  journal = {Phys. Rev.},
  volume  = {136},
  pages   = {A1500},
  year    = {1964},
  doi     = {10.1103/PhysRev.136.A1500}
}

@article{SigristUeda1991,
  author  = {Sigrist, M. and Ueda, K.},
  title   = {Phenomenological theory of unconventional superconductivity},
  journal = {Rev. Mod. Phys.},
  volume  = {63},
  pages   = {239},
  year    = {1991},
  doi     = {10.1103/RevModPhys.63.239}
}

@article{ArmitageMeleVishwanath2018,
  author  = {Armitage, N. P. and Mele, E. J. and Vishwanath, A.},
  title   = {{Weyl} and {Dirac} semimetals in three-dimensional solids},
  journal = {Rev. Mod. Phys.},
  volume  = {90},
  pages   = {015001},
  year    = {2018},
  doi     = {10.1103/RevModPhys.90.015001}
}

@article{LiHaldane2018,
  author  = {Li, Yi and Haldane, F. D. M.},
  title   = {Topological Nodal {Cooper} Pairing in Doped {Weyl} Metals},
  journal = {Phys. Rev. Lett.},
  volume  = {120},
  pages   = {067003},
  year    = {2018},
  doi     = {10.1103/PhysRevLett.120.067003}
}

@article{MunozSotoGarridoJuricic2020,
  author  = {Mu\~{n}oz, E. and Soto-Garrido, R. and Juri\v{c}i\'{c}, V.},
  title   = {Monopole versus spherical harmonic superconductors: Topological repulsion, coexistence, and stability},
  journal = {Phys. Rev. B},
  volume  = {102},
  pages   = {195121},
  year    = {2020},
  doi     = {10.1103/PhysRevB.102.195121}
}

@article{MunozEsparza2024,
  author  = {Mu\~{n}oz, E. and Esparza, J. and Braun, J. and Soto-Garrido, R.},
  title   = {Topological versus conventional superconductivity in a {Weyl} semimetal: A microscopic approach},
  journal = {Superconductivity},
  volume  = {12},
  pages   = {100132},
  year    = {2024},
  doi     = {10.1016/j.supcon.2024.100132}
}

@article{TapiaMunoz2026,
  author  = {Tapia, A. and Mu\~{n}oz, E.},
  title   = {Superconductivity in multi-{Weyl} semimetals: Conditions for the coexistence of topological and conventional phases},
  journal = {Phys. Rev. B},
  volume  = {113},
  pages   = {024501},
  year    = {2026},
  doi     = {10.1103/8r9k-zf4v}
}

@article{KlierGornyiMirlin2019,
  author  = {Klier, J. and Gornyi, I. V. and Mirlin, A. D.},
  title   = {From weak to strong disorder in {Weyl} semimetals: Self-consistent {Born} approximation},
  journal = {Phys. Rev. B},
  volume  = {100},
  pages   = {125160},
  year    = {2019},
  doi     = {10.1103/PhysRevB.100.125160}
  }

@article{ChoBardarsonLuMoore2012,
  author  = {Cho, G. Y. and Bardarson, J. H. and Lu, Y.-M. and Moore, J. E.},
  title   = {Superconductivity of doped {Weyl} semimetals: Finite-momentum pairing and electronic analog of the $^3$He-A phase},
  journal = {Phys. Rev. B},
  volume  = {86},
  pages   = {214514},
  year    = {2012},
  doi     = {10.1103/PhysRevB.86.214514}
}

@article{AndersenRamiresAndo2020,
  author  = {Andersen, L. and Ramires, A. and Wang, Z. and Lorenz, T. and Ando, Y.},
  title   = {Generalized {Anderson}'s theorem for superconductors derived from topological insulators},
  journal = {Sci. Adv.},
  volume  = {6},
  pages   = {eaay6502},
  year    = {2020},
  doi     = {10.1126/sciadv.aay6502}
}

@article{BeraSauRoy2016,
  author  = {Bera, S. and Sau, J. D. and Roy, B.},
  title   = {Dirty {Weyl} semimetals: Stability, phase transition, and quantum criticality},
  journal = {Phys. Rev. B},
  volume  = {93},
  pages   = {201302(R)},
  year    = {2016},
  doi     = {10.1103/PhysRevB.93.201302}
}

@article{SyzranovRadzihovsky2018,
  author  = {Syzranov, S. V. and Radzihovsky, L.},
  title   = {High-Dimensional Disorder-Driven Phenomena in {Weyl} Semimetals, Semiconductors, and Related Systems},
  journal = {Annu. Rev. Condens. Matter Phys.},
  volume  = {9},
  pages   = {35},
  year    = {2018},
  doi     = {10.1146/annurev-conmatphys-033117-054037}
}

@article{KuibarovBorisenko2024,
  author  = {Kuibarov, A. and Suvorov, O. and Vocaturo, R. and others},
  title   = {Evidence of superconducting {Fermi} arcs},
  journal = {Nature},
  volume  = {626},
  pages   = {294},
  year    = {2024},
  doi     = {10.1038/s41586-023-06977-7}
}

@article{TimmonsPdTe2_2020,
  author  = {Timmons, E. I. and Teknowijoyo, S. and Ko\'{n}czykowski, M. and others},
  title   = {Electron irradiation effects on superconductivity in PdTe$_2$: An application of a generalized {Anderson} theorem},
  journal = {Phys. Rev. Research},
  volume  = {2},
  pages   = {023140},
  year    = {2020},
  doi     = {10.1103/PhysRevResearch.2.023140}
}

@article{LiNdAlSi2025,
  author  = {Li, C. and Wang, Y. and Zhang, J. and others},
  title   = {Disorder-driven non-{Anderson} transition in a {Weyl} semimetal},
  journal = {Proc. Natl. Acad. Sci. U.S.A.},
  volume  = {122},
  pages   = {e2508569122},
  year    = {2025},
  doi     = {10.1073/pnas.2508569122}
}

@article{Balatsky2006,
  title = {Impurity-induced states in conventional and unconventional superconductors},
  author = {Balatsky, A. V. and Vekhter, I. and Zhu, Jian-Xin},
  journal = {Rev. Mod. Phys.},
  volume = {78},
  pages = {373--433},
  year = {2006},
  doi     = {10.1103/RevModPhys.78.373}
}

@article{Hirschfeld1993,
  title = {Effect of strong scattering on the low-temperature penetration depth of a d-wave superconductor},
  author = {Hirschfeld, P. J. and Goldenfeld, N.},
  journal = {Phys. Rev. B},
  volume = {48},
  pages = {4219--4222},
  year = {1993},
  doi     = {10.1103/PhysRevB.48.4219}
}

@article{Gorkov1985,
  title = {Defects and an unusual superconductivity},
  author = {Gor'kov, L. P. and Kalugin, P. A.},
  journal = {JETP Lett.},
  volume = {41},
  pages = {253},
  year = {1985}
}

@article{GolubovMazin1997,
  title = {Effect of magnetic and nonmagnetic impurities on highly anisotropic superconductivity},
  author = {Golubov, A. A. and Mazin, I. I.},
  journal = {Phys. Rev. B},
  volume = {55},
  pages = {15146--15152},
  year = {1997},
  doi     = {10.1103/PhysRevB.55.15146}
}

@article{Kogan2009,
  title = {Pair breaking in iron pnictides},
  author = {Kogan, V. G.},
  journal = {Phys. Rev. B},
  volume = {80},
  pages = {214532},
  year = {2009},
  doi     = {10.1103/PhysRevB.80.214532}
}

@article{Efremov2011,
  title = {Disorder-induced transition between $s_\pm$ and $s_{++}$ states in two-band superconductors},
  author = {Efremov, D. V. and Korshunov, M. M. and Dolgov, O. V. and Golubov, A. A. and Hirschfeld, P. J.},
  journal = {Phys. Rev. B},
  volume = {84},
  pages = {180512},
  year = {2011},
  doi     = {10.1103/PhysRevB.84.180512}
}

@article{Serbyn2013,
  title = {Interplay of disorder and interactions in {D}irac and {W}eyl semimetals},
  author = {Nandkishore, Rahul and Huse, David A. and Sondhi, S. L.},
  journal = {Phys. Rev. B},
  volume = {89},
  pages = {245110},
  year = {2014},
  doi     = {10.1103/PhysRevB.89.245110}
}

@article{WanTurnerVishwanathSavrasov2011,
  title = {Topological semimetal and Fermi-arc surface states in the electronic structure of pyrochlore iridates},
  author = {Wan, Xiangang and Turner, Ari M. and Vishwanath, Ashvin and Savrasov, Sergey Y.},
  journal = {Phys. Rev. B}, volume = {83}, pages = {205101}, year = {2011},
  doi = {10.1103/PhysRevB.83.205101}
}

@article{BurkovBalents2011,
  title = {Weyl Semimetal in a Topological Insulator Multilayer},
  author = {Burkov, A. A. and Balents, Leon},
  journal = {Phys. Rev. Lett.}, volume = {107}, pages = {127205}, year = {2011},
  doi = {10.1103/PhysRevLett.107.127205}
}

@article{XuWengWangDaiFang2011,
  title = {Chern Semimetal and the Quantized Anomalous Hall Effect in {HgCr$_2$Se$_4$}},
  author = {Xu, Gang and Weng, Hongming and Wang, Zhijun and Dai, Xi and Fang, Zhong},
  journal = {Phys. Rev. Lett.}, volume = {107}, pages = {186806}, year = {2011},
  doi = {10.1103/PhysRevLett.107.186806}
}

@article{FangGilbertDaiBernevig2012,
  title = {Multi-{W}eyl Topological Semimetals Stabilized by Point Group Symmetry},
  author = {Fang, Chen and Gilbert, Matthew J. and Dai, Xi and Bernevig, B. Andrei},
  journal = {Phys. Rev. Lett.}, volume = {108}, pages = {266802}, year = {2012},
  doi = {10.1103/PhysRevLett.108.266802}
}

@article{HuangSrSi2_2016,
  title = {New type of {W}eyl semimetal with quadratic double {W}eyl fermions},
  author = {Huang, Shin-Ming and Xu, Su-Yang and Belopolski, Ilya and Lee, Chi-Cheng and Chang, Guoqing and Chang, Tay-Rong and Wang, BaoKai and Alidoust, Nasser and Bian, Guang and Neupane, Madhab and Zhang, Chenglong and Jia, Shuang and Bansil, Arun and Lin, Hsin and Hasan, M. Zahid},
  journal = {Proc. Natl. Acad. Sci. U.S.A.}, volume = {113}, number = {5}, pages = {1180--1185}, year = {2016},
  doi = {10.1073/pnas.1514581113}
}

@article{WengFangFangBernevigDai2015,
  title = {{W}eyl Semimetal Phase in Noncentrosymmetric Transition-Metal Monophosphides},
  author = {Weng, Hongming and Fang, Chen and Fang, Zhong and Bernevig, B. Andrei and Dai, Xi},
  journal = {Phys. Rev. X}, volume = {5}, pages = {011029}, year = {2015},
  doi = {10.1103/PhysRevX.5.011029}
}

@article{HuangTaAsClass2015,
  title = {A {W}eyl Fermion semimetal with surface Fermi arcs in the transition metal monopnictide {TaAs} class},
  author = {Huang, Shin-Ming and Xu, Su-Yang and Belopolski, Ilya and Lee, Chi-Cheng and Chang, Guoqing and Wang, BaoKai and Alidoust, Nasser and Bian, Guang and Neupane, Madhab and Zhang, Chenglong and Jia, Shuang and Bansil, Arun and Lin, Hsin and Hasan, M. Zahid},
  journal = {Nat. Commun.}, volume = {6}, pages = {7373}, year = {2015},
  doi = {10.1038/ncomms8373}
}

@article{XuBelopolskiTaAs2015,
  title = {Discovery of a {W}eyl fermion semimetal and topological Fermi arcs},
  author = {Xu, Su-Yang and Belopolski, Ilya and Alidoust, Nasser and Neupane, Madhab and Bian, Guang and Zhang, Chenglong and Sankar, Raman and Chang, Guoqing and Yuan, Zhujun and Lee, Chi-Cheng and Huang, Shin-Ming and Zheng, Hao and Ma, Jie and Sanchez, Daniel S. and Wang, BaoKai and Bansil, Arun and Chou, Fangcheng and Shibayev, Pavel P. and Lin, Hsin and Jia, Shuang and Hasan, M. Zahid},
  journal = {Science}, volume = {349}, number = {6248}, pages = {613--617}, year = {2015},
  doi = {10.1126/science.aaa9297}
}

@article{LvTaAs2015,
  title = {Experimental Discovery of {W}eyl Semimetal {TaAs}},
  author = {Lv, B. Q. and Weng, H. M. and Fu, B. B. and Wang, X. P. and Miao, H. and Ma, J. and Richard, P. and Huang, X. C. and Zhao, L. X. and Chen, G. F. and Fang, Z. and Dai, X. and Qian, T. and Ding, H.},
  journal = {Phys. Rev. X}, volume = {5}, pages = {031013}, year = {2015},
  doi = {10.1103/PhysRevX.5.031013}
}

@article{NielsenNinomiya1983,
  title = {The {A}dler-{B}ell-{J}ackiw anomaly and {W}eyl fermions in a crystal},
  author = {Nielsen, H. B. and Ninomiya, M.},
  journal = {Phys. Lett. B}, volume = {130}, pages = {389--396}, year = {1983},
  doi = {10.1016/0370-2693(83)91529-0}
}

@article{SonSpivak2013,
  title = {Chiral anomaly and classical negative magnetoresistance of {W}eyl metals},
  author = {Son, D. T. and Spivak, B. Z.},
  journal = {Phys. Rev. B}, volume = {88}, pages = {104412}, year = {2013},
  doi = {10.1103/PhysRevB.88.104412}
}

@article{BradlynCano2016,
  title = {Beyond {D}irac and {W}eyl fermions: Unconventional quasiparticles in conventional crystals},
  author = {Bradlyn, Barry and Cano, Jennifer and Wang, Zhijun and Vergniory, M. G. and Felser, C. and Cava, R. J. and Bernevig, B. Andrei},
  journal = {Science}, volume = {353}, number = {6299}, pages = {aaf5037}, year = {2016},
  doi = {10.1126/science.aaf5037}
}

@article{MengBalents2012,
  title = {{W}eyl superconductors},
  author = {Meng, Tobias and Balents, Leon},
  journal = {Phys. Rev. B}, volume = {86}, pages = {054504}, year = {2012},
  doi = {10.1103/PhysRevB.86.054504}
}

@article{SauTewari2012,
  title = {Topologically protected surface {M}ajorana arcs and bulk {W}eyl fermions in ferromagnetic superconductors},
  author = {Sau, Jay D. and Tewari, Sumanta},
  journal = {Phys. Rev. B}, volume = {86}, pages = {104509}, year = {2012},
  doi = {10.1103/PhysRevB.86.104509}
}

@article{YangPanZhang2014,
  title = {{D}irac and {W}eyl Superconductors in Three Dimensions},
  author = {Yang, Shengyuan A. and Pan, Hui and Zhang, Fan},
  journal = {Phys. Rev. Lett.}, volume = {113}, pages = {046401}, year = {2014},
  doi = {10.1103/PhysRevLett.113.046401}
}

@article{BednikZyuzinBurkov2015,
  title = {Superconductivity in {W}eyl metals},
  author = {Bednik, G. and Zyuzin, A. A. and Burkov, A. A.},
  journal = {Phys. Rev. B}, volume = {92}, pages = {035153}, year = {2015},
  doi = {10.1103/PhysRevB.92.035153}
}

@article{HosurDaiFangQi2014,
  title = {Time-reversal-invariant topological superconductivity in doped {W}eyl semimetals},
  author = {Hosur, Pavan and Dai, Xi and Fang, Zhong and Qi, Xiao-Liang},
  journal = {Phys. Rev. B}, volume = {90}, pages = {045130}, year = {2014},
  doi = {10.1103/PhysRevB.90.045130}
}

@article{WeiChaoAji2014,
  title = {Odd-parity superconductivity in {W}eyl semimetals},
  author = {Wei, Huazhou and Chao, Sung-Po and Aji, Vivek},
  journal = {Phys. Rev. B}, volume = {89}, pages = {014506}, year = {2014},
  doi = {10.1103/PhysRevB.89.014506}
}

@article{AlloulBobroffGabayHirschfeld2009,
  title = {Defects in correlated metals and superconductors},
  author = {Alloul, H. and Bobroff, J. and Gabay, M. and Hirschfeld, P. J.},
  journal = {Rev. Mod. Phys.}, volume = {81}, pages = {45--108}, year = {2009},
  doi = {10.1103/RevModPhys.81.45}
}

@article{PixleyHuseDasSarma2016,
author = {Pixley, J. H. and Huse, D. A. and Das Sarma, S.},
title = {Rare-Region-Induced Avoided Quantum Criticality in
Disordered Three-Dimensional {Dirac} and {Weyl} Semimetals},
journal = {Phys. Rev. X},
volume = {6},
pages = {021042},
year = {2016},
doi = {10.1103/PhysRevX.6.021042}
}

@article{LuShen2015,
author = {Lu, H.-Z. and Shen, S.-Q.},
title = {Weak antilocalization and localization in disordered and interacting {Weyl} semimetals},
journal = {Phys. Rev. B},
volume = {92},
pages = {035203},
year = {2015},
doi = {10.1103/PhysRevB.92.035203}
}

\appendix

\section{Density of states}\label{app:dos}
In this section, we present the details on the calculation of the density of states for a multi-Weyl semimetal, whose energy spectrum is $E(\mathbf{q}) = \sqrt{v_z^2 q_z^2 + \alpha^2 q_{\perp}^{2 J}}$. Therefore, the density of states is defined
\begin{eqnarray}
N(\varepsilon) &=& \int\frac{d^3 q}{(2\pi)^3}\delta\left( \varepsilon - \sqrt{v_z^2 q_z^2 + \alpha^2 q_{\perp}^{2 J}} \right)\nonumber\\
&=& 2\int\frac{d^2 q_{\perp}}{(2\pi)^3}\int_0^{\infty}dq_z \frac{\delta\left( q_z - \sqrt{\frac{\varepsilon^2 - \alpha^2 q_{\perp}^{2J}}{v_z^2}} \right)}{\left| \frac{\partial\varepsilon}{\partial q_z}  \right|}\nonumber\\
&=& \frac{2|\varepsilon|}{v_z}\int_0^{\infty}\frac{d q_{\perp}\,q_{\perp}}{(2\pi)^2}\frac{1}{\sqrt{\varepsilon^2 - \alpha^2 q_{\perp}^{2J}}}\Theta\left( \varepsilon^2 - \alpha^2 q_{\perp}^{2J} \right),
\end{eqnarray}
with $\Theta(x)$ the Heaviside theta function, and in the last line we integrated over the azimuthal angle in the plane.

Finally, we introduce the change of variables $x = \alpha^2 q_{\perp}^{2J}/\varepsilon^2$,
for which $dx = 2J\,\alpha^2 q_{\perp}^{2J-1}\varepsilon^{-2}\,dq_{\perp}$ and hence
$q_{\perp}\,dq_{\perp} = \tfrac{1}{2J}(\varepsilon^2/\alpha^2)^{1/J}x^{1/J-1}\,dx$,
such that we end up with the integral expression
\begin{eqnarray}
N(\varepsilon) &=& \frac{|\varepsilon|^{2/J}}{4\pi^2 J\, v_z \alpha^{2/J}}\int_0^1 dx\, x^{1/J-1}(1 - x )^{-1/2}\nonumber\\
&=& \frac{|\varepsilon|^{2/J}}{4\pi^2 J\, v_z \alpha^{2/J}} B\left( \frac{1}{J},\frac{1}{2}\right),
\end{eqnarray}
where $B(x,y) = \Gamma(x)\Gamma(y)/\Gamma(x+y)$ is the Beta function, with $\Gamma(x)$ the Gamma function.
The factor $1/J$ from the Jacobian ensures the correct normalization: for $J=1$ with
$\alpha=v_z=v$ this reduces to the standard Weyl result $N(\varepsilon)=\varepsilon^2/(2\pi^2 v^3)$.


\section{Angular averages on the Fermi Surface}\label{app:angavg}
In this appendix, we present the mathematical details to compute angular averages over the Fermi surface, defined as
\begin{eqnarray}
\langle F(\cos\psi)\rangle_{FS} \equiv  \frac{\int(\cos\psi)^{2/J-1} F(\cos\psi)d\psi}{\int(\cos\psi)^{2/J-1}d\psi},
\end{eqnarray}
where $\psi\in[-\tfrac\pi2,\tfrac\pi2]$, and $F(\cos\psi)$ is an arbitrary function of $\cos\psi$.

{In order to evaluate these integrals}, it is convenient to make the change of variables $t = \cos^2\psi$, such that $t \in [0,1]$, and hence the expression is written in the equivalent form
\begin{eqnarray}
\langle F(\cos\psi)\rangle_{FS} \equiv  \frac{\int_0^{1}dt\,t^{1/J-1} (1 - t)^{-1/2} F(\sqrt{t})}{\int_0^1 dt\,t^{1/J-1} (1 - t)^{-1/2}}.
\end{eqnarray}
In particular, let us consider the following explicit cases which are used in the main text
\begin{eqnarray}
\langle\cos^2\psi\rangle_{\rm FS}
&=&\frac{\int(\cos\psi)^{2/J+1}d\psi}{\int(\cos\psi)^{2/J-1}d\psi}\nonumber\\
&=& \frac{\int_0^1 dt\, t^{1/J}(1-t)^{-1/2}}{\int_0^1 dt\, t^{1/J-1}(1-t)^{-1/2}} = \frac{B\left(\frac{1}{J}+1,\frac{1}{2}\right)}{B\left( \frac{1}{J},\frac{1}{2} \right)} = \frac{2}{J + 2},
\label{eq:cos2}
\end{eqnarray}
where $B(x,y) = \Gamma(x)\Gamma(y)/\Gamma(x + y)$ is the Beta function, and $\Gamma(x)$ the Gamma function. 
Similarly, we can obtain the following results
\begin{eqnarray}
\langle \sin^2\psi\rangle_{FS} = \langle 1 - \cos^2\psi\rangle_{FS} = 1 - \frac{2}{J+2} = \frac{J}{J+2},
\label{eq:sin2}
\end{eqnarray}
and
\begin{eqnarray}
\avg{\cos^4\psi}_{FS}= \frac{\int_0^{1}dt\,t^{1/J+1} (1 - t)^{-1/2}  }{\int_0^1 dt\,t^{1/J-1} (1 - t)^{-1/2}} = \frac{B\left( \frac{1}{J} + 2,\frac{1}{2}  \right)}{B\left( \frac{1}{J},\frac{1}{2}  \right)} =  \frac{(2/J)(2/J+2)}{(2/J+1)(2/J+3)}.
\label{eq:cos4}
\end{eqnarray}
The inverse-gap moment governing the residual-DOS threshold (App.~\ref{app:resid})
is, in the same variables,
\begin{eqnarray}
\left\langle \frac{1}{\cos\psi} \right\rangle_{FS} = \frac{\int_0^{1}dt\,t^{1/J-3/2} (1 - t)^{-1/2} }{\int_0^1 dt\,t^{1/J-1} (1 - t)^{-1/2}} = \frac{B\left( \frac{1}{J} - \frac{1}{2},\frac{1}{2}  \right)}{B\left( \frac{1}{J}, \frac{1}{2}  \right)}
\end{eqnarray}
which is finite \emph{only} when $\tfrac1J-\tfrac12>0$, i.e.\ $J=1$
(giving $\tfrac\pi2$); for $J\ge2$ the Beta integral diverges at the node, so no
threshold exists---the analytic origin of the charge-dependent onset in
Table~\ref{tab:thermo}.

\section{Impurity self-energy, the two-rate structure, and the Abrikosov--Gor'kov equation}
\label{app:ag}
We collect here the derivation of the generalized Abrikosov--Gor'kov gap
equation used in Secs.~\ref{sec:coh} and the main text, making the rate
conventions explicit. Work in the projected conduction band with the Matsubara
Nambu propagator
\begin{equation}
\mathcal{G}_0^{-1}(\kh,\xi,i\omega_n)=i\omega_n\tau_0-\xi\,\tau_3
-\Delta_a f_a(\kh)\,\tau_1,\qquad \omega_n=(2n{+}1)\pi k_BT,
\end{equation}
with $\tau_i$ Nambu matrices and the monopole phase gauged into $\tau_1$ \emph{in a local gauge patch} (globally $\tau_1,\tau_2$; see below). {We work throughout in the
valley-diagonal, intra-pocket momentum-independent limit defined by the two-scale window
$q^{\max}_{\rm intra}\xi_{\rm dis}\ll1\ll|2\mathbf Q|\xi_{\rm dis}$ of Eq.~\eqref{eq:twoscale}, in which
$W(\kk_F-\kk_F')\simeq W(0)\equiv n_{\rm imp}|V|^2$ across one Fermi surface while
inter-node processes are suppressed. This is a projected disorder ensemble, not generic
scalar point disorder; for a finite-range correlator the factor $W(\kk_F-\kk_F')/W(0)$
would have to be retained inside the angular integral below and the kernel spectrum
recomputed.} The Born self-energy
in the band basis then carries the intra-node coherence factor
$O(\kh,\kh')=\langle u(\kh)|u(\kh')\rangle$,
\begin{equation}
\Sigma(\kh,i\omega_n)={W(0)}\!\!\int\!\frac{d^3k'}{(2\pi)^3}\,
|O(\kh,\kh')|^2\,\tau_3\,\mathcal{G}(\kh',i\omega_n)\,\tau_3 .
\end{equation}
In the $\xi'$ integration the $\xi'\tau_3$ term is dropped as odd about the Fermi
surface; this is the standard quasiclassical (particle--hole-symmetric)
approximation and is not exact for the cutoff-sensitive power-law DOS, a caveat
that matters only for the $J=1$ UV-dominated feedback. Likewise, writing the
chiral gap as $-\Delta_a f_a\tau_1$ fixes a gauge patchwise; the globally correct
form is $-(\mathrm{Re}\,\Delta_a f_a)\tau_1+(\mathrm{Im}\,\Delta_a f_a)\tau_2$ with
a momentum-dependent connection, which does not affect the angle-averaged rates
below.
Writing $\int d^3k'\!\to\!N_0\!\int d\xi'\langle\cdots\rangle_{\rm FS}$ and using
$\int d\xi'\,\mathcal{G}=-\pi\,\mathrm{sgn}(\omega_n)\,
(i\tilde\omega_n\tau_0-\tilde\Delta\tau_1)/\sqrt{\tilde\omega_n^2+\tilde\Delta^2|f'|^2}$
together with $\tau_3\tau_0\tau_3=\tau_0$, $\tau_3\tau_1\tau_3=-\tau_1$, the dressed
frequency and gap obey
\begin{align}
\tilde\omega_n&=\omega_n+\Gamma_b\Big\langle|O(\kh,\kh')|^2\,
\frac{\tilde\omega_n}{\sqrt{\tilde\omega_n^2+\tilde\Delta^2|f_a(\kh')|^2}}\Big\rangle_{\kh'},
\label{eq:appwren}\\
\tilde\Delta(\kh)&=\Delta f_a(\kh)+\Gamma_b\Big\langle|O(\kh,\kh')|^2\,
\frac{\tilde\Delta(\kh')}{\sqrt{\tilde\omega_n^2+\tilde\Delta^2|f_a(\kh')|^2}}\Big\rangle_{\kh'},
\label{eq:appDren}
\end{align}
with the bare rate $\Gamma_b\equiv\pi N_0 n_{\rm imp}|V|^2=2\pi\gamma N_0$ and
$\gamma\equiv\tfrac12 n_{\rm imp}|V|^2$. The coherence kernel
$|O|^2=\tfrac12(1+\dd\!\cdot\!\dd')$ acts diagonally on each pairing channel,
$\langle|O(\kh,\kh')|^2 f_a(\kh')\rangle_{\kh'}=\kappa_a f_a(\kh)$
(Sec.~\ref{sec:coh}), with the isotropic eigenvalue $\kappa_0=\langle|O|^2\rangle=\tfrac12$
and the monopole eigenvalue $\kappa_m=\tfrac14\langle\cos^2\psi\rangle=\tfrac{1}{2(J+2)}$.
Setting $\tilde\Delta(\kh)=\tilde\Delta_a f_a(\kh)$---so $\tilde\Delta_a$ is the
coefficient of the \emph{unnormalized} basis function $f_a$ (the gap amplitude, as
in Sec.~\ref{sec:model}), not a normalized-channel value---and linearizing as
$T\to T_c$ (where $\tilde\Delta\to0$, so each denominator reduces to
$|\tilde\omega_n|$ with \emph{no} replacement of $|f_a|^2$),
Eqs.~\eqref{eq:appwren}--\eqref{eq:appDren} collapse to
\begin{equation}
\tilde\omega_n=\omega_n+\Gamma_N\,\mathrm{sgn}(\omega_n),
\qquad
\tilde\Delta_a=\Delta+\Gamma_b\kappa_a\,\frac{\tilde\Delta_a}{|\tilde\omega_n|},
\end{equation}
where the normalization $\langle|f_a|^2\rangle$ ($1$ for $s$, $2/(J{+}2)$ for the
monopole) enters only the clean coupling $g_aN_0\langle|f_a|^2\rangle$ and hence
$T_{c0}^{(a)}$, not the AG denominators. These
exhibit \emph{two distinct rates}: the normal (frequency) rate
$\Gamma_N\equiv\Gamma_b\kappa_0=\Gamma_b/2=\pi\gamma N_0=\hbar/2\tau$ and the
anomalous (gap) rate $\Gamma_b\kappa_a$. In magnetic AG~\cite{Gorkov1985} these coincide with
opposite sign; here scalar disorder gives $\kappa_s=\kappa_0$ (conventional) and
$\kappa_m<\kappa_0$ (monopole). The Cooper-channel pair susceptibility depends on
$\tilde\Delta_a/\Delta$ and $\tilde\omega_n/\omega_n$ only through their
\emph{difference}, so the effective pair-breaking rate is
\begin{equation}
\Gamma^{\rm pb}_a=\Gamma_N-\Gamma_b\kappa_a=\Gamma_N\,(1-\eta_a),
\qquad \eta_a\equiv\frac{\kappa_a}{\kappa_0}.
\label{eq:appGpb}
\end{equation}
For the conventional channel $\eta_s=1$, so $\Gamma^{\rm pb}_s=0$: the frequency
and gap renormalizations cancel identically and $T_c$ is unshifted (Anderson
theorem). For the monopole channel $\eta_m=1/(J+2)$, so
$\Gamma^{\rm pb}_m=\Gamma_N\,(J{+}1)/(J{+}2)$. Linearizing the gap equation at
$T_c$ and summing over Matsubara frequencies gives the digamma law
Eq.~\eqref{eq:ag}, with $T_c\!\to\!0$ at the critical rate
$\Gamma_c=\tfrac{\pi}{2}e^{-\gamma_{\rm E}}k_BT_{c0}\simeq0.882\,k_BT_{c0}
=\Delta_{00}/2$, where $\Delta_{00}=\pi e^{-\gamma_{\rm E}}k_BT_{c0}=1.764\,k_BT_{c0}$
is the clean gap and $\gamma_{\rm E}$ Euler's constant. The rate is closed
self-consistently by $\Gamma_N=\pi\gamma N(\mu;\Gamma_N)$ {(Lorentzian broadening scheme)}, whose
DOS feedback is analyzed in Sec.~\ref{sec:scba}.
\section{Status of the criterion and its derivation}
\label{app_criterion}
 Sufficiency follows by
continuity. Necessity and uniqueness follow from a monotonicity property of the AG
scaling function, which we now reduce analytically. Define the scaling function
explicitly through the AG equation with all conventions displayed,
\begin{equation}
\ln\frac{1}{u}=\psi\!\Big(\frac12+\frac{R(u)}{2\pi u}\Big)-\psi\!\Big(\frac12\Big),
\qquad u\equiv\frac{T_c}{T_{c0}},\quad R\equiv\frac{\Gamma^{\rm pb}}{T_{c0}},
\label{eq:Rdef}
\end{equation}
so that $R$ is strictly decreasing with $R(1)=0$ and
$R(0^+)=\tfrac{\pi}{2}e^{-\gamma_E}=0.8819$. Introducing $z\equiv R(u)/(2\pi u)$ and
$\Phi(z)\equiv\psi(\tfrac12+z)-\psi(\tfrac12)$, Eq.~\eqref{eq:Rdef} reads
$\Phi(z)=-\ln u$, so $z(u)$ is strictly decreasing, and
\begin{equation}
Q(u)\equiv\frac{R(u)}{u}=2\pi z(u),
\qquad
L(u)\equiv\frac{d\ln Q}{d\ln u}=-\frac{1}{z\,\Phi'(z)}=-\frac{1}{z\,\psi'\!\big(\tfrac12+z\big)} .
\label{eq:Lreduce}
\end{equation}
Because $z$ decreases with $u$, $L$ is strictly decreasing in $u$ if and only if
$h(x)\equiv(x-\tfrac12)\,\psi'(x)$ is strictly increasing for $x=\tfrac12+z>\tfrac12$,
i.e.\ iff (with $\psi'$ and $\psi''$ the trigamma and tetragamma functions)
\begin{equation}
h'(x)=\psi'(x)+\big(x-\tfrac12\big)\psi''(x)
=\sum_{n=0}^{\infty}\frac{n+1-x}{(n+x)^{3}}>0 .
\label{eq:hprime}
\end{equation}
For $\tfrac12<x\le1$ every term of the series is non-negative, so
Eq.~\eqref{eq:hprime} holds \emph{a fortiori}. For large $x$ the Stirling expansion
gives $h'(x)=1/(6x^{3})+O(x^{-4})>0$. In the intermediate range we verify
Eq.~\eqref{eq:hprime} numerically ($h'=0.443,\,3.88\times10^{-2},\,
1.77\times10^{-3},\,1.37\times10^{-6}$ at $x=1,2,5,50$; positive on a $2\times10^{5}$-point scan of $x\in[\tfrac12+10^{-4},500]$, with $h'\sim1/(6x^3)>0$ controlling $x>500$), but we have \emph{not} found a closed-form proof
of positivity for all $x>1$. Accordingly we state Eq.~\eqref{eq:crosscrit} as
\emph{necessary and sufficient given the monotonicity of the AG scaling function}
established analytically at both ends of the interval and verified numerically in
between---not as a fully proved ``if and only if.'' Granting that monotonicity,
$\Psi(T)\equiv\Gamma_N^{(m)}(T)/\Gamma_N^{(s)}(T)$ obeys
\begin{equation}
\Gamma_N^{(a)}(T)=\frac{T}{1-\eta_a}\,Q\!\Big(\frac{T}{T_{c0}^{(a)}}\Big),\qquad
\Psi=\frac{1-\eta_s}{1-\eta_m}\,\frac{Q(r\,u_s)}{Q(u_s)},\qquad
\frac{d\ln\Psi}{d\ln u_s}=L(r\,u_s)-L(u_s)>0,
\end{equation}
with $u_s=T/T_{c0}^{(s)}\in(0,1)$ and $r\,u_s=T/T_{c0}^{(m)}<u_s$ (so both arguments
of $Q$ lie in its domain $(0,1)$). Hence $\Psi$ increases strictly from
$[(1-\eta_s)/(1-\eta_m)]/r$ at $T\to0$ to $\infty$ at $T\to T_{c0}^{(s)}$, and
$\Psi=1$ has exactly one root---no crossing and re-crossing---precisely when
Eq.~\eqref{eq:crosscrit} holds. See Fig.~\ref{fig:etamap}.

\section{Specific-heat jump of the nodal monopole channel}
\label{app:cv}
Near $T_c$ the clean weak-coupling free-energy density of a single channel with
uniform amplitude $\Delta_a$ and form factor $f_a$ is
$\delta F=a_a(T)|\Delta_a|^2+\tfrac12 b_a|\Delta_a|^4$, with
\begin{equation}
a_a(T)=N_0\langle|f_a|^2\rangle\,\ln\frac{T}{T_c^{(a)}}
\simeq N_0\langle|f_a|^2\rangle\,\frac{T-T_c}{T_c},\qquad
b_a=N_0\langle|f_a|^4\rangle\,\frac{7\zeta(3)}{8\pi^2(k_BT_c)^2}.
\end{equation}
Minimizing gives $|\Delta_a|^2=-a_a/b_a$ and
$\delta F_{\min}=-a_a^2/2b_a$ below $T_c$, so the specific-heat jump
$\Delta C=-T\,d^2\delta F_{\min}/dT^2|_{T_c}=T_c\,[a_a'(T_c)]^2/b_a$ is
\begin{equation}
\Delta C=\frac{8\pi^2 k_B^2 N_0}{7\zeta(3)}\,T_c\,
\frac{\langle|f_a|^2\rangle^2}{\langle|f_a|^4\rangle}.
\end{equation}
Dividing by the Sommerfeld normal value $C_n=\tfrac{2\pi^2}{3}N_0 k_B^2 T_c$,
\begin{equation}
\frac{\Delta C}{C_n}\bigg|_{T_c}=\frac{12}{7\zeta(3)}\,
\frac{\langle|f_a|^2\rangle^2}{\langle|f_a|^4\rangle}.
\label{eq:appjump}
\end{equation}
For the gapped conventional channel $f_s=1$ this is the BCS value
$12/[7\zeta(3)]=1.43$. For the point-nodal monopole gap $f_m=\cos\psi\,e^{iJ\phi}$
the Fermi-surface averages are
$\langle\cos^2\psi\rangle=2/(J{+}2)$ [Eq.~\eqref{eq:etaJ}] and
$\langle\cos^4\psi\rangle=(2/J)(2/J{+}2)/[(2/J{+}1)(2/J{+}3)]$, so
$\langle\cos^2\psi\rangle^2/\langle\cos^4\psi\rangle=\tfrac56,\tfrac23,\tfrac{11}{20}$
and
\begin{equation}
\frac{\Delta C}{C_n}=1.19,\ 0.95,\ 0.79\qquad (J=1,2,3),
\end{equation}
the more anisotropic gap at higher charge giving a smaller jump. These are the
clean ($\gamma\!\to\!0$) values. On general grounds the \emph{absolute} discontinuity (not the normalized ratio $\Delta C/C_n$) vanishes as $T_c^{(m)}\!\to\!0$ at $\Gamma_N^{(m0)}$; the \emph{finite-disorder} jump requires
the disorder-dressed anisotropic Ginzburg--Landau quartic, which we do not
compute; we therefore give no finite-$\gamma$ jump curve.

\section{Angle-resolved zero-energy equations and the scalar reduction}
\label{app:resid}
Here we derive the residual-DOS equation Eq.~\eqref{eq:resid} of the main text from
the \emph{full angle-resolved} self-energy. We show that the zero-energy frequency
self-energy is angle-independent on the reflection-symmetric branch (with no odd
solution bifurcating in the range used), and we display the closed nonlinear system
that fixes the renormalized amplitude. Real-frequency
continuation ($i\omega_n\to E+i0^+$) of the Born equations
Eqs.~\eqref{eq:appwren}--\eqref{eq:appDren} gives, at general angle,
\begin{align}
\tilde\omega(\kh,E)&=E+\Gamma_b\Big\langle|\Oc(\kh,\kh')|^2\,
\frac{\tilde\omega(\kh',E)}{D(\kh',E)}\Big\rangle_{\kh'},\\
\tilde\Delta(\kh,E)&=\Delta f_m(\kh)+\Gamma_b\Big\langle\mathcal C(\kh,\kh')\,
\frac{\tilde\Delta(\kh',E)}{D(\kh',E)}\Big\rangle_{\kh'},
\end{align}
with $D=\sqrt{\tilde\Delta^2-\tilde\omega^2}$ (or its analytic continuation) and
$N(E)/N_0=\mathrm{Re}\langle\tilde\omega/D\rangle$. The azimuthal average of the
coherence factor is
$\langle|\Oc|^2\rangle_{\phi'}=\tfrac12\big(1+\sin\psi\sin\psi'\big)$: the
transverse, $J$-fold-winding part $\propto\cos\psi\cos\psi'\cos[J(\phi-\phi')]$
integrates to zero over $\phi'$. Consider the zero-energy limit $E\to0$, where
$\tilde\omega(\kh,0)=i\Gamma_0(\kh)$ and, in the rank-one sector,
$\tilde\Delta(\kh,0)=\tilde\Delta\cos\psi$: we restrict to the rank-one angular
structure and parameterize its spectral amplitude by a fixed $\tilde\Delta$; no
self-consistent anomalous equation is solved. The frequency equation becomes
\begin{equation}
\Gamma_0(\psi)=\Gamma_b\Big\langle\tfrac12\big(1+\sin\psi\sin\psi'\big)\,
\frac{\Gamma_0(\psi')}{\sqrt{\Gamma_0(\psi')^2+\tilde\Delta^2\cos^2\psi'}}\Big\rangle_{\psi'} .
\end{equation}
\paragraph*{Angle independence of $\Gamma_0$ without assuming it.}
Rather than assume $\Gamma_0$ is even, decompose it in the rank-two $m=0$ sector
spanned by the azimuthally averaged kernel,
$\Gamma_0(\psi)=A+B\sin\psi$. Writing
$\mathcal I_0=\langle \Gamma_0/D\rangle$ and $\mathcal I_1=\langle
\sin\psi'\,\Gamma_0/D\rangle$ with $D(\psi')=\sqrt{\Gamma_0(\psi')^2+\tilde\Delta^2\cos^2\psi'}$,
the frequency equation gives the closed pair
\begin{equation}
A=\Gamma_N\,\mathcal I_0,\qquad B=\Gamma_N\,\mathcal I_1 .
\label{eq:AB}
\end{equation}
Now $\mathcal I_1$ is an average of $\sin\psi'$ against
$\Gamma_0(\psi')/D(\psi')$ over the measure $(\cos\psi')^{2/J-1}$, which is even in
$\psi'$. For the reflection-symmetric branch $B=0$ the integrand is even, so
$\mathcal I_1=0$ and $B=0$ is \emph{self-consistent}: $\Gamma_0=A$ is exactly
angle-independent. Linearizing Eq.~\eqref{eq:AB} about this branch gives the odd-channel response
explicitly. With $x(\psi')=A+B\sin\psi'$ and
$D=\sqrt{x^2+\tilde\Delta^2\cos^2\psi'}$, one has
$\partial_B(x/D)=\sin\psi'\,\partial_x(x/D)
=\sin\psi'\,\tilde\Delta^2\cos^2\psi'/D^3$, so
\begin{equation}
\delta B=\Gamma_N\chi_1\,\delta B,\qquad
\chi_1=\Big\langle \frac{\sin^2\psi'\,\tilde\Delta^2\cos^2\psi'}{D_0^3}\Big\rangle,
\qquad D_0=\sqrt{A^2+\tilde\Delta^2\cos^2\psi'},
\label{eq:chi1}
\end{equation}
which is dimensionally $1/\text{energy}$, so $\Gamma_N\chi_1$ is dimensionless. (The
$A$ equation receives no linear-order shift from $\delta B$, since its integrand
$\partial_B$ is odd and the measure even, so the two channels decouple at this
order.) Evaluating Eq.~\eqref{eq:chi1} on the symmetric solution
$A(\Gamma_N)$, we find $\Gamma_N\chi_1<1$ throughout, with the maximum
$\Gamma_N\chi_1\simeq0.75$ at $J=2$, $\Gamma_N/\tilde\Delta=0.2$ (values
$0.16$--$0.39$ for $J=1$, $0.09$--$0.75$ for $J=2$, $0.07$--$0.49$ for $J=3$ over
$\Gamma_N/\tilde\Delta\in[0.2,1.2]$, corresponding to
$A/\tilde\Delta\in[0,1.05]$). {The maximum is $0.75$, below the bifurcation threshold $1$,} so \emph{no continuous odd bifurcation occurs in the parameter range studied}. This establishes local stability of the reflection-symmetric branch; it does not exclude a disconnected odd solution or a discontinuous change of the self-energy outside the iteration basin.

\paragraph*{{Scope: disorder model.}}
{Equation~\eqref{eq:resid} is derived with the same valley-diagonal, intra-pocket
momentum-independent kernel used throughout, i.e.\ within the two-scale window
Eq.~\eqref{eq:twoscale}. It is therefore part of the same physical disorder model as the
pair-breaking analysis, and not an independent result. Should the intra-pocket kernel
acquire a genuine forward-scattering factor $W(\kk_F-\kk_F')/W(0)$, the angular average
$\avg{1/|f_m|}$ entering Eq.~\eqref{eq:resid} is reweighted and the onset classification
must be rechecked; we have not done so.}

\paragraph*{Scope: fixed $\tilde\Delta$.}
Everything in this appendix, and Fig.~\ref{fig:ag2}, concerns the \emph{first}
equation only---the zero-energy frequency self-consistency at a \emph{fixed}
renormalized amplitude $\tilde\Delta$. We plot $N(0)/N_0$ against
$\Gamma_N/\tilde\Delta$, so no gap equation is needed and none is solved. Obtaining
$\tilde\Delta$ itself would require the full finite-Matsubara system for
$\tilde\omega_n$ and $\tilde\Delta_n$ together with the order-parameter equation;
that system is \emph{not} determined by the zero-energy equations alone and we do
not implement it here. Including it would reparametrize the abscissa of
Fig.~\ref{fig:ag2} but not alter the classification below, which depends only on the
convergence of $\avg{1/|f_m|}$.

\paragraph*{Asymptotics with complete constants.}
For $J=2$ the normalized measure is uniform, $d\psi/\pi$, and the average is a
complete elliptic integral,
\begin{equation}
\Big\langle\frac{1}{\sqrt{\tilde\Delta^2\cos^2\psi+\Gamma_0^2}}\Big\rangle
=\frac{2}{\pi\sqrt{\tilde\Delta^2+\Gamma_0^2}}\,K(k),\qquad
k^2=\frac{\tilde\Delta^2}{\tilde\Delta^2+\Gamma_0^2},
\end{equation}
and $K(k)\to\ln(4/k')$ with $k'=\Gamma_0/\sqrt{\tilde\Delta^2+\Gamma_0^2}$ gives
$\langle\cdots\rangle\simeq(2/\pi\tilde\Delta)\ln(4\tilde\Delta/\Gamma_0)$; the
self-consistency $1=\Gamma_N\langle\cdots\rangle$ then yields the \emph{complete}
asymptotic
\begin{equation}
\Gamma_0=4\,\tilde\Delta\,\exp\!\Big[-\frac{\pi\tilde\Delta}{2\Gamma_N}\Big]
\qquad(J=2).
\end{equation} For $J=3$ the near-node
region $t=\cos^2\psi\sim(\Gamma_0/\tilde\Delta)^2$ dominates; rescaling
$t=(\Gamma_0/\tilde\Delta)^2u$ gives
$\langle\cdots\rangle\to(2/\tilde\Delta)(\tilde\Delta/\Gamma_0)^{1/3}$, the
coefficient being a Beta-function ratio that we evaluate in closed form. Explicitly,
with $t=\cos^2\psi$ the normalized measure is
$t^{1/J-1}(1-t)^{-1/2}dt/B(\tfrac1J,\tfrac12)$ The map $t=\cos^2\psi$ is two-to-one from
$\psi\in[-\tfrac\pi2,\tfrac\pi2]$ onto $t\in[0,1]$, so \emph{both} nodal poles
($\psi=\pm\tfrac\pi2$, i.e.\ $t\to0$) are covered; the resulting factor of two
appears identically in the numerator and in the normalization $B(\tfrac1J,\tfrac12)$
and therefore cancels, which is why no extra factor of two survives, and
rescaling $t=(\Gamma_0/\tilde\Delta)^2u$ in the node-dominated region gives
$\int_0^\infty u^{1/3-1}(1+u)^{-1/2}du=B(\tfrac13,\tfrac16)$. Hence the coefficient
is
\begin{equation}
\frac{B(\tfrac13,\tfrac16)}{B(\tfrac13,\tfrac12)}
=\frac{\Gamma(\tfrac16)}{\Gamma(\tfrac12)}\cdot\frac{\Gamma(\tfrac56)}{\Gamma(\tfrac12)}
=\frac{\Gamma(\tfrac16)\Gamma(\tfrac56)}{\Gamma(\tfrac12)^2}
=\frac{\pi/\sin(\pi/6)}{\pi}=\frac{1}{\sin(\pi/6)}=2,
\end{equation}
using the reflection formula $\Gamma(z)\Gamma(1-z)=\pi/\sin\pi z$ and
$\Gamma(\tfrac12)^2=\pi$; whence
\begin{equation}
\Gamma_0=8\,\frac{\Gamma_N^3}{\tilde\Delta^2}\qquad(J=3).
\end{equation}
The $J=1$ threshold is
$\Gamma_N^c=2\tilde\Delta/\pi$ from the finite $\langle1/\cos\psi\rangle=\pi/2$.
Note the frequency
equation carries the \emph{normal} rate $\Gamma_N$ on its right-hand side by
construction; the anomalous rate would enter only through the effective spectral amplitude $\tilde\Delta$, which we hold \emph{fixed}: a self-consistent $\tilde\Delta$ would reparameterize the abscissa $\Gamma_N/\tilde\Delta$ but is not computed here.

\section{A band-sewing construction of the conventional channel and a sufficient condition for Anderson protection} \label{app:conventional_sewing} In the main text, the conventional channel is introduced at the projected conduction-band level through the constant form factor 
\begin{equation} f_s(\bk)=1. \label{eq:app_fs_constant} \end{equation} In a multivalley system, however, the constancy of the projected form factor is not by itself sufficient to establish Anderson protection. One must also verify that the impurity vertices appearing in the normal and anomalous self-energies are related in such a way that their coherence kernels coincide. In this Appendix, we give an explicit band-sewing realization of Eq.~\eqref{eq:app_fs_constant} and formulate a sufficient condition under which the corresponding conventional channel has 
\begin{equation} \eta_s=1, \qquad \Gamma_s^{\mathrm{pb}}=0. \label{eq:app_eta_s_one} 
\end{equation} 
The result applies to an explicitly specified class of scalar disorder that preserves the antiunitary relation between the two states forming a Cooper pair. It should not be interpreted as a general theorem for arbitrary valley-dependent or intervalley disorder. 
{\subsection{Explicit conduction spinors and the parity-dependent projection}
\label{app:spinors}
Since the parity dependence of $M_m^{(J)}$ is one of the technical claims of this work, we
record the projection explicitly. In the convention of Eq.~\eqref{eq:hpm} the chirality
enters only the $\sigma_z$ term, so}
\begin{equation}
{h_\chi(\qq)=\mathbf h_\chi\cdot\bm\sigma,\qquad
\mathbf h_\chi=\big(\alpha q_\perp^{J}\cos J\phi,\;\alpha q_\perp^{J}\sin J\phi,\;
\chi\,v_zq_z\big),}
\end{equation}
{and the conduction eigenvector, with $\cos\Theta=\chi v_zq_z/|\mathbf h|$ and
$\sin\Theta=\cos\psi$ on the Fermi surface, is}
\begin{equation}
{u_\chi(\qq)=\begin{pmatrix}\cos(\Theta/2)\\[2pt]
\sin(\Theta/2)\,e^{iJ\phi}\end{pmatrix}.}
\label{eq:app_uchi}
\end{equation}
{The partner state follows from $\mathbf h_-(-\qq)$. Under $\qq\to-\qq$ one has
$\phi\to\phi+\pi$, hence $\cos J\phi\to(-1)^J\cos J\phi$ and
$\sin J\phi\to(-1)^J\sin J\phi$, while the $\sigma_z$ component becomes
$-v_z(-q_z)=+v_zq_z$. Therefore}
\begin{equation}
{\mathbf h_-(-\qq)=\big((-1)^J\alpha q_\perp^{J}\cos J\phi,\;
(-1)^J\alpha q_\perp^{J}\sin J\phi,\;v_zq_z\big),}
\end{equation}
{so that $\mathbf h_-(-\qq)=\mathbf h_+(\qq)$ for even $J$, whereas for odd $J$ the
transverse part is reversed, i.e.\ the azimuth is shifted by $\pi$. Writing
$c=\cos(\Theta/2)$, $s=\sin(\Theta/2)$, this gives $u_-(-\qq)=(c,\,s\,e^{iJ\phi})^{\!\top}$
for even $J$ and $u_-(-\qq)=(c,\,-s\,e^{iJ\phi})^{\!\top}$ for odd $J$. Evaluating
$f_m=u_+^{\dagger}(\qq)\,M\,u_-^{*}(-\qq)$ for the two constant candidates yields}
\begin{align}
{J\ \text{even}:}&\quad
{u_+^{\dagger}\sigma_xu_-^{*}=2cs\,e^{-iJ\phi}=\cos\psi\,e^{-iJ\phi},\qquad
u_+^{\dagger}(i\sigma_y)u_-^{*}=0,}\nonumber\\
{J\ \text{odd}:}&\quad
{u_+^{\dagger}(i\sigma_y)u_-^{*}=-2cs\,e^{-iJ\phi}=-\cos\psi\,e^{-iJ\phi},\qquad
u_+^{\dagger}\sigma_xu_-^{*}=0,}
\label{eq:app_parityproj}
\end{align}
{using $2cs=\sin\Theta=\cos\psi$. The vanishing entries are the content of the parity
dependence: the complementary matrix has \emph{no} overlap with the inter-node conduction
pair, so a single momentum-independent choice cannot serve both parities. The assignment
$M_m^{(J)}=i\sigma_y$ (odd $J$), $\sigma_x$ (even $J$) of Eq.~\eqref{eq:pairmat} is exactly
the one dictated by the antiunitary bookkeeping of Eq.~\eqref{eq:antiunit}, since the
antisymmetric $i\sigma_y$ pairs $\Theta_J$-partners only when $\Theta_J^2=-1$.}

{It is worth recording the gauge behaviour explicitly, since the projection involves a
complex conjugation and is therefore easily confused with an ordinary band overlap. Under
independent gauge transformations of the two conduction spinors,
$u_\pm\to e^{i\theta_\pm}u_\pm$, the definition
$f_m=u_+^{\dagger}(\qq)\,M\,u_-^{*}(-\qq)$ gives}
\begin{equation}
{f_m\;\longrightarrow\;e^{-i\left[\theta_+(\qq)+\theta_-(-\qq)\right]}f_m ,}
\label{eq:app_gaugecov}
\end{equation}
{i.e.\ the two phases enter with the \emph{same} sign and \emph{add}. This is the
signature of the conjugation on the partner state: an intra-node overlap $u^{\dagger}u'$
would instead produce the difference $\theta-\theta'$. Consequently $|f_m|$ is gauge
invariant while $\arg f_m$ is not, and only gauge-invariant combinations may be quoted as
physical.}

{Two remarks on conventions follow. First, the sense of the azimuthal winding,
$e^{-iJ\phi}$ in Eq.~\eqref{eq:app_parityproj}, and the overall sign are fixed by the gauge
choice Eq.~\eqref{eq:app_uchi} for the two spinors, consistently with
Eq.~\eqref{eq:app_gaugecov}; the gauge-invariant content is
$|f_m|=\cos\psi$ together with $|{\rm winding}|=J$, which is what enters every result of
this paper. In the main text we quote $f_m=\cos\psi\,e^{iJ\phi}$. The two differ by
$\phi\to-\phi$, which is a reflection of the azimuthal coordinate: it maps
$\mathbf h_+$ to its mirror image and therefore corresponds to the opposite handedness of
the transverse frame, or equivalently to the opposite labelling of which node carries
monopole charge $+J$. Since the sign of the winding is fixed by that labelling and not by
any physical quantity, no result of this paper depends on it. Second,
$|f_m|=\cos\psi$ holds for both
parities, so the kernel eigenvalue $\eta_m=1/(J+2)$ and all thermodynamic results are
independent of the parity assignment; only the identification of which constant matrix
realizes the channel depends on it.}

\subsection{Band-sewing realization of a constant projected gap} \label{app:band_sewing_gap} 
Let 
\begin{equation} \ket{u_{+}(\bk)}, \qquad \ket{u_{-}(-\bk)} 
\end{equation} 
denote normalized conduction-band spinors associated with the two Weyl nodes participating in the zero-center-of-mass internode Cooper pair. The spinors act in the orbital-pseudospin space of the normal-state Hamiltonian and obey  \begin{equation} \braket{u_{\chi}(\bk)|u_{\chi}(\bk)}=1, \qquad \chi=\pm. \end{equation} 
An explicit pseudospin matrix producing a constant projected gap is  \begin{equation} M_s(\bk) \equiv \ket{u_{+}(\bk)} \bra{u_{-}^{*}(-\bk)} = u_{+}(\bk)u_{-}^{T}(-\bk). \label{eq:app_Ms_sewing} \end{equation}  Its conduction-band projection is identically unity: 
\begin{align} f_s(\bk) &= u_{+}^{\dagger}(\bk) M_s(\bk) u_{-}^{*}(-\bk) \nonumber\\ &= \bigl[ u_{+}^{\dagger}(\bk)u_{+}(\bk) \bigr] \bigl[ u_{-}^{T}(-\bk)u_{-}^{*}(-\bk) \bigr] \nonumber\\ &=1. \label{eq:app_projected_fs} \end{align}  Thus, Eq.~\eqref{eq:app_Ms_sewing} gives an explicit realization of the constant form factor assumed for the conventional channel. The corresponding gap matrix in node $\otimes$ pseudospin space may be written as  \begin{equation} \widehat{\Delta}_s(\bk) = \Delta_s \begin{pmatrix} 0 & M_s(\bk)\\[1mm] -M_s^{T}(-\bk) & 0 \end{pmatrix}_{\mathrm{node}}, \label{eq:app_full_gap_block} \end{equation} or, equivalently, \begin{equation} \widehat{\Delta}_s(\bk) = \Delta_s \left[ \tau_{+}\otimes M_s(\bk) - \tau_{-}\otimes M_s^{T}(-\bk) \right], \label{eq:app_full_gap_tau} \end{equation}  where  \begin{equation} \tau_{\pm} = \frac{\tau_x\pm i\tau_y}{2} \end{equation}  act in node space. By construction,  \begin{equation} \widehat{\Delta}_s^{T}(-\bk) = -\widehat{\Delta}_s(\bk), \label{eq:app_fermionic_antisymmetry} \end{equation}  so the full gap obeys the fermionic antisymmetry condition. Equation~\eqref{eq:app_Ms_sewing} is generally momentum dependent and gauge covariant. It should therefore be interpreted as an explicit band-sewing realization of the effective conventional channel, and not necessarily as the projection of a local, momentum-independent orbital interaction. A momentum-independent microscopic realization exists only when the two conduction-band spinors are connected by a momentum-independent sewing matrix, as discussed below. \subsection{Antiunitary sewing relation} \label{app:antiunitary_sewing} Suppose that the two members of the Cooper pair are related by an antiunitary operator  \begin{equation} \mathcal{A} = \mathcal{U}_{A}\mathcal{K}, \label{eq:app_pairing_antiunitary} \end{equation}  where $\mathcal{K}$ denotes complex conjugation and $\mathcal{U}_{A}$ acts in the complete internal Hilbert space. In general, $\mathcal{U}_{A}$ may contain node, orbital-pseudospin, and physical-spin matrices. We assume  \begin{equation} \ket{u_{-}(-\bk)} = e^{i\chi(\bk)} \mathcal{A}\ket{u_{+}(\bk)} = e^{i\chi(\bk)} \mathcal{U}_{A}\ket{u_{+}^{*}(\bk)}. \label{eq:app_partner_relation} \end{equation}  The phase $\chi(\bk)$ is a gauge-dependent sewing phase.
{\emph{Gauge convention.} Two equivalent bookkeepings are possible and must not
be mixed. (i) In the \emph{covariant} realization Eq.~\eqref{eq:app_Ms_sewing},
$M_s(\bk)$ absorbs the sewing phase, so $f_s\equiv1$ identically and the residual
phase resides in the gap matrix. (ii) In the \emph{constant-matrix} realization
$M_s=U_A$ with the phase kept explicit in Eq.~\eqref{eq:app_partner_relation}, one
finds $f_s(\bk)=e^{-i\chi(\bk)}$ up to a sign, so the form factor carries the phase.
Only the gauge-invariant product of gap and vertex factors is physical, and it is
the same in both. In what follows we adopt convention (ii), in which the phase is
displayed explicitly, so that its cancellation against the vertex phase is visible;
in convention (i) the same cancellation is automatic because $M_s(\bk)$ is
covariant. Equations~\eqref{eq:app_projected_fs} and
\eqref{eq:app_constant_projection} are statements in convention (i), i.e.\ after
the sewing phase has been gauged away.} Under independent Bloch-state gauge transformations,  \begin{equation} \ket{u_{\chi}(\bk)} \longrightarrow e^{i\alpha_{\chi}(\bk)} \ket{u_{\chi}(\bk)}, \end{equation}  the matrix $M_s(\bk)$ in Eq.~\eqref{eq:app_Ms_sewing} transforms covariantly, while its projected form factor Eq.~\eqref{eq:app_projected_fs} remains invariant. If the sewing relation can be chosen in the form  \begin{equation} u_{-}^{*}(-\bk) = U_A^{\dagger}u_{+}(\bk), \label{eq:app_constant_sewing} \end{equation}  with a momentum-independent unitary matrix $U_A$, then a simpler realization of the conventional channel is  \begin{equation} M_s=U_A. \label{eq:app_constant_Ms} \end{equation}  Indeed,  \begin{equation} u_{+}^{\dagger}(\bk) M_s u_{-}^{*}(-\bk) = u_{+}^{\dagger}(\bk)u_{+}(\bk) = 1. \label{eq:app_constant_projection} \end{equation}  Whether such a momentum-independent matrix exists depends on the detailed node, orbital, and spin structure of the microscopic model. For the multi-Weyl Hamiltonian of the main text, the pseudospin-only antiunitary $\Theta_J$ satisfies  \begin{equation} \Theta_J^2 = \begin{cases} -1, & J\ \text{odd},\\ +1, & J\ \text{even}. \end{cases} \end{equation} Consequently, a single momentum-independent pseudospin matrix need not provide a fermionically antisymmetric conventional pair for both charge parities. The required antisymmetry may instead reside in node space, physical-spin space, or a combination of internal sectors. This is one reason why the full node $\otimes$ pseudospin construction is essential.

{\subsection{No-go for a momentum-independent orbital sewing matrix}\label{app:orbital_nogo}
For the specific two-band model of the main text we can be sharper: \emph{no}
momentum-independent pseudospin-only $U_A$ exists, for any $J$. Requiring
Eq.~\eqref{eq:app_constant_sewing} for all $\bk$ is equivalent to
$U_A\,h_+(\bk)^{*}\,U_A^{\dagger}=h_-(-\bk)$. Using
$h_+(\bk)^{*}=\mathrm{Re}\,z\,\sigma_x-\mathrm{Im}\,z\,\sigma_y+k_z\sigma_z$ and
$h_-(-\bk)=-(-1)^{J}\big[\mathrm{Re}\,z\,\sigma_x+\mathrm{Im}\,z\,\sigma_y\big]
+k_z\sigma_z$ with $z=(k_x+ik_y)^J$, this demands, for even $J$,
$U_A\sigma_xU_A^{\dagger}=-\sigma_x$ together with $\sigma_y,\sigma_z$ left
invariant; but invariance of $\sigma_y$ and $\sigma_z$ forces
$U_A\propto\mathbb{1}$, which contradicts the sign reversal of $\sigma_x$. For odd
$J$ the same argument applies with $\sigma_x\leftrightarrow\sigma_y$. The covariant construction
Eq.~\eqref{eq:app_Ms_sewing} therefore remains available and exact, but a
\emph{local, momentum-independent} realization requires internal structure beyond
the orbital pseudospin. Whether enlarging the internal space by physical spin
supplies it is examined in {App.~\ref{app:spin_singlet_realization}}; we find that it
does not.} \subsection{Normal and anomalous impurity vertices} \label{app:normal_anomalous_vertices} Let the disorder potential in the full internal Hilbert space be denoted by  \begin{equation} \widehat{V}_{\bk\bk'}. \end{equation}  The projected particle-line matrix element at the positive-chirality node is  \begin{equation} U_{+}(\bk,\bk') = \bra{u_{+}(\bk)} \widehat{V}_{\bk\bk'} \ket{u_{+}(\bk')}. \label{eq:app_Uplus} \end{equation}  The corresponding matrix element for the partner state is  \begin{equation} U_{-}(-\bk,-\bk') = \bra{u_{-}(-\bk)} \widehat{V}_{-\bk,-\bk'} \ket{u_{-}(-\bk')}. \label{eq:app_Uminus} \end{equation}  
The normal coherence kernel is proportional to \begin{equation} K_N(\bk,\bk') = \overline{ \left| U_{+}(\bk,\bk') \right|^2 }, \label{eq:app_normal_kernel} \end{equation}  where the overline denotes the disorder average. { The anomalous Born self-energy contains the particle vertex at the positive node and the hole vertex at the negative node. In the Nambu convention used here, the projected anomalous kernel is proportional to \begin{equation} K^{\rm an}_s(\bk,\bk') = U_+(\bk,\bk')\, U_-(-\bk,-\bk')\, \frac{f_s(\bk')}{f_s(\bk)}. \end{equation} Equivalently, since the band-isotropic sewing form factor has unit magnitude, one may write \begin{equation} K^{\rm an}_s(\bk,\bk') = U_+(\bk,\bk')\, U_-(-\bk,-\bk')\, f_s(\bk')f_s^*(\bk). \end{equation} The absence of a complex conjugate on \(U_-\) follows after combining the hole vertex with the reversed momentum indices and using Hermiticity: \begin{equation} U_-^*(-\bk',-\bk)=U_-(-\bk,-\bk'). \end{equation} If the disorder preserves the pairing antiunitary, \begin{equation} \mathcal A \hat V_{\bk\bk'}\mathcal A^{-1} = \hat V_{-\bk,-\bk'}, {\label{eq:app_disorder_invariance}} \end{equation} then \begin{equation} U_-(-\bk,-\bk') = e^{i[\chi(\bk')-\chi(\bk)]} U_+^*(\bk,\bk'). \end{equation} In the constant-matrix sewing convention, \(f_s(\bk)=e^{-i\chi(\bk)}\), up to a momentum-independent phase. Consequently, \begin{align}\label{eq:app_equal_kernels} K^{\rm an}_s(\bk,\bk') &= e^{i[\chi(\bk')-\chi(\bk)]} |U_+(\bk,\bk')|^2 e^{-i[\chi(\bk')-\chi(\bk)]} \nonumber\\ &= |U_+(\bk,\bk')|^2 = K_N(\bk,\bk'). \end{align} Thus the normal and anomalous impurity kernels coincide pointwise.}

Equation~\eqref{eq:app_equal_kernels} is the pointwise coherence-kernel identity required for Anderson protection. An equivalent full-matrix sufficient condition may be written schematically as \begin{equation} \boxed{ \widehat{V}_{\bk\bk'} \widehat{\Delta}_s(\bk') \widehat{V}^{T}_{-\bk,-\bk'} = W_N(\bk,\bk') \widehat{\Delta}_s(\bk) }, \label{eq:app_matrix_identity} \end{equation} where $W_N(\bk,\bk')$ is the same scalar weight entering the normal self-energy. Equation~\eqref{eq:app_matrix_identity} should be understood after projection onto the paired low-energy subspace. A related superconducting-fitness condition is  \begin{equation} F_V(\bk,\bk') \equiv \widehat{V}_{\bk\bk'} \widehat{\Delta}_s(\bk') - \widehat{\Delta}_s(\bk) \widehat{V}^{*}_{-\bk,-\bk'} = 0. \label{eq:app_disorder_fitness} \end{equation} The vanishing of $F_V$ expresses compatibility between the internal structure of the impurity potential and that of the Cooper pair. It is a sufficient condition for the equality of the normal and anomalous impurity renormalizations. 

\subsection{Cancellation in the linearized gap equation} \label{app:anderson_cancellation} Let $\kappa_0$ denote the eigenvalue of the normal impurity kernel acting on the constant form factor, and let $\kappa_s$ denote the corresponding anomalous eigenvalue. Equation~\eqref{eq:app_equal_kernels} implies 
\begin{equation} \kappa_s=\kappa_0. \label{eq:app_equal_eigenvalues} \end{equation}  Therefore,  \begin{equation} \eta_s \equiv \frac{\kappa_s}{\kappa_0} = 1. \label{eq:app_eta_s_proof} \end{equation}  The linearized Born equations then have the same impurity rate in the frequency and gap channels:  \begin{align} \widetilde{\omega}_n &= \omega_n + \Gamma_N \operatorname{sgn}(\omega_n), \label{eq:app_omega_renorm} \\ \widetilde{\Delta}_{s,n} &= \Delta_s + \Gamma_N \frac{ \widetilde{\Delta}_{s,n} }{ |\widetilde{\omega}_n| }. \label{eq:app_gap_renorm} \end{align}  Solving Eq.~\eqref{eq:app_gap_renorm} gives  \begin{equation} \frac{ \widetilde{\Delta}_{s,n} }{ |\widetilde{\omega}_n| } = \frac{ \Delta_s }{ |\omega_n| }. \label{eq:app_anderson_ratio} \end{equation} The impurity rate consequently drops out of the linearized pairing susceptibility. At fixed normal-state density of states, 
\begin{equation} T_c^{(s)} = T_{c0}^{(s)}, \qquad \Gamma_s^{\mathrm{pb}} = \Gamma_N(1-\eta_s) = 0. \label{eq:app_anderson_result} \end{equation}  This is the generalized Anderson cancellation for the conventional band-sewn channel. 

\subsection{Example: valley-even scalar disorder} \label{app:valley_even_example} A simple disorder class satisfying the required condition is a scalar potential that acts identically on the two nodes,  \begin{equation} \widehat{V}_{0}(\br) = V_0(\br)\, \tau_0\otimes\sigma_0. \label{eq:app_valley_even_disorder} \end{equation} If the full pairing antiunitary exchanges the two paired nodes and leaves $\widehat{V}_0$ invariant, then Eq.~\eqref{eq:app_disorder_invariance} holds. The particle and partner-state matrix elements are antiunitary conjugates, and the normal and anomalous coherence kernels coincide. Under these conditions, \begin{equation} \eta_s=1 \end{equation} follows exactly within the Born and projected-band framework. This result requires the same random scalar field to act on the two paired nodes. It need not hold if the disorder contains valley-dependent components. For example, consider valley-odd disorder,  \begin{equation} \widehat{V}_{z}(\br) = V_z(\br)\, \tau_z\otimes\sigma_0. \label{eq:app_valley_odd_disorder} \end{equation}  A node-exchanging antiunitary generally changes the sign of $\widehat{V}_z$:  \begin{equation} \mathcal{A} \widehat{V}_{z} \mathcal{A}^{-1} = -\widehat{V}_{z}. \end{equation}  The two members of the Cooper pair then experience opposite random potentials. The identity in Eq.~\eqref{eq:app_equal_kernels} need not hold, and the conventional channel may acquire a finite pair-breaking rate,  \begin{equation} \eta_s<1, \qquad \Gamma_s^{\mathrm{pb}}>0. \end{equation}  The same caution applies to intervalley disorder proportional to $\tau_x$ or $\tau_y$, for which the complete valley-space impurity ladder must be calculated. 

{\subsection{On a possible realization with explicit physical spin, and why we do not claim one}} \label{app:spin_singlet_realization}
{The previous subsection showed that no momentum-independent pseudospin-only sewing
matrix exists for this model. It is natural to ask whether enlarging the internal space
by physical spin repairs this. If physical spin is retained as an additional
two-component degree of freedom, a momentum-independent candidate gap is}
\begin{equation} \widehat{\Delta}_s = \Delta_s\, \tau_x\otimes\sigma_0\otimes i s_y, \label{eq:app_spin_singlet_gap} \end{equation}
{where $s_i$ acts in physical-spin space. Since}
\begin{equation} \tau_x^{T}=\tau_x, \qquad \sigma_0^{T}=\sigma_0, \qquad (is_y)^{T}=-is_y, \end{equation}
{the full matrix is antisymmetric, $\widehat{\Delta}_s^{T} = -\widehat{\Delta}_s$, for
\emph{both} charge parities, since the antisymmetry now resides in the spin sector rather
than in the pseudospin sector. This removes the even/odd-$J$ \emph{antisymmetry}
obstruction encountered above.}

{We emphasize, however, that this does \emph{not} by itself constitute a microscopic
realization, and we withdraw any such claim. The projection of
Eq.~\eqref{eq:app_spin_singlet_gap} onto the conduction bands is}
\begin{equation}
f_s(\bk)=u_{+}^{\dagger}(\bk)\,\sigma_0\,u_{-}^{*}(-\bk),
\end{equation}
{which equals unity only if the \emph{orbital} sewing relation
$u_{-}(-\bk)=u_{+}^{*}(\bk)$ holds. The spin-singlet factor $is_y$ supplies fermionic
antisymmetry, but it is a spectator in the orbital sector and therefore cannot establish
that relation; indeed, the no-go argument of {App.~\ref{app:orbital_nogo}} shows that for the
two-band Hamiltonian Eq.~\eqref{eq:hpm} the required momentum-independent orbital sewing
does \emph{not} exist. Asserting $u_{-}(-\bk)=u_{+}^{*}(\bk)$ would therefore contradict
that no-go rather than evade it. A genuine local realization would require an enlarged
normal-state Hamiltonian in which a momentum-independent antiunitary $\mathcal A$ satisfies
$\mathcal A\,\mathcal H(\bk)\,\mathcal A^{-1}=\mathcal H(-\bk)$ \emph{and} maps the
positive-node conduction state onto its negative-node partner globally, including all gauge
patches. We have not constructed such a model, and we do not claim one.}

{Accordingly, throughout this work the band-isotropic conventional channel is introduced
\emph{phenomenologically at the projected-band level}, and $\eta_s=1$ is retained as an
explicit assumption whose consequences, when relaxed, are quantified by
Eq.~\eqref{eq:crosscrit}. The band-sewing construction of
{App.~\ref{app:band_sewing_gap}} remains available as an exact but momentum-dependent,
gauge-covariant realization of a constant projected form factor; it should not be read as a
local orbital $s$-wave order parameter.}

\subsection{Scope of the result} \label{app:eta_s_scope} The construction above establishes two logically distinct statements. First, the band-sewing matrix  \begin{equation} M_s(\bk) = u_{+}(\bk)u_{-}^{T}(-\bk) \end{equation}  provides an explicit gap matrix whose conduction-band projection is the constant form factor $f_s=1$. Second, exact Anderson protection additionally requires the impurity potential to preserve the antiunitary sewing relation between the paired states. A sufficient condition is the equality of the normal and anomalous coherence kernels in Eq.~\eqref{eq:app_equal_kernels}, or equivalently the vanishing disorder-fitness condition Eq.~\eqref{eq:app_disorder_fitness}. When this condition holds, 
\begin{equation} \eta_s=1. \end{equation} When it does not hold, the constancy of $f_s$ alone does not prevent pair breaking, and the conventional channel must instead be described by an effective protection factor $\eta_s<1$. This is the situation parameterized by the general two-channel crossing criterion in the main text.


\section{Continuum Ginzburg--Landau estimate}
\label{app:gl}
Writing $\Delta(\kh)=\Delta_sf_s+\Delta_mf_m$ and expanding the free energy to
quartic order in the continuum ($C_\infty$) approximation, {the quartic term is the
Fermi-surface average $\mathcal F_4=\tfrac{b}{2}\avg{|\Delta(\kh)|^4}$ with the single
positive weak-coupling coefficient $b=7\zeta(3)N_0/(8\pi^2T_c^2)$. Because the two channels
carry different azimuthal winding, every term odd in $e^{iJ\phi}$ averages to zero and}
\begin{equation}
{\avg{|\Delta|^4}=\beta_s|\Delta_s|^4+\beta_m|\Delta_m|^4
+4\beta_{sm}|\Delta_s|^2|\Delta_m|^2,}
\label{eq:app_gl_quartic}
\end{equation}
{with}
\begin{equation}
{\beta_s=\avg{|f_s|^4}=1,\qquad \beta_m=\avg{|f_m|^4},\qquad
\beta_{sm}=\avg{|f_s|^2|f_m|^2}=\avg{|f_m|^2}.}
\label{eq:app_gl_coeffs}
\end{equation}
{We emphasize the cross coefficient $4$, which is the origin of the factor of two below
and is easily mis-stated: expanding $|\Delta|^4=(|\Delta_s f_s|^2+|\Delta_m f_m|^2
+2\,{\rm Re}\,\Delta_s^*\Delta_mf_s^*f_m)^2$, the term $2|\Delta_sf_s|^2|\Delta_mf_m|^2$ and
the surviving half of $4({\rm Re}\,\Delta_s^*\Delta_mf_s^*f_m)^2$ each contribute
$2\beta_{sm}$. Comparing Eq.~\eqref{eq:app_gl_quartic} with the canonical two-component form
$\tfrac{\tilde\beta_s}{2}|\Delta_s|^4+\tfrac{\tilde\beta_m}{2}|\Delta_m|^4
+\tilde\beta_{sm}|\Delta_s|^2|\Delta_m|^2$ gives $\tilde\beta_{s,m}=b\beta_{s,m}$ and
$\tilde\beta_{sm}=2b\beta_{sm}$. Homogeneous coexistence of the two condensates is a stable
minimum iff $\tilde\beta_{sm}<\sqrt{\tilde\beta_s\tilde\beta_m}$, i.e.\ iff}
\begin{equation}
R_{\rm clean}\equiv\frac{2\avg{|f_m|^2}}{\sqrt{\avg{|f_m|^4}}}<1 ,
\end{equation}
{so that $R_{\rm clean}>1$ disfavors simple homogeneous coexistence. The ratio} is
invariant under rescaling $f_m\to c\,f_m$ and therefore independent of the
normalization convention. Using the raw moments $\avg{|f_m|^2}=2/(J+2)$ and
$\avg{|f_m|^4}=(2/J)(2/J+2)/[(2/J+1)(2/J+3)]$ (App.~\ref{app:angavg}) gives
\begin{equation}
R_{\rm clean}=1.83,\;1.63,\;1.48 \qquad (J=1,2,3),
\end{equation}
all exceeding unity. {The correct reading of this is narrow: the continuum quartic
overlap disfavors simple coexistence, but \emph{no} conclusion concerning the order of the
transition or the presence of intermediate phases can be drawn without the full
point-group-resolved, disorder-dressed quartic functional.} The continuum expansion drops
the phase-locking term $\propto e^{2iJ\phi}$ that a discrete $C_n$ restores
whenever $n\mid 2J$ (e.g.\ $C_4$ at $J=2$), a complete treatment requires the
representation content of the two order parameters under the actual point group, and disorder renormalizes all quartic coefficients, which we do not compute. The order of the ordered-state transition is left open.

\section{{Lorentzian spectral broadening of the density of states and its
feedback on the crossing}}\label{app:dosfeedback}

This appendix collects the disorder-induced density-of-states feedback. It is kept out of
the main text because it is a phenomenological spectral convolution rather than a
self-consistent Born solution, because it is ultraviolet sensitive for $J=1,2$, and because
none of the primary results uses it.

The scattering rate itself is fixed self-consistently {within a Lorentzian
spectral-broadening scheme},
$\Gamma_N\equiv\hbar/2\tau=\pi\gamma\,N(\mu;\Gamma_N)$ with
$N(\mu;\Gamma_N)=\int d\varepsilon\,N(\varepsilon)\,
(\Gamma_N/\pi)/[(\mu-\varepsilon)^2+\Gamma_N^2]$. {We should be precise about what this
is. Retaining only the imaginary part of the self-energy---i.e.\ a pure Lorentzian
broadening of the density of states---is exact only if the real part
$\mathrm{Re}\,\Sigma$ and the associated shift of the chemical potential vanish, which is
guaranteed by particle--hole symmetry. That symmetry is precisely what the
asymmetric, cutoff-sensitive power-law density of states Eq.~\eqref{eq:dos} violates:
with $N(\varepsilon)\propto|\varepsilon|^{2/J}$ regulated at $\Lambda$, the
Kramers--Kronig partner of the broadened spectrum generically produces a nonzero
$\mathrm{Re}\,\Sigma$ and hence a disorder-induced shift $\delta\mu$, which we do not
compute. We therefore do \emph{not} call this a self-consistent Born calculation. It should be
read as a \emph{phenomenological spectral (Lorentzian) convolution} that captures the
disorder-induced filling of the pseudogap while omitting the real self-energy and the
attendant chemical-potential renormalization. A consistent treatment would compute
$\mathrm{Re}\,\Sigma$ and $\mathrm{Im}\,\Sigma$ with the same ultraviolet regularization
and re-fix $\mu$ at fixed density; we have not done so, and this is one reason the
feedback is kept out of the primary phase diagrams rather than propagated into them. A
further restriction should be noted: for finite-range disorder the normal self-energy is
generally momentum dependent, especially in the forward-scattering regime, so the scalar
convolution written here is consistent only within the intra-pocket momentum-independent
limit of Eq.~\eqref{eq:twoscale}.}
Two cautions apply to this
convolution and we treat it accordingly. First, the Lorentzian (Cauchy) kernel
has no finite second moment and the power law $N(\varepsilon)=N_J|\varepsilon|^{2/J}$
[Eq.~\eqref{eq:dos}] is unbounded, so $N(\mu;\Gamma_N)$ \emph{cannot} be reduced to a
local moment expansion $N(\mu)+\tfrac12\Gamma^2N''(\mu)+\dots$; such an expansion
is non-convergent here. We therefore regulate every {broadening} integral with the
ultraviolet cutoff $\Lambda$ of the effective model of Eq.~\eqref{eq:hpm} and
solve the self-consistency \emph{numerically at fixed $\mu$} {(Sec.~\ref{sec:scba})}. The result corrects the naive picture: $N(\mu;\Gamma_N)$ \emph{rises}
monotonically with disorder for \emph{all} $J=1,2,3$ {within this cutoff-regulated
continuum convolution} (broadening fills in the
pseudogap), not in the rise/flat/fall pattern that a curvature expansion would suggest.
The strength and cutoff-dependence, however, differ sharply by $J$, controlled by
the large-energy tail exponent $2/J-2$ of the integrand
$N(\varepsilon)/[(\mu-\varepsilon)^2+\Gamma^2]$:
\begin{itemize}\itemsep2pt
\item $J=1$ ($2/J-2=0$): the integral is \emph{ultraviolet divergent},
$N(\mu;\Gamma_N)\!\sim\!(\Gamma_N/\pi)N_1\cdot2\Lambda\propto\Lambda$, i.e.\
cutoff-dominated. This is \emph{not} a controlled low-energy quantity---it
signals the breakdown of the unbounded continuum-DOS model---so we quote \emph{no}
numerical $J=1$ DOS-feedback value and disable this feedback everywhere below (as
for the phase diagram). The divergence is an artifact of the unbounded linear
dispersion, not of the physics: on a lattice the Weyl cone terminates in a band of
finite width $W$, the density of states turns over and vanishes at the band edges,
and the Lorentzian convolution is cut off by the band rather than by an arbitrary
$\Lambda$. A tight-binding regularization therefore replaces the arbitrary continuum cutoff by
material-specific spectral weight distributed over a finite band: the convolution
becomes a finite integral controlled by the full lattice density of states and the
bandwidth, rather than by $\Lambda$. We deliberately do not reduce this to a scaling
formula, since the result depends on the entire band structure and its DOS
normalization, not on the nodal theory. The point is simply that the $J=1$ pathology
is a continuum artifact with a known lattice cure, and {that a material specific lattice-based} evaluation is the natural next step for quantitative $J=1$ predictions.
\item $J=2$ ($2/J-2=-1$): \emph{logarithmically} cutoff-divergent,
$N(\mu;\Gamma_N)-N(\mu)\propto\Gamma_N\ln(\Lambda/\Gamma_N)$---mild but cutoff-sensitive
($N/N_0\simeq1.16$--$1.29$ over the same $\Lambda$ range).
\item $J=3$ ($2/J-2=-4/3<-1$): ultraviolet \emph{convergent}, a small
cutoff-insensitive rise ($N/N_0\simeq1.06$--$1.10$).
\end{itemize}
We emphasize that this feedback is \emph{not} numerically negligible even for
$J=2$: a $16$--$29\%$ change in $N_0$ enters $T_{c0}\propto e^{-1/\lambda}$
exponentially and can shift $T_{c0}$ substantially. We therefore do not claim it is
small; we \emph{isolate} the pair-breaking mechanism by disabling the feedback in
all phase diagrams, and report the DOS calculation separately. We also do \emph{not} claim that the feedback cancels in the clean ratio
$r=T_{c0}^{(s)}/T_{c0}^{(m)}$: in weak coupling
$T_{c0}^{(a)}=A\Lambda\exp[-1/(g_aN\avg{|f_a|^2})]$, so
\begin{equation}
\ln r(N)=\frac{1}{\lambda_m(N)}-\frac{1}{\lambda_s(N)},
\label{eq:lnr}
\end{equation}
which is \emph{exponentially sensitive} to a common change of $N$ whenever the two
channel couplings differ---a $16$--$29\%$ shift in $N$ can move $r$ appreciably.
The decomposition used here is therefore a \emph{deliberate simplification}, not a
cancellation: we hold $r$ fixed by construction in order to isolate the
pair-breaking mechanism, and we report the DOS calculation separately. If the
feedback were included, $r$ would become $\gamma$-dependent and the crossing
criterion Eq.~\eqref{eq:crosscrit} would hold locally, as an implicit condition on
$r(\Gamma_N)$, rather than at the displayed constant $r$. For $J=1$ we make no
quantitative DOS-feedback claim at all. (The near-node
non-Anderson physics at $\varepsilon\to0$~\cite{BeraSauRoy2016,SyzranovRadzihovsky2018}
is regulated here by the finite Fermi energy, $\Gamma_N\ll\mu$; Sec.~\ref{sec:scba}.
All results below are at fixed $\mu$; fixing the carrier density instead would
shift $\mu(\gamma)$, a refinement we do not pursue here.)

\paragraph*{{Coupled Lorentzian-DOS-plus-AG test}: does the crossing survive the DOS feedback?}
Because the disorder-induced change of $N(\mu;\Gamma_N)$ is not physically separable
from the pair-breaking it accompanies, the decoupling used above must be tested. We
therefore solve a \emph{coupled} problem for $J=2,3$: the {self-consistent broadening} rate
$\Gamma_N=\pi\gamma N(\mu;\Gamma_N)$ self-consistently, with $N(\mu;\Gamma_N)$ fed
back into \emph{both} clean scales through
$T_{c0}^{(a)}(\Gamma_N)=\tfrac{2e^{\gamma_E}}{\pi}\Lambda\,
e^{-1/[g_aN(\mu;\Gamma_N)\avg{|f_a|^2}]}$, each channel then suppressed by its own
AG rate. We are explicit about the scope of the word ``coupled'': the normal-state
DOS and the linearized $T_c$ equations are solved together, whereas the
ordered-state gap equation, the impurity $T$-matrix, and a fixed-density
chemical potential are \emph{not}. Couplings are fixed by matching the clean
$T_{c0}^{(m)}=0.133\,\mu$ and $r=0.625$ at each cutoff, at fixed $\mu$.

\begin{table}[h]
\caption{{Coupled Lorentzian-DOS-plus-AG} diagnostics. $\Gamma_N^\ast$ normalized both to the
\emph{clean} $T_{c0}^{(m)}$ and to the \emph{local} disorder-dressed scale
$T_{c0}^{(m)}(\Gamma_N^\ast)$, which is the coordinate actually entering the AG
equation; $r(\Gamma_N^\ast)$ is the disorder-dressed clean ratio; $T^\ast$ is the
crossing temperature. {The crossing rate is also given in units of $\mu$, since that is
the ratio controlling the validity of the projected {Born treatment}. We deliberately do \emph{not}
quote a coupled monopole-destruction rate: in every case tested the formal destruction
point lies at $\Gamma_N\gtrsim0.18\,\mu$---and for the largest cutoffs at
$\Gamma_N\sim8\,\mu$---i.e.\ far outside the regime $\Gamma_N\ll\mu$ in which the
calculation is meaningful, so those numbers would be extrapolations of the scheme beyond
its own domain rather than predictions. $^{\dagger}$The $J=2$, $\Lambda/\mu=20$ entry has
$\Gamma_N^\ast/\mu=0.169$, which is no longer comfortably within $\Gamma_N/\mu\ll1$; that
parameter set should not be used to support claims of controlled metallic behavior.} All crossings occur at $T_c^{(m)}>0$.}
\label{tab:feedback}
\begin{ruledtabular}
\begin{tabular}{cccccccc}
$J$ & $\Lambda/\mu$ & $\dfrac{\Gamma_N^\ast}{T_{c0}^{(m)}}$ &
$\dfrac{\Gamma_N^\ast}{T_{c0}^{(m)}(\Gamma_N^\ast)}$ & $N/N_0$ &
$r(\Gamma_N^\ast)$ & {$\dfrac{\Gamma_N^\ast}{\mu}$} & {monopole destruction}\\
\hline
2 & 10 & 0.795 & 0.558 & 1.087 & 0.649 & {0.106} & {outside controlled range}\\
2 & 20 & 1.269 & 0.514 & 1.214 & 0.679 & {0.169$^{\dagger}$} & {outside controlled range}\\
3 & 10 & 0.602 & 0.547 & 1.022 & 0.631 & {0.080} & {outside controlled range}\\
3 & 20 & 0.659 & 0.540 & 1.040 & 0.636 & {0.088} & {outside controlled range}\\
3 & 40 & 0.740 & 0.533 & 1.060 & 0.642 & {0.098} & {outside controlled range}\\
\end{tabular}
\end{ruledtabular}
\end{table}

\begin{figure}[t]
\centering
\includegraphics[width=0.99\columnwidth]{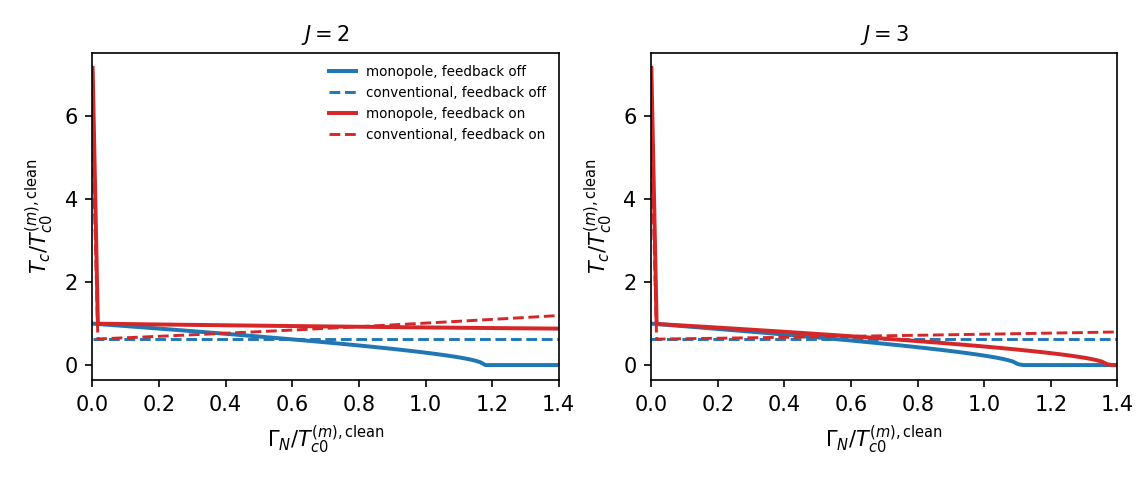}
\caption{{Coupled Lorentzian-DOS-plus-AG} instability crossing (red) compared with the
decoupled treatment (blue), for $J=2$ (left) and $J=3$ (right) at $\Lambda/\mu=10$.
{The calculation is self-consistent only within the restricted DOS-broadening scheme of
Sec.~\ref{sec:scba}, which retains the imaginary self-energy alone.}
Solid: monopole channel; dashed: conventional channel. Including the
disorder-dressed $N(\mu;\Gamma_N)$ in both clean scales raises both curves and
displaces the crossing to a larger rate, but the crossing persists. Temperatures
and rates are normalized to the \emph{clean} $T_{c0}^{(m)}$. Couplings are fixed by
matching $T_{c0}^{(m)}=0.133\,\mu$ and $r=0.625$ in the clean limit at each cutoff,
at fixed $\mu$; see Table~\ref{tab:feedback} for the diagnostics.}
\label{fig:selfcons}
\end{figure}

The crossing survives in every case tested (Fig.~\ref{fig:selfcons}). {Two features of the table matter.} First, in the
\emph{local} coordinate the crossing barely moves, $\Gamma_N^\ast/
T_{c0}^{(m)}(\Gamma_N^\ast)=0.51$--$0.56$ throughout: most of the apparent
displacement in the clean-normalized column is the rise of the pairing scale itself,
not a change in the pair-breaking competition. Second, and consistently with
Eq.~\eqref{eq:lnr}, the feedback \emph{does} change the clean ratio, $r$ drifting
from $0.625$ to $0.63$--$0.68$.

We therefore state the conclusion narrowly. \emph{These coupled solutions verify
that the crossing survives for the chosen $\lambda_m,\lambda_s$, at fixed $\mu$, for
$J=2,3$, over the cutoff ranges listed.} We do \emph{not} claim that a common DOS
factor protects the crossing in general: because $\ln r(N)=1/\lambda_m(N)-
1/\lambda_s(N)$ [Eq.~\eqref{eq:lnr}] is exponentially sensitive whenever the channel
couplings differ, a sufficiently strong feedback could in principle move $r$ across
the boundary Eq.~\eqref{eq:crosscrit}. Note also that the rise of both $T_{c0}$'s is
a property of the regulated convolution used here, not a universal consequence of
disorder. The \emph{location} is moreover cutoff sensitive for the logarithmically
divergent $J=2$ case ($0.795\to1.269$ as $\Lambda/\mu$ doubles) and mild for the
convergent $J=3$ case; this is why the cutoff-independent decoupled presentation is
retained for the quantitative phase diagrams, and why no such calculation is
attempted for $J=1$.

\end{document}